\documentclass[12pt]{article}

\usepackage{amsmath}
\usepackage{amssymb}
\usepackage{a4wide}
\usepackage{amsmath,amsthm,amssymb,xspace}
\newtheorem{Theorem}{Theorem}[section]
\newtheorem{Lemma}{Lemma}[section]
\newtheorem{Remark}{Remark}[section]
\newtheorem{Proposition}{Proposition}[section]
\newtheorem{Definition}{Definition}[section]

\newcommand{\oneder}[2]{\ensuremath{{\mathop{{%
            \Rightarrow}}\limits^{{#1}}}_{\!\!_{#2}}\!}}
\newcommand{\multider}[1]{\ensuremath{{\mathop{{ %
            \Rightarrow}}\limits^{{#1}}}_{\!_\Re} %
            \raisebox{2pt}{\!\!\!\scriptsize *}}}
\newcommand{\multidernorm}[2]{\ensuremath{{\mathop{{%
          \Rightarrow}}\limits^{{#1}}}%
       \raisebox{2pt}{\!\scriptsize *}}\!_{\!_{#2}}\,}
\newcommand{\prsoneder}[1]{\ensuremath{\oneder{#1}{\Re}}}
\newcommand{\onederparK}[1]{\ensuremath{\oneder{#1}{\Re^{K}_{PAR}}}}
\newcommand{\onederparKomega}[1]{\ensuremath{\oneder{#1}{\Re^{K,K^{\omega}}_{PAR}}}}
\newcommand{\onederseqK}[1]{\ensuremath{\oneder{#1}{\Re^{K}_{SEQ}}}}

\newcommand{\multiderparK}[1]{\ensuremath{\multidernorm{#1}{\Re^{K}_{PAR}}}}
\newcommand{\multiderparKomega}[1]{\ensuremath{\multidernorm{#1}{\Re^{K,K^{\omega}}_{PAR}}}}
\newcommand{\multiderseqK}[1]{\ensuremath{\multidernorm{#1}{\Re^{K}_{SEQ}}}}
\newcommand{\Rule}[1]{\ensuremath{{\mathop{{%
            \rightarrow}}\limits^{{#1}}}}}

\def\prsder#1#2#3{{#1\,\multidernorm{#2}{\Re}#3}}
\def\prsderpar#1#2#3{{#1\,\multidernorm{#2}{\Re_{PAR}^{K}}#3}}

\def\prsrule#1#2#3{{#1\mathop{{\rightarrow}}\limits^{{#2}}#3}}
\def\prslongrule#1#2#3{{#1\mathop{{\rightarrow}}\limits^{{#2}}#3}}

\def\prsdernorm#1#2#3#4{{\,#1\,\multidernorm{#2}{#3}#4}}

\newcommand{\maximal}[1]{$\Upsilon^{f}_{M}({#1})$}
\newcommand{\maximalG}[2]{$\Upsilon^{f}_{#1}({#2})$}
\newcommand{\maximalGInf}[2]{$\Upsilon^{\infty}_{#1}({#2})$}
\newcommand{\maximalseq}[1]{$\Upsilon^{f}_{M_{SEQ,K}}({#1})$}

\newcommand{\maximalpar}[1]{$\Upsilon^{f}_{M_{PAR,K}}({#1})$}
\newcommand{\maximalparK}[1]{$\Upsilon^{f}_{M^{K}_{PAR}}({#1})$}
\newcommand{\maximalparKomega}[1]{$\Upsilon^{f}_{M^{K,K^{\omega}}_{PAR}}({#1})$}
\newcommand{\maximalparKomegaPlus}[1]{$\Upsilon^{f}_{M^{K,K^{\omega}}_{PAR,\infty}}({#1})$}
\newcommand{\maximalparKomegaInf}[1]{$\Upsilon^{\infty}_{M^{K,K^{\omega}}_{PAR}}({#1})$}
\newcommand{\maximalseqK}[1]{$\Upsilon^{f}_{M^{K}_{SEQ}}({#1})$}
\newcommand{\maximalseqInfK}[1]{$\Upsilon^{\infty}_{M_{SEQ}^{K}}({#1})$}
\newcommand{\maximalInf}[1]{$\Upsilon^{\infty}_{M}({#1})$}

\def\parcomp#1#2{{#1\!\parallel\!#2}}
\def\seqcomp#1#2{{#1.(#2)}}

\def\PRS{\text{{\em PRS}}\xspace}
\def\PRSs{\text{{\em PRS\/}s}\xspace}
\def\BRS{\text{{\em BRS}}\xspace}

\def\ALTL{\text{{\em ALTL}}\xspace}

\def\MBRS{\text{{\em MBRS}}\xspace}
\def\MBRSs{\text{{\em MBRS\/}s}\xspace}

\def\npla#1{{\langle #1\rangle}}

\def\ldsrule#1#2#3{\ensuremath{
   \cfrac[c]{#1}
    {\ #2\ }\text{\footnotesize \ \ensuremath{#3}}}}

    \def\suffix#1{\ensuremath{\text{\emph{suffix}}}(#1)}
\def\firstact#1{\ensuremath{\text{\emph{firstact}}}(#1)}

\title{Verification of Process Rewrite Systems in normal form}
\author{
L.~Bozzelli$^1$ \\[10pt]
{\small\begin{tabular}{c@{\hspace{0cm}}c}
$\!\!^1$Dept. of Mathematics and Applications \\
Universit\`a di Napoli ``Federico II'' \\
Napoli, Italy \\
{\tt laura.bozzelli@dma.unina.it}
\end{tabular}
}}

\date{}

\begin{document}
\maketitle
\begin{abstract}
  We consider the problem of model--checking
  for Process Rewrite Systems (\PRSs) in normal form. In a \PRS in normal form
  every rewrite rule either only deals with procedure calls and procedure
  termination, possibly with value return,  (this kind of rules
  allows to capture Pushdown Processes), or only deals with dynamic
 activation of processes and synchronization (this kind of rules
  allows to capture Petri Nets). The model-checking problem for
  \PRSs and action-based linear temporal logic (\ALTL) is
  undecidable. However, decidability of model--checking for \PRSs and
  some interesting fragment of \ALTL remains an open question.
   In this paper  we state decidability results concerning
   generalized acceptance properties about infinite derivations (infinite term
  rewritings) in \PRSs in normal form. As a consequence, we obtain
  decidability of the model-checking (restricted to infinite runs)
  for \PRSs in normal form and a meaningful fragment of \ALTL.
\end{abstract}

\section{Introduction}
Automatic verification of systems is nowadays one of the most
investigated topics. A major difficulty to face when considering
this problem comes to the fact that, reasoning about systems in
general may require dealing with infinite state models. For
instance, software systems may introduce infinite states both
manipulating data ranging over infinite domains, and having
unbounded control structures such as recursive procedure calls
and/or dynamic creation of concurrent processes (e.g.
multi--treading). Many different formalisms have been proposed for
the description of infinite state systems. Among the most popular
are the well known formalisms of Context Free Processes, Pushdown
Processes, Petri Nets, and Process Algebras. The first two are
models of sequential computation, whereas Petri Nets and Process
Algebra explicitly take into account concurrency.  The model
checking problem for these infinite state formalisms have been
studied in the literature.  As far as Context Free Processes and
Pushdown Automata are concerned (see
\cite{ay98,bouajjani97,BS94,burkart01,hun94,ehrs00,walukiewicz96}),
decidability of the modal $\mu$--calculus, the most powerful of
the modal and temporal logics used for verification, has been
established (e.g. see \cite{BS94,hun94}). In
\cite{esp97,esparza94}, model checking for Petri nets has been
studied. The branching temporal logic as well as the state-based
linear temporal logic are undecidable even for restricted logics.
Fortunately, the model checking for action-based linear temporal
logic (\ALTL) is decidable.\newline

Verification of formalisms which accommodate both parallelism and
recursion is a challenging problem. To formally study this kind of
systems, recently the formal framework of Process Rewrite Systems
(\PRSs) has been introduced \cite{mayr98}.  This framework, which
is based on term rewriting, subsumes many common infinite states
models such us Pushdown Systems, Petri Nets, Process Algebra, etc.
 The decidability results already
known in the literature for the general framework of \PRSs
concerns reachability analysis. However, the model checking of
action-based temporal logic becomes undecidable. It remains
un\-de\-ci\-dable even for restricted models such as those
presented in
 \cite{bouajjani96}.\newline

In this paper we extend the known decidability results, for a
relevant syntactic fragment of \PRSs, to properties of infinite
derivations, thus allowing for automatic verification of some
interesting classes of action-based linear time properties. The
fragment we consider is that of \PRSs in normal form, where every
rewrite rule either only deals with procedure calls and procedure
termination, possibly with value return,  (this kind of rules
allows to capture Pushdown Processes), or only deals with dynamic
activation of processes and synchronization (this kind of rules
allows to capture Petri Nets).\newline Our result extends our
previous result established in \cite{Bozz03}, and regards the
decidability of two problems: the first (resp., the second)
concerns generalized acceptance properties of finite derivations
(resp., infinite derivations) in \PRSs in normal form. As a
consequence we obtain decidability of the model-checking
(restricted to infinite executions) for \PRSs in normal form and a
meaningful \ALTL fragment.\newline
 The rest
of the paper is structured as follows.    In Section 2, we recall
the framework of Process Rewrite Systems, we summarize some
decidability results for reachability problems, and our previous
result for \PRSs in normal form. In Section 3, it is shown how our
decidability results about generalized acceptance properties of
infinite derivations in \PRSs in normal form can be used in
model-checking for a meaningful \ALTL fragment. In Section 4, we
prove decidability of the two problems about finite and infinite
derivations in $\PRSs$ in normal form, mentioned above. Appendix
contains detailed proof of our results.

\section{Process Rewrite Systems}

In this section we recall the framework of \emph{Process Rewrite
Systems} (\PRSs). We also recall the notion of \emph{B\"{u}chi
Rewrite System} (\BRS) introduced in \cite{Bozz03} to prove
decidability of the model--checking problem for some classes of
linear time properties
 and \PRSs \emph{in normal form}. We conclude
this section by summarizing some decidability results on \PRSs,
known in the literature,  that will be exploited in further
sections of the paper.

\subsection{Process Rewrite Systems and B\"{u}chi Rewrite Systems}

In this subsection we recall the notion of Process Rewrite System,
as introduced in~\cite{mayr98}. The idea is that a process (and
its current state) is described by a term. The behavior of a
process is given by rewriting the corresponding term by means of a
finite set of rewrite rules.

\begin{Definition}[Process Term] Let $Var$ be a finite set of process
  variables. The set $T$ of {\em process terms} over $Var$ is
  inductively defined as follows:
\begin{itemize}
\item $Var\subseteq T$
 \item $\varepsilon\in{}T$
  \item
 $t_{1}$$\parallel$$t_{2}\in{}T$, for all $t_{1},t_{2}\in{}T$
\item  $X.(t)\in{}T$, for all $X\in{}Var$ and $t\in{}T$
\end{itemize}
where $\varepsilon$ denotes the empty term, ``\,$\parallel$''
denotes parallel composition, and ``$.()$'' denotes sequential
composition\footnote{\cite{mayr98} also allows terms of the form
  $t_{1}.(t_{2})$, where $t_{1}$ is a parallel composition of
  variables. In the current context this generalization is not
  relevant.}.
\end{Definition}

We denote by $T_{SEQ}$ the subset of terms in $T$ devoid of any
occurrence of parallel composition operator, and by $T_{PAR}$ the
subset of terms in $T$ devoid of any occurrence of the sequential
composition operator. Notice that we have $T_{PAR}\cap T_{SEQ} =
Var\cup\{\varepsilon\}$.

In the rest of the paper we only consider process terms modulo
commutativity and associativity of ``$\parallel$'', moreover
$\varepsilon$ will act as the identity for both parallel and
sequential composition.
 Therefore, we introduce the relation
$\approx_{T}$, which is the smallest equivalence relation on $T$
such that for all $t_1,t_2, t_3\in{}T$ and $X\in{}Var$:
\begin{itemize}
\item $t_{1}$$\parallel$$t_{2}\approx_{T}t_{2}$$\parallel$$t_{1}$,
  $t_{1}$$\parallel$$(t_{2}$$\parallel$$t_3)\approx_{T}(t_{1}$$\parallel$$t_{2})$$\parallel$$t_3$,
  and
    $t_{1}$$\parallel$$\varepsilon\approx_{T}t_{1}$.
\item $X.(\varepsilon)\approx_{T}X$, and if
$t_{1}\approx_{T}t_{2}$,
  then $X.(t_{1})\approx_{T}X.(t_{2})$.
\end{itemize}

In the paper, we always confuse terms and their equivalence
classes (w.r.t.  $\approx_{T}$). In particular, $t_1 = t_2$
(resp., $t_1 \not= t_2$) will be used to mean that $t_1$ is
equivalent (resp., not equivalent) to $t_2$.

\begin{Definition}[Process Rewrite System]
  A {\em Process Rewrite System} $\mathrm{(}$or \PRS, or {\em Rewrite System}$\mathrm{)}$ over the
  alphabet $\Sigma$ and the set of process variables $Var$ is a finite
  set of rewrite rules $\Re\subseteq{}T\times\Sigma\times{}T$ of the
  form $\prsrule{t}{a}{t'}$, where $t$ $\mathrm{(}$$\neq\varepsilon$$\mathrm{)}$ and $t'$
  are terms in $T$, and $a\in\Sigma$.
\end{Definition}

The semantics of a \PRS $\Re$ is given by a Labelled Transition
System $\npla{T,\Sigma,\prslongrule{}{}{}}$, where the set of
states is the set of terms $T$ of $\Re$, the set of actions is the
alphabet $\Sigma$ of $\Re$, and the transition relation
$\prslongrule{}{}{} \subseteq T \times \Sigma \times T$ is the
smallest relation satisfying the following inference rules:

\ldsrule{}{\prslongrule{t}{a}{t'}}{(\prsrule{t}{a}{t'})\in\Re}\hspace*{1cm}
\ldsrule{\prslongrule{t_1}{a}{t_1'}}
     {\prslongrule{\parcomp{t_1}{t}}{a}{\parcomp{t_1'}{t}}}{\forall{}t\in{}T}
     \hspace*{1cm}
\ldsrule{\prslongrule{t_1}{a}{t_1'}}
     {\prslongrule{\seqcomp{X}{t_1}}{a}{\seqcomp{X}{t_1'}}}{\forall{}X\in{}Var}
\newline\newline

 For a \PRS $\Re$ with set of terms $T$ and LTS
$\npla{T,\Sigma,\prslongrule{}{}{}}$, a \emph{path in} $\Re$ from
$t\in{}T$ is a path in $\npla{T,\Sigma,\prslongrule{}{}{}}$ from
$t$, i.e. a (finite or infinite) sequence of LTS edges
$\prsrule{t_{0}}{a_{0}}{t_{1}} \prsrule{}{a_{1}}{t_{2}}
\prsrule{}{a_{2}}{}$ such that $t_0=t$ and
$\prsrule{t_{j}}{a_{j}}{t_{j+1}} \in \prsrule{}{}{}$ for any $j$.
A \emph{run in $\Re$} from $t$ is a maximal path from $t$, i.e. a
path from $t$ which is either infinite or has the form
$\prsrule{t_{0}}{a_{0}}{t_{1}} \prsrule{}{a_{1}}{\ldots}
\prsrule{}{a_{n-1}}{t_n}$ and there is no edge
$\prsrule{t_n}{a_n}{t'} \in \prsrule{}{}{}$, for any $a_n\in
\Sigma$ and $t' \in T$. We write $runs_{\Re}(t)$ (resp.,
$runs_{\Re,\infty}(t)$) to refer to the set of runs (resp.,
infinite runs) in $\Re$ from $t$, and $runs(\Re)$ to refer to the
set of all the runs in $\Re$.

\bigskip

The LTS semantics induces, for a rule $r\in \Re$, the following
notion of \emph{one--step derivation} by $r$. The \emph{one--step
derivation} by $r$ relation, \prsoneder{r}, is the least relation
such that:
\begin{itemize}
\item $t$ \prsoneder{r} $t'$, for $r=\prsrule{t}{a}{t'}$ \item
$t_{1}$$\parallel$$t$ \prsoneder{r} $t_{2}$$\parallel$$t$, if
$t_{1}$ \prsoneder{r} $t_{2}$ and $t\in{}T$ \item $X.(t_{1})$
\prsoneder{r} $X.(t_{2})$, if $t_{1}$ \prsoneder{r} $t_{2}$ and
$X\in{}Var$
\end{itemize}

A \emph{finite derivation} in $\Re$ from a term $t$ to a term $t'$
(through a finite sequence $\sigma = r_{1}r_{2}\ldots{}r_{n}$ of
rules in $\Re$), is a sequence $d$ of one--step derivations $t_0$
\prsoneder{r_1} $t_1$ \prsoneder{r_2} $t_2\ldots$
\prsoneder{r_{n-1}} $t_{n-1}$ \prsoneder{r_n} $t_n$, with $t_0=t$,
$t_n=t'$ and $t_i$ \prsoneder{r_{i+1}} $t_{i+1}$ for all
$i=0,\ldots,n-1$. The derivation $d$ is a \emph{$n$--step
derivation} (or a \emph{derivation of length $n$}), and for
succinctness is denoted by $t$ \multider{\sigma} $t'$. Moreover,
we say that $t'$ is \emph{reachable} in $\Re$ from  term $t$
(through derivation $d$). If $\sigma$ is empty, we say that $d$ is
a {\em null derivation\/}.
\newline
A \emph{infinite derivation} in $\Re$ from a term $t$ (through an
infinite sequence $\sigma = r_{1}r_{2}\ldots $ of rules in $\Re$),
is an infinite sequence of one step derivations $t_0$
\prsoneder{r_1} $t_1$ \prsoneder{r_2} $t_2\ldots$ such that
$t_0=t$ and $t_i$ \prsoneder{r_{i+1}} $t_{i+1}$  for all $i \geq
0$. For succinctness such derivation is denoted by $t$
\multider{\sigma}.
\newline
Notice that there is a strict correspondence between the notion of
derivation from a term $t$ and that of path from the term $t$. In
fact, there exists a path $\prsrule{t_{0}}{a_{0}}{t_{1}}
\prsrule{}{a_{1}}{t_{2}\ldots}$  from $t_0$ in $\Re$ iff there
exists a derivation $t_0$ \prsoneder{r_1} $t_1$ \prsoneder{r_2}
$t_2\ldots$ from $t_0$ in $\Re$, with $a_i = label(r_i)$, for any
$i$ (where for a rule $r \in \Re$ with $r = \prsrule {t}{a}{t'}$,
$label(r)$ denotes the label $a$ of $r$).

\bigskip

In the following, we shall consider \PRSs in a syntactical
restricted form called \emph{normal form}.

\begin{Definition}[Normal Form] A \PRS $\Re$ is said to be in {\em
    normal form} if every rule $r\in\Re$ has one of the following forms:
\begin{description}
\item[PAR rules:] Any rule devoid of sequential composition;
\item[SEQ rules:] $\prsrule{X}{a}{Y.(Z)}$, $\prsrule{X.(Y)}{a}{Z}$
or
  $\prsrule{X}{a}{Y}$, or $\prsrule{X}{a}{\varepsilon}$.
\end{description}
\noindent with $X,Y,Z\in{}Var$. A \PRS where all the rules are SEQ
rules is called \emph{sequential} \PRS. Similarly, a \PRS where
all the rules are PAR rules is called \emph{parallel} \PRS.
\end{Definition}

The sequential and parallel fragments of $PRS$ are significant: in
\cite{mayr98} it is shown that sequential \PRS\/s are semantically
equivalent (via \emph{bisimulation equivalence}) to Pushdown
Processes, while parallel \PRS\/s are semantically equivalent to
Petri Nets. Moreover, from the fact that Pushdown systems and
Petri Nets are not comparable (see \cite{M96,BCS96}) it follows
that \PRSs in normal form are strictly more expressive than both
their sequential and parallel fragment. So, the following result
holds:

\begin{Proposition}
\PRSs in normal form are strictly more expressive than Petri nets
and Pushdown Processes.
\end{Proposition}

Now, let us extend the notion of \PRS to that of B\"{u}chi Process
Rewrite System (\BRS). Intuitively, a \BRS is a \PRS where we can
distinguish between non--accepting rewrite rules and accepting
rewrite rules.

\begin{Definition}[B\"{u}chi Rewrite System]
  A \emph{B\"{u}chi Rewrite System} $\mathrm{(BRS)}$ over
  a finite set of process variables $Var$ and an alphabet $\Sigma$ is
  a pair $\npla{\Re,\Re_{F}}$, where $\Re$ is a \PRS over $Var$
  and $\Sigma$, and
  $\Re_{F}\subseteq\Re$ is the set of accepting rules.
\end{Definition}

A B\"{u}chi Rewrite System $\npla{\Re,\Re_{F}}$ is called a \BRS
\emph{in normal form} (resp., \emph{sequential} \BRS,
\emph{parallel} \BRS), if the underlying process rewrite system
$\Re$ is a \PRS in normal form (resp., parallel \PRS, sequential
\PRS).
\newline
\begin{Definition}[Acceptance in B\"{u}chi Rewrite Systems]\label{Def:OldAcceptance}
  Let us consider a \BRS $M=\npla{\Re,\Re_{F}}$. An \emph{infinite}
  derivation $t$ \multider{\sigma} in $\Re$ from $t$ is said to
  be \emph{accepting} $\mathrm{(}$in $M\mathrm{)}$ if
  $\sigma$ contains infinite occurrences of accepting
  rules.\newline
  A \emph{finite}
  derivation $t$ \multider{\sigma} $t'$ in $\Re$ from $t$ is said to
  be \emph{accepting} $\mathrm{(}$in $M\mathrm{)}$ if
  $\sigma$ contains some occurrence of accepting rule.
\end{Definition}

\subsection{Decidability results for \PRSs}\label{sec:OldResults}

In this section we will summarize decidability results on \PRSs
which are known in the literature, and which will be exploited in
further sections of the paper.

\paragraph{Verification of ALTL (Action--based LTL)}\

Given a finite set $\Sigma$ of atomic propositions, the set of
formulae $\varphi$ of ALTL over $\Sigma$ is defined as follows:
\begin{displaymath}
\textrm{$\varphi::= true$ $|$ $\neg\varphi$ $|$
$\varphi_{1}\wedge\varphi_{2}$ $|$
  $\npla{a}\varphi$
  $|$ $\varphi_{1}U\varphi_{2}$ $|$ $G\varphi$ $|$ $F\varphi$}
\end{displaymath}

\noindent where $a\in\Sigma$.

In order to give semantics to ALTL formulae on a \PRS $\Re$, we
need some additional notation. Given a path $\pi=t_{0}$
$\mathop{\rightarrow}\limits^{a_{0}}$ $t_{1}$
$\mathop{\rightarrow}\limits^{a_{1}}$ $t_{2}$
$\mathop{\rightarrow}\limits^{a_{2}}\ldots$ in $\Re$, $\pi^{i}$
denotes the suffix of $\pi$ starting from the $i$--th term in the
sequence, i.e. the path $t_{i}$
$\mathop{\rightarrow}\limits^{a_{i}}$ $t_{i+1}$
$\mathop{\rightarrow}\limits^{a_{i+1}}\ldots$. The set of all the
suffixes of $\pi$ is denoted by $\suffix{\pi}$ (notice that if
$\pi$ is a run in $\Re$, then $\pi^{i}$ is also a run in $\Re$,
for each $i$.)  If the path $\pi=t_{0}$
$\mathop{\rightarrow}\limits^{a_{0}}$ $t_{1}$
$\mathop{\rightarrow}\limits^{a_{1}}\ldots$ is \emph{non--trivial}
(i.e., the sequence contains at least two terms) $\firstact{\pi}$
denotes $a_{0}$, otherwise we set $\firstact{\pi}$ to an element
non in $\Sigma$, say it 0.

\ALTL formulae over a \PRS $\Re$ are interpreted in terms of the
set of the runs in $\Re$ satisfying the given \ALTL formula.  The
\emph{denotation of a formula $\varphi$} relative to $\Re$, in
symbols $[[\varphi]]_{\Re}$, is  defined inductively as follows:
\begin{itemize}
\item $[[true]]_{\Re}= runs(\Re)$ \item
$[[\neg\varphi]]_{\Re}=runs(\Re)\setminus[[\varphi]]_{\Re}$ \item
$[[\varphi_{1}\wedge\varphi_{2}]]_{\Re}=
  [[\varphi_{1}]]_{\Re}\,\cap\,[[\varphi_{2}]]_{\Re}$
\item $[[\npla{a}\varphi]]_{\Re}=\{\pi\in{}runs(\Re)\ \mid\
  \firstact{\pi}=a \text{ and } \pi^{1}\in{}[[\varphi]]_{\Re}\}$
\item $[[\varphi_{1}U\varphi_{2}]]_{\Re} = \{\begin{array}[t]{l}
    \pi\in{}runs(\Re)\ \mid\ \text{for some } i\geq 0,
    \pi^{i} \text{ is defined and }
 \pi^{i}\in{}[[\varphi_{2}]]_{\Re}, \text{ and }\\ %
    \text{for all } j<i, \pi^{j}\in{}[[\varphi_{1}]]_{\Re}\
    \}\end{array}$
\item $[[G\varphi]]_{\Re}=\{\pi\in{}runs(\Re)\ \mid\ \suffix{\pi}
  \subseteq [[\varphi]]_{\Re}\}$
\item $[[F\varphi]]_{\Re}=\{\pi\in{}runs(\Re)\ \mid\
  \suffix{\pi}\cap[[\varphi]]_{\Re}\not=\emptyset\}$
\end{itemize}

For any term $t\in{}T$ and \ALTL formula $\varphi$, we say that
$t$ satisfies $\varphi$ (resp., satisfies $\varphi$ restricted to
infinite runs) (w.r.t $\Re$), in symbols $t \models_{\Re}\varphi$
(resp., $t \models_{\Re,\infty}\varphi$), if
$runs_{\Re}(t)\subseteq[[\varphi]]_{\Re}$ (resp.,
$runs_{\Re,\infty}(t)\subseteq[[\varphi]]_{\Re}$).

The model-checking problem (resp., model--checking problem
restricted to infinite runs) for \ALTL and \PRSs is the problem of
deciding if, given a \PRS $\Re$, a \ALTL formula $\varphi$ and a
term $t$ of $\Re$, $t\models_{\Re}\varphi$ (resp.,
$t\models_{\Re,\infty}\varphi$). The following are well--known
results:

\begin{Proposition}[see~\cite{mayr98,esparza94,esp97}]\label{prop:parallel-altl}
  The model--checking problem for \ALTL and  parallel \PRSs, possibly
  restricted to infinite runs, is decidable.
\end{Proposition}

\begin{Proposition}[see~\cite{alur01,bouajjani97,mayr98}]\label{prop:sequential-altl}
The model--checking problem for \ALTL and  sequential \PRSs,
possibly restricted to infinite runs, is decidable.
\end{Proposition}

The model--checking problem for \ALTL and unrestricted \PRSs is
known
undecidable (see~\cite{mayr98}).\\
 In ~\cite{Bozz03} we showed that the model--checking problem for \PRSs in
 normal form (that are more expressive than parallel and sequential
 \PRSs) and a small fragment of \ALTL is decidable. In particular,
 we established the following result.

\begin{Theorem}[see~\cite{Bozz03}]\label{Theorem:OldResults}
\noindent Given a \BRS $\npla{\Re,\Re_F}$ in normal form and a
        process variable $X$ it is decidable if
\begin{enumerate}
\item  there exists an infinite accepting derivation in
       $\Re$ from $X$.
 \item there exists an infinite derivation in $\Re$ from
        $X$, not containing occurrences of accepting rules.
  \item there exists an infinite derivation from $X$,
  containing a finite non--null number of occurrences of accepting
  rules.
  \end{enumerate}
\end{Theorem}

This result implies the decidability of the model--checking
problem (restricted to infinite runs) for \PRSs in normal form and
the following fragment of \ALTL

\setcounter{equation}{0}
\begin{equation}\label{Eq:OldFragmentALTL}
  \varphi ::= F\,\psi\ |\ GF\,\psi \ |\ \neg \varphi
\end{equation}
where $\psi$ denotes a \ALTL \emph{propositional}
formula\footnote{The set of  \ALTL propositional formulae $\psi$
over the set
$\Sigma$ of atomic propositions is so defined:\\
\hspace*{4cm}$\psi ::= <$$a$$>true$$ \ | \psi \wedge \psi \ |\
\neg \psi$ (where $a\in{}\Sigma$)}.

\noindent{}Thus, the following result holds

\begin{Theorem}[see~\cite{Bozz03}]\label{Theorem:OldResultOnALTL}
\noindent{}The model--checking problem for  \PRSs in normal form
and the fragment \ALTL
$\mathrm{(}$\ref{Eq:OldFragmentALTL}$\mathrm{)}$, restricted to
infinite runs from process variables, is decidable.
\end{Theorem}

\section{Multi B\"{u}chi Rewrite Systems}

In this section we generalize the notion of acceptance in \PRSs,
as defined in \ref{Def:OldAcceptance}, in\-tro\-ducing the notion
of \emph{Multi B\"{u}chi Rewrite System} (\MBRS). Intuitively, a
\MBRS is a \PRS  with a finite number of accepting components,
where each component is a subset of the \PRS. The goal is to
extend the decidability result of theorem
\ref{Theorem:OldResults}. As a consequence, we obtain decidability
of  model--checking for \PRSs in normal form and a meaningful
fragment of \ALTL, that includes strictly the fragment
(\ref{Eq:OldFragmentALTL}) defined in subsection
\ref{sec:OldResults}.\newline

\begin{Definition}[Multi B\"{u}chi Rewrite System]
  A \emph{Multi B\"{u}chi Rewrite System} $\mathrm{(}$\emph{MBRS}$\mathrm{)}$
  $\mathrm{(}$with $n$ \emph{accepting components}$\mathrm{)}$ over a finite set of
  process variables $Var$ and an alphabet $\Sigma$ is a tuple
  $M=\npla{\Re,\npla{\Re_{1}^{A},\ldots,\Re_{n}^{A}}}$, where $\Re$ is a \PRS over $Var$
  and $\Sigma$, and
  $\forall{}i=1,\ldots,n$ $\Re_{i}^{A}\subseteq\Re$. $\Re$ is called the \emph{support} of $M$.
\end{Definition}

A \MBRS $M=\npla{\Re,\npla{\Re_{1}^{A},\ldots,\Re_{n}^{A}}}$ is a
\MBRS \emph{in
  normal form} (resp., \emph{sequential}  \MBRS, \emph{parallel} \MBRS),
if the underlying \PRS $\Re$ is in normal form (resp., is
sequential, is parallel).

\begin{Definition}
$\forall{}n\in{}N\setminus\{0\}$ let us denote by $P_{n}$ the set
$2^{\{1,\ldots,n\}}$ $\mathrm{(}$i.e., the set of the subsets of
$\{1,\ldots,n\}$$\mathrm{)}$.
\end{Definition}

\begin{Definition}[Finite Maximal]
 Let
  $M=\npla{\Re,\npla{\Re_{1}^{A},\ldots,\Re_{n}^{A}}}$ be
  a  \MBRS, and let $\sigma$ be a  rule sequence
  in $\Re$.  The
  \emph{finite maximal of}  $\sigma$ \emph{as to} $M$,  denoted by \maximal{\sigma}, is the
  set $\{i\in\{1,\ldots,n\}|$ $\sigma$ contains some occurrence of rule in $\Re_{i}^{A}\}$.
\end{Definition}

\begin{Definition}[Infinite Maximal]
 Let
  $M=\npla{\Re,\npla{\Re_{1}^{A},\ldots,\Re_{n}^{A}}}$ be
  a  \MBRS, and let $\sigma$ be a  rule sequence
  in $\Re$.
    The
  \emph{infinite maximal of}  $\sigma$ \emph{as to} $M$,  denoted by \maximalInf{\sigma}, is the
  set $\{i\in\{1,\ldots,n\}|$ $\sigma$ contains infinite  occurrences of some rule in $\Re_{i}^{A}\}$.
\end{Definition}

Now, we give a generalized notion of accepting derivation in a
\PRS.

\begin{Definition}[Acceptance]
 Let
  $M=\npla{\Re,\npla{\Re_{1}^{A},\ldots,\Re_{n}^{A}}}$ be
  a  \MBRS, and let $t$ \multider{\sigma} be a derivation
  in $\Re$. Given $K,K^{\omega}\in{}P_{n}$, we say that $t$ \multider{\sigma}
  is a $(K,K^{\omega})-$\emph{accepting} derivation in $M$ if  \maximal{\sigma} $=K$ and
  \maximalInf{\sigma} $=K^{\omega}$.
\end{Definition}

\begin{Definition}
Let $\sigma_{1}$ and $\sigma_{2}$ be  sequences of rules in a \PRS
$\Re$.\\ We denote by $Interleaving(\sigma_{1},\sigma_{2})$ the
set of rule sequences  in  $\Re$ defined inductively in the
following way $\mathrm{(}$we denote by $\varepsilon$ the empty
sequence$\mathrm{)}$:
\begin{itemize}
    \item $Interleaving(\varepsilon,\sigma)=\{\sigma\}$
    \item $Interleaving(\sigma,\varepsilon)=\{\sigma\}$
    \item $Interleaving(r_{1}\sigma_{1},r_{2}\sigma_{2})=\{r_{1}\sigma|\sigma\in{}
     Interleaving(\sigma_{1},r_{2}\sigma_{2})\}\bigcup$\\
     $\{r_{2}\sigma|\sigma\in{}
     Interleaving(r_{1}\sigma_{1},\sigma_{2})\}$ where $r_{1}$ and
     $r_{2}$ are rules in $\Re$.
\end{itemize}
\end{Definition}

The above definition can be extended in obvious way to an
arbitrary number of rule sequences.\newline

Now, we  establish (through propositions \ref{Prop:Maximal})
simple properties of rule sequences in \MBRSs, important in the
following. We need the following definition.

\begin{Definition}
 Let $\{K_{h}\}_{h\in{}N}$ be a succession of sets in $P_{n}$. Let us
 denote by $\bigoplus_{h\in{}N}K_{h}$ the subset of
 $P_{n}$ given by $\{i|$ $\forall{}j\in{}N$ there exists a $h>j$
 such that $i\in{}K_{h}\}$.
\end{Definition}

\begin{Proposition}\label{Prop:Maximal}
Given a \MBRS
  $M=\npla{\Re,\npla{\Re_{1}^{A},\ldots,\Re_{n}^{A}}}$ and two
  rule sequences $\sigma$ and $\sigma'$ in $\Re$, the following properties hold:
\begin{enumerate}
    \item If $\sigma$ is finite, then
            \maximalInf{\sigma} = $\emptyset$.
    \item If $\sigma'$ is a subsequence of
            $\sigma$, then
            \maximal{\sigma'} $\subseteq$ \maximal{\sigma} and
            \maximalInf{\sigma'} $\subseteq$ \maximalInf{\sigma}.
    \item If
           $\lambda\in{}Interleaving(\sigma,\sigma')$, then
           \maximal{\lambda} = \maximal{\sigma} $\cup$ \maximal{\sigma'}  and
           \maximalInf{\lambda} = \maximalInf{\sigma} $\cup$ \maximalInf{\sigma'}.
     \item If
           $\sigma=\sigma_0\sigma_1\sigma_2\ldots$, then
           \maximal{\sigma} = $\bigcup_{h\in{}N}$\maximal{\sigma_{h}} and
           \maximalInf{\sigma} = $\bigoplus_{h\in{}N}$\maximal{\sigma_{h}}.
    \item If $\sigma'$ is a reordering of $\sigma$, then
            \maximal{\sigma} = \maximal{\sigma'} and \maximalInf{\sigma} = \maximalInf{\sigma'}.
\end{enumerate}
\end{Proposition}

In the following subsection we enunciate the two main results of
the paper (proved in section \ref{sec:Result}): the one regarding
acceptance properties of finite derivations in \MBRSs in normal
form, the second regarding acceptance properties of infinite
derivations in \MBRSs in normal form. Moreover, we show that the
second result can be exploited for  automatic verification of some
meaningful (action-based) linear time properties of infinite runs
in \PRSs in normal form.

\subsection{Model-checking of \PRSs in normal form}\label{sec:Model-checkingPRS}

\noindent{}The main result of the paper is the following:

\noindent \emph{Given a \MBRS in normal form
    $M=\npla{\Re,\npla{\Re_{1}^{A},\ldots,\Re_{n}^{A}}}$ over Var and
    the alphabet $\Sigma$, given a variable $X\in{}Var$ and two
    sets
    $K,K^{\omega}\in{}P_{n}$ it is decidable if}:
\begin{description}
\item[Problem 1:] \emph{There exists a $(K,\emptyset)$-accepting
finite derivation in $M$ from $X$}.
 \item[Problem 2:] \emph{There exists a $(K,K^{\omega})$-accepting infinite
derivation in $M$ from $X$}.
  \end{description}

The decidability of Problem 1 is used mainly, as we'll see, to
prove decidability of Pro\-blem 2.\\
 Before proving these results in Sections \ref{sec:finite} and
\ref{sec:infinite}, we show how a solution to these pro\-blems can
be effectively employed to perform model checking of some linear
time properties of infinite runs (from process variables) in \PRSs
in normal form. In particular we consider the following  \ALTL
fragment, that includes strictly the fragment
(\ref{Eq:OldFragmentALTL}) defined in subsection
\ref{sec:OldResults}

 \setcounter{equation}{0}
\begin{equation}\label{Eq:NewFragmentALTL}
  \varphi ::= F\,\psi\ |\ GF\,\psi \ |\ \neg \varphi \ |\  \varphi
  \wedge \varphi \ |\ \varphi
  \vee \varphi \
\end{equation}
where $\psi$ denotes a \ALTL \emph{propositional} formula. For
succinctness, we denote a \ALTL propositional formula of the form
$<$$a$$>true$ (with $a\in\Sigma$) simply by $a$.
\newline
The difference with fragment (\ref{Eq:OldFragmentALTL}) defined in
subsection \ref{sec:OldResults} is that, now, we allow boolean
combinations of formulae of the form $T\psi$, where $T$ denotes a
temporal operator in $\{F,G,FG,$ $GF\}$ and $\psi$ is a \ALTL
propositional formula.\newline Within the fragment above, property
patterns frequent in system verification can be expressed. In
particular, we can express \emph{safety properties} (e.g.,
$G\,p$), \emph{guarantee properties} (e.g., $F\,p$),
\emph{obligation properties} (e.g., $F\,p \rightarrow F\,q$, or
$G\,p \rightarrow G\,q$), \emph{response properties} (e.g.,
$GF\,p$), \emph{persistence properties} (e.g.,  $FG\,p$), and
finally \emph{reactivity properties} (e.g., $GF\,p \rightarrow
GF\,q$). Notice that important classes of properties like
invariants, as well as strong and weak fairness constraints, can
be expressed.\newline
 To prove  decidability of the
model--checking problem restricted to infinite runs for this
fragment of \ALTL we need some
definitions.\\
Given a propositional formula $\psi$ over $\Sigma$, we denote by
$[[\psi]]_{\Sigma}$ the subset of $\Sigma$ inductively defined as
follows
\begin{itemize}
    \item $\forall{a}\in\Sigma$  $[[a]]_{\Sigma}=\{a\}$
\item $[[\neg\psi]]_{\Sigma}=\Sigma\setminus[[\psi]]_{\Sigma}$
\item $[[\psi_{1}\wedge\psi_{2}]]_{\Sigma}=
  [[\psi_{1}]]_{\Sigma}\,\cap\,[[\psi_{2}]]_{\Sigma}$
\end{itemize}
Evidently, given a \PRS $\Re$ over $\Sigma$, a \ALTL propositional
formula $\psi$ and an infinite run $\pi$ of $\Re$ we have that
$\pi\in[[\psi]]_{\Re}$ iff $firstact(\pi)\in[[\psi]]_{\Sigma}$.\\
Given a rule $r=t$\Rule{a}$t'\in\Re$, we say that $r$
\emph{satisfies} the propositional formula $\psi$ if
$a\in[[\psi]]_{\Sigma}$. We denote by $AC_{\Re}(\psi)$  the set of
 rules in $\Re$ that satisfy $\psi$. \newline Now, we introduce a
new temporal operator, denoted by $F^{+}$, whose semantic is so
defined:
\begin{itemize}
    \item $[[F^{+}\varphi]]_{\Re}=\{\pi\in{}runs(\Re)\ \mid\
  \suffix{\pi}\cap[[\varphi]]_{\Re}\not=\emptyset$, and either
  $\pi$ is finite or
  there exists a $j\geq{}0$ such that $\forall{}h\geq{}j$ $\pi^{h}\notin{}[[\varphi]]_{\Re}\}$
\end{itemize}

\noindent{}Now, we can prove the following result

\begin{Theorem}\label{Theorem:NewResultOnALTL}
\noindent{}The model--checking problem for  \PRSs in normal form
and the fragment \ALTL
$\mathrm{(}$\ref{Eq:NewFragmentALTL}$\mathrm{)}$, restricted to
infinite runs from process variables, is decidable.
\end{Theorem}

\begin{proof}
Given a \PRS $\Re$ in normal form, a variable $X$ and a formula
$\varphi$ belonging to \ALTL fragment (\ref{Eq:NewFragmentALTL}),
we have to prove that it's decidable if
\begin{equation}\label{Eq:1}
 X \models_{\Re,\infty} \varphi
\end{equation}
This problem is reducible to the problem of deciding if the
following property is satisfied
\begin{description}
    \item[A.] There exists an infinite run $\pi$, with
    $\pi\in{}runs_{\Re,\infty}(X)$, satisfying the formula
    $\neg\varphi$, i.e. with $\pi\in[[\neg\varphi]]_{\Re}$.
\end{description}
Pushing negation inward, and using the following logic
equivalences
\begin{itemize}
    \item $G\varphi_{1}\wedge{}G\varphi_{2}\equiv{}G(\varphi_{1}\wedge\varphi_{2})$
    \item $\neg{}F\varphi_{1}\equiv{}G\neg\varphi_{1}$
    \item $\neg{}G\varphi_{1}\equiv{}F\neg\varphi_{1}$
    \item $F\varphi_{1}\equiv{}F^{+}\varphi_{1}\vee{}GF\varphi_{1}$
    \item
    $FG\varphi_{1}\equiv{}F^{+}\neg\varphi_{1}\vee{}G\varphi_{1}$
    (this equivalence holds for infinite runs)
\end{itemize}
the formula $\neg\varphi$ can be written in the following
disjunctive normal form
\begin{equation}\label{Eq:2}
 \neg\varphi \equiv \bigvee_{i} \Bigr( \bigwedge_{j} F^{+}\psi_{j} \wedge
 \bigwedge_{k}  GF\eta_{k} \wedge G\zeta \Bigr)
\end{equation}
where $\psi_{j}$, $\eta_{k}$ and $\zeta$ are \ALTL propositional
formulae. Evidently, we can restrict ourselves to consider a
single disjunct in (\ref{Eq:2}). In other words,   problem in
equation (\ref{Eq:1})  is reducible to the problem of deciding,
given a formula having the following form
\begin{equation}\label{Eq:3}
   F^{+}\psi_{1} \wedge\ldots\wedge F^{+}\psi_{m_{1}}\wedge
   GF\eta_{1} \wedge\ldots\wedge GF\eta_{m_{2}} \wedge
   G\zeta\footnote{$\psi_{j}$, $\eta_{k}$ and $\zeta$ are \ALTL propositional
formulae}
\end{equation}
 if the following property is satisfied
\begin{description}
    \item[B.] There exists an infinite run $\pi$, with
    $\pi\in{}runs_{\Re,\infty}(X)$, satisfying  formula
    (\ref{Eq:3}).
\end{description}
Let us consider the \MBRS in normal form
$M=\npla{\Re,\npla{\Re_{1}^{A},\ldots,\Re_{n}^{A}}}$ where
$n=m_{1}+m_{2}+1$ and
\begin{eqnarray}
\textrm{for all $i=1,\ldots,m_{1}\quad\quad$
$\Re_{i}^{A}=AC_{\Re}(\psi_{i})$}\nonumber\\
 \textrm{for all ${}j=1,\ldots,m_{2}\quad\quad$
 $\Re_{j+m_{1}}^{A}=AC_{\Re}(\eta_{j})$}\nonumber\\
 \textrm{ $\Re_{m_{1}+m_{2}+1}^{A}=AC_{\Re}(\neg\zeta)$}\nonumber
\end{eqnarray}
Let $K=\{1,\ldots,m_{1}+m_{2}\}$ and
$K^{\omega}=\{m_{1}+1,\ldots,m_{1}+m_{2}\}$. It is easy to show
that property B is satisfied if, and only if, there exists a
$(K,K^{\omega})$-accepting infinite derivation in $M$ from
variable $X$.
 From decidability of Problem 2, we obtain the assertion.
\end{proof}

\section{Decidability results on  \MBRSs in normal form}\label{sec:Result}
In this section we prove the main results of the paper, namely the
decidability of  problems about  derivations in \MBRSs stated in
subsection~\ref{sec:Model-checkingPRS}. Therefore, in
subsection~\ref{sec:SimpleResults} we report some preliminary
results on the decidability of some properties about  derivations
of parallel and sequential \MBRSs which are necessary to carry out
the proof of the main results, which are given in
subsection~\ref{sec:finite} and \ref{sec:infinite}.

\subsection{Decidability results on parallel and sequential \MBRSs}\label{sec:SimpleResults}

In this section we establish simple decidability results on
properties of derivations in parallel and sequential \MBRSs. These
results are the basis for the decidability proof of the problems
1-2 stated in subsection~\ref{sec:Model-checkingPRS}.

\begin{Proposition}\label{Prop:ParMBRS-F}
 Given a parallel \MBRS
  $M_{P}=\npla{\Re_{P},\npla{\Re_{P,1}^{A},\ldots,\Re_{P,n}^{A}}}$ over $Var$ and the alphabet
   $\Sigma$,  given  two variables $X,Y\in Var$ and $K\in{}P_{n}$,
   it is decidable if
\begin{enumerate}
\item there exists a finite derivation
   in $\Re_{p}$ of the form
  $\prsdernorm{X}{\sigma}{\Re_{P}}{\parcomp{t}{Y}}$, for some term
  $t$, with $|\sigma|>0$ and  \maximalG{M_{P}}{\sigma} $=K$.
\item there exists a finite derivation
 in $\Re_{P}$ of the form
  $\prsdernorm{X}{\sigma}{\Re_{P}}{\varepsilon}$ such that \maximalG{M_{P}}{\sigma} $=K$.
\item there exists a finite derivation  in $\Re_{P}$ of the form
$\prsdernorm{X}{\sigma}{\Re_{P}}{Y}$ such that
\maximalG{M_{P}}{\sigma} $=K$. \item there exists a
$(K,\emptyset)$-accepting finite derivation  in $M_{P}$ from $X$.
\end{enumerate}
\end{Proposition}

\begin{proof}
  We exploit  the decidability of the model-checking problem
  for  \ALTL and parallel \PRSs (see prop. \ref{prop:parallel-altl}).\\
 For all $r\in\Re_{P}$ let us denote by $label(r)$ the set
  $\{h\in{}\{1,\ldots,n\}|r\in\Re^{A}_{P,h}\}$. Let us denote by
  $\zeta$ the set $\{label(r)|r\in\Re_{P}\}$.\\
Let us consider the first problem. Starting from $\Re_{P}$, we
build a new parallel \PRS $\Re'_{P}$ over the alphabet
$\overline{\Sigma}=\{Y\}\cup\zeta$, as follows. At first, we
substitute every  rule $r$  in $\Re_{P}$ of the form
$t$\Rule{a}$t'$ with the rule $t$\Rule{label(r)}$t'$. Finally, we
add the rule $Y$\Rule{Y}$Y$. The reason to add this rule is to
allow to
express reachability of variable $Y$ as a \ALTL formula.\\
Now, let us assume that $K\neq\emptyset$. A similar reasoning
applies if $K=\emptyset$.
 Let us indicate by $\varphi_{1}$ the following \ALTL formula,
\begin{displaymath}
\textrm{$F(<Y>true)$ $\bigwedge$
  $G\Bigr(\neg(<Y>true)$ $\vee$  $<Y>(G(<Y>true))\Bigr)$}
\end{displaymath}
This formula is satisfied by infinite runs $\pi$ in $\Re'_{P}$
having the form $\pi_{1}\pi_{2}$, where $\pi_{2}$ contains only
occurrences of label $Y$, and $\pi_{1}$ doesn't contain
occurrences of label $Y$.\\
It's easy to deduce that  property 1 is satisfied if, and only if,
there exists a run  $\pi$ in $\Re'_{P}$ with
$\pi\in{}runs_{\Re'_{P}}(X)$ satisfying the following \ALTL
formula:
\begin{equation}
\textrm{$\varphi:= \varphi_{1} \wedge$
 $\bigr(\bigwedge_{i\in{}K}\bigvee_{r\in{}\Re^{A}_{P,i}}F(<label(r)>true)\bigr)\wedge$}
\textrm{$\bigr(\bigwedge_{i\notin{}K}\bigwedge_{r\in{}\Re^{A}_{P,i}}G(\neg<label(r)>true)\bigr)$}\nonumber
\nonumber\footnote{for all $i\in{}K$ if $\Re^{A}_{P,i}=\emptyset$,
then
 $\bigvee_{r\in{}\Re^{A}_{P,i}}F(<label(r)>true)$ denotes $false$}
\end{equation}

Therefore,  property 1 isn't satisfied if, and only if,
$\forall\pi\in{}runs_{\Re'_{P}}(X)$
$\pi\notin[[\varphi]]_{\Re'_{P}}$, that
is if, and only if,  $X\models_{\Re'_{P}}\neg\varphi$.\\

Now, let us consider the second problem. Similarly to the problem
above, starting from  $\Re_{P}$, we build a new parallel \PRS
$\Re'_{P}$, this time on the alphabet
$\overline{\Sigma}=Var\cup{}\zeta$, as follows. At first,  we
substitute every  rule $r$  in $\Re_{P}$ of the form
$t$\Rule{a}$t'$ with the rule $t$\Rule{label(r)}$t'$. Finally,
$\forall{}Y\in{}Var$ we add the rule $Y$\Rule{Y}$Y$. Notice that,
by construction,  a term $t$ has no successor in $\Re'_{P}$ if,
and only if, $t=\varepsilon$. Let us indicate by $\varphi_{1}$ the
following \ALTL propositional formula,
\begin{displaymath}
        \bigvee_{Y\in{}Var}\Bigr(<Y>true\Bigr)\vee\bigvee_{l\in\zeta}\Bigr(<l>true\Bigr)
\end{displaymath}
The negation of $\varphi_{1}$ means that no rule can be applied,
in other words the system
has terminated.\\
It's easy to deduce that  property 2 is satisfied if, and only if,
there exists a run  $\pi$ in $\Re'_{P}$ with
$\pi\in{}runs_{\Re'_{P}}(X)$ satisfying the following \ALTL
formula:
\begin{eqnarray}
\textrm{$\varphi:= F(\neg\varphi_{1}) \wedge$
 $\bigr(\bigwedge_{i\in{}K}\bigvee_{r\in{}\Re^{A}_{P,i}}F(<label(r)>true)\bigr)
\wedge$}\nonumber\\
        \textrm{$\bigr(\bigwedge_{i\notin{}K}\bigwedge_{r\in{}\Re^{A}_{P,i}}G(\neg<label(r)>true)\bigr)$}\nonumber
\end{eqnarray}
Therefore,  property 2 isn't satisfied if, and only if,
$\forall\pi\in{}runs_{\Re'_{P}}(X)$
$\pi\notin[[\varphi]]_{\Re'_{P}}$, that
is if, and only if,  $X\models_{\Re'_{P}}\neg\varphi$.\\

Now, let us consider the third problem. Starting from $\Re_{P}$,
we build a new parallel \PRS $\Re'_{P}$ over the alphabet
$\overline{\Sigma}=\{\varepsilon\}\cup{}Var\cup{}\zeta$, as
follows. At first, we  substitute every  rule $r$ in $\Re_{P}$ of
the form $t$\Rule{a}$t'$ with the rule
 $t$\Rule{label(r)}$t'$.  $\forall{}Z\in{}Var$ we add the rule
$Z\mathop{\rightarrow}\limits^{Z}Z$. Finally, we add the rule
$Y$\Rule{\varepsilon}$\varepsilon$.  Notice that, by construction,
a term $t$ has no successor in $\Re'_{P}$ if, and only if,
$t=\varepsilon$. Let us indicate by
 $\varphi_{1}$ the following \ALTL propositional
formula,
\begin{displaymath}
        \bigvee_{Y\in{}Var}\Bigr(<Y>true\Bigr)\vee<\varepsilon>true\vee\bigvee_{l\in\zeta}\Bigr(<l>true\Bigr)
\end{displaymath}
Moreover, let us indicate by $\varphi_{2}$ the following \ALTL
formula,
\begin{displaymath}
        \textrm{$\bigvee_{l\in\zeta}\Bigr(<l>true\Bigr)$ $U$ $(<\varepsilon> \neg\varphi_{1})$}
\end{displaymath}
This formula is satisfied by runs in $\Re'_{P}$ that end in
$\varepsilon$ such that the last label is $\varepsilon$, with each
other label
 in $\zeta$, and the last but one term is $Y$.\\
It's easy to deduce that  property 3 is satisfied if, and only if,
there exists a run  $\pi$ in $\Re'_{P}$ with
$\pi\in{}runs_{\Re'_{P}}(X)$ satisfying the following \ALTL
formula:
\begin{eqnarray}
\textrm{$\varphi:= \varphi_{2} \wedge$
 $\bigr(\bigwedge_{i\in{}K}\bigvee_{r\in{}\Re^{A}_{P,i}}F(<label(r)>true)\bigr)
\wedge$}\nonumber\\
        \textrm{$\bigr(\bigwedge_{i\notin{}K}\bigwedge_{r\in{}\Re^{A}_{P,i}}G(\neg<label(r)>true)\bigr)$}\nonumber
\end{eqnarray}
Therefore,  property 3 isn't satisfied if, and only if,
$\forall\pi\in{}runs_{\Re'_{P}}(X)$
$\pi\notin[[\varphi]]_{\Re'_{P}}$,
that is if, and only if,  $X\models_{\Re'_{P}}\neg\varphi$.\\

Finally, it's easy to prove the decidability of the fourth problem
applying a reasoning similar to previous ones.
\end{proof}

\begin{Proposition}\label{Prop:ParMBRS-Inf} Let us consider two parallel
 \MBRSs
  $M_{P_{1}}=\npla{\Re_{P},\npla{\Re_{P_{1},1}^{A},\ldots,\Re_{P_{1},n}^{A}}}$ and
  $M_{P_{2}}=\npla{\Re_{P},\npla{\Re_{P_{2},1}^{A},\ldots,\Re_{P_{2},n}^{A}}}$ over $Var$ and the alphabet
  $\Sigma$, and with the same support $\Re_{P}$.  Given  a variable $X\in Var$,
  two sets $K,K^{\omega}\in{}P_{n}$, and a subset $\Re^{*}_{P}$ of $\Re_{P}$
  it's decidable if the following condition is satisfied:
\begin{enumerate}
\item There exists a  derivation in $\Re_{P}$ of
    the form $\prsdernorm{X}{\sigma}{\Re_{P}}{}$ such that \maximalG{M_{P_{1}}}{\sigma} $=K$
    and \maximalGInf{M_{P_{1}}}{\sigma} $\cup$ \maximalG{M_{P_{2}}}{\sigma}
    $=K^{\omega}$. Moreover, either $\sigma$ is
    infinite or $\sigma$ contains some occurrence of rule in
    $\Re_{P}\setminus\Re^{*}_{P}$.
\end{enumerate}
\end{Proposition}

\begin{proof}
The proof relies on the decidability of the model-checking problem
for \ALTL and
  parallel \PRSs (see prop. \ref{prop:parallel-altl}).\\
Let us consider the tuple
$\npla{\overline{\Re}_{P,1}^{A},\ldots,\overline{\Re}_{P,2n+1}^{A}}$
where $\forall{}i=1,\ldots,n$
$\overline{\Re}_{P,i}^{A}=\Re_{P_{1},i}^{A}$ and
$\overline{\Re}_{P,i+n}^{A}=\Re_{P_{2},i}^{A}$, and
$\overline{\Re}_{P,2n+1}^{A}=\Re_{P}\setminus\Re^{*}_{P}$.\\
Let us denote by $S$ the set
$\{(K_{1},K_{2})\in{}P_{n}\times{}P_{n}|$
$K_{1}\cup{}K_{2}=K^{\omega}\}$. \\
Evidently,  property 1 is equivalent to the following property:
\begin{enumerate}
\item[2.] There exists a  derivation in $\Re_{P}$ of
    the form $\prsdernorm{X}{\sigma}{\Re_{P}}{}$ such that
    \begin{enumerate}
        \item[2.1] $\forall{}i\in{}K$ $\sigma$ contains some occurrence
        of rule in $\overline{\Re}_{P,i}^{A}$, and
        $\forall{}j\in{}\{1,\ldots,n\}\setminus{}K$  $\sigma$
        doesn't contain occurrences of rules in
        $\overline{\Re}_{P,j}^{A}$.
        \item[2.2] There exists a $(K_{1},K_{2})\in{}S$ such that
        $\forall{}i\in{}K_{1}$ (resp., $\forall{}i\in{}K_{2}$)
             $\sigma$ contains infinite occurrences
        of rules in $\overline{\Re}_{P,i}^{A}$ (resp., contains
        some occurrence of rule in $\overline{\Re}_{P,i+n}^{A}$),
        and
        $\forall{}j\in{}\{1,\ldots,n\}\setminus{}K_{1}$ (resp., $\forall{}j\in{}\{1,\ldots,n\}\setminus{}K_{2}$)
        $\sigma$
        doesn't contain infinite occurrences of rules in
        $\overline{\Re}_{P,j}^{A}$ (resp., doesn't contain occurrences of rules in
        $\overline{\Re}_{P,j+n}^{A}$).
        \item[2.3] Either $\sigma$ is infinite or $\sigma$ contains
        some occurrence of rule in $\overline{\Re}_{P,2n+1}^{A}$.
    \end{enumerate}
\end{enumerate}
  For all $r\in\Re_{P}$ let us denote by $label(r)$ the set
  $\{h\in{}\{1,\ldots,2n+1\}|$ $r\in\overline{\Re}^{A}_{P,h}\}$. Moreover, let us denote
  by
  $\zeta$ the set $\{label(r)|r\in\Re_{P}\}$. Now, we construct
  a new parallel \PRS $\overline{\Re}_{P}$ over the alphabet
  $\zeta\cup{}Var$, as follows.
  At first, we replace every rule $r$  in
  $\Re_{P}$ of the form $t$\Rule{a}$t'$ with the rule
  $t$\Rule{label(r)}$t'$. Finally, $\forall{}Y\in{}Var$
  we add the rule $Y$\Rule{Y}$Y$. Let us consider the following  \ALTL propositional
formula,
\begin{displaymath}
        \psi=\bigvee_{l\in\zeta}\Bigr(<l>true\Bigr)
\end{displaymath}
 Now, let us consider the following \ALTL formula.
\begin{displaymath}
\textrm{$\varphi_{3}:=GF(\bigvee_{label(r)\in\zeta}<label(r)>true)$
 $\vee$ $\bigvee_{r\in\overline{\Re}^{A}_{2n+1}}F\bigr(<label(r)>(FG\neg\psi)\bigr)$}
\end{displaymath}
This formula is satisfied either from infinite runs in
$\overline{\Re}_{P}$ containing infinite occurrences of labels
associated to rules in $\Re_{P}$, or runs containing some
occurrence of a label associated to a rule belonging to
$\Re_{P}\setminus\Re^{*}_{P}$, and containing a finite number of
occurrences of labels related to rules in $\Re_{P}$. So,  formula
$\varphi_{3}$ expresses property 2.3.\\
Now, let us consider the following two \ALTL formulae
\begin{displaymath}
\textrm{$\varphi_{1}:=(
\bigwedge_{i\in{}K}\bigvee_{r\in{}\overline{\Re}^{A}_{P,i}}F(<label(r)>true))
\wedge(\bigwedge_{i\in{}\{1,\ldots,n\}\setminus{}K}\bigwedge_{r\in{}\overline{\Re}^{A}_{P,i}}G(\neg<label(r)>true))$}
\footnote{for all $i\in{}K$ if
$\overline{\Re}^{A}_{P,i}=\emptyset$, then
 $\bigvee_{r\in{}\overline{\Re}^{A}_{P,i}}F(<label(r)>true)$ denotes $false$}
\end{displaymath}
\begin{displaymath}
\textrm{$\varphi_{2}:=\bigvee_{(K_{1},K_{2})\in{}S}\Bigr((
\bigwedge_{i\in{}K_{1}}\bigvee_{r\in{}\overline{\Re}^{A}_{P,i}}GF(<label(r)>true))\wedge$}
\footnote{for all $i\in{}K_1$ if
$\overline{\Re}^{A}_{P,i}=\emptyset$, then
 $\bigvee_{r\in{}\overline{\Re}^{A}_{P,i}}GF(<label(r)>true)$ denotes $false$}
\end{displaymath}
\begin{displaymath}
\textrm{$
(\bigwedge_{i\in{}\{1,\ldots,n\}\setminus{}K_{1}}\bigwedge_{r\in{}\overline{\Re}^{A}_{P,i}}FG(\neg<label(r)>true))\wedge$}
\end{displaymath}
\begin{displaymath}
\textrm{$(
\bigwedge_{i\in{}K_{2}}\bigvee_{r\in{}\overline{\Re}^{A}_{P,i+n}}F(<label(r)>true))
\wedge(\bigwedge_{i\in{}\{1,\ldots,n\}\setminus{}K_{2}}\bigwedge_{r\in{}\overline{\Re}^{A}_{P,i+n}}G(\neg<label(r)>true))\Bigr)$}
\nonumber \footnote{for all $i\in{}K_2$ if
$\overline{\Re}^{A}_{P,i+n}=\emptyset$, then
 $\bigvee_{r\in{}\overline{\Re}^{A}_{P,i+n}}F(<label(r)>true)$ denotes $false$}
\end{displaymath}

Evidently,  formula $\varphi_{1}$ (resp., $\varphi_{2}$) expresses
property 2.1 (resp., 2.2). So, property 2 is satisfied
 if, and only if, there
exists a  run $\pi$ in $\overline{\Re}_{P}$ with
$\pi\in{}runs_{\overline{\Re}_{P}}(X)$ satisfying the following
\ALTL formula:
\begin{displaymath}
\textrm{$\varphi:=\varphi_{1}\wedge\varphi_{2}\wedge\varphi_{3}$}
\end{displaymath}
Therefore,  property 2 isn't satisfied if, and only if,
$\forall\pi\in{}runs_{\overline{\Re}_{P}}(X)$
$\pi\notin[[\varphi]]_{\overline{\Re}_{P}}$, that is if, and only
if, $X\models_{\overline{\Re}_{P}}\neg\varphi$.
\end{proof}

Now, let us give an additional notion of reachability in
sequential \emph{PRSs}.
\begin{Definition} Given a sequential \PRS $\Re_{S}$ over $Var$, and given $X,Y\in{}Var$,
we say that  $Y$ is \emph{reachable from} $X$ \emph{in} $\Re_{S}$
whether there exists a term
 $t\in{}T\setminus\{\varepsilon\}$ of the form
$X_{1}.(X_{2}.(\ldots{}X_{n}.(Y)\ldots))$ $\mathrm{(}$with $n$
possibly equals to  zero$\mathrm{)}$ such that
$\prsdernorm{X}{}{\Re_{S}}{t}$.
\end{Definition}

\begin{Proposition}\label{Prop:SeqMBRS} Let us consider a sequential \MBRS
  $M_{S}=\npla{\Re_{S},\npla{\Re_{S,1}^{A},\ldots,\Re_{S,n}^{A}}}$ over $Var$ and the alphabet
   $\Sigma$.  Given  two variables $X,Y\in Var$ and two sets
   $K,K^{\omega}\in{}P_{n}$, it is decidable if
\begin{enumerate}
\item $Y$ is reachable from $X$ in $\Re_{S}$ through a derivation
having finite maximal $K$ as to $M_{S}$.
 \item  there exists a
$(K,K^{\omega})$-accepting infinite derivation in  $M_{S}$ from
$X$.
\end{enumerate}
\end{Proposition}

\begin{proof}
The proof relies on the decidability of the model-checking problem
for \ALTL and
  sequential \PRSs (possibly restricted to infinite runs) (see prop. \ref{prop:sequential-altl}).\\
  $\forall{}r\in\Re_{S}$ let us denote by $label(r)$ the set
  $\{h\in{}\{1,\ldots,n\}|r\in\Re^{A}_{S,h}\}$. Moreover, let us denote by
  $\zeta$ the set $\{label(r)|r\in\Re_{S}\}$.\\
Let us consider the first problem. Starting from $\Re_{S}$ we
build a new sequential \PRS $\Re'_{S}$ over the alphabet
$\overline{\Sigma}=\{Y\}\cup\zeta$, as follows. At first, we
substitute every  rule $r$  in $\Re_{S}$ of the form
$t$\Rule{a}$t'$ with the rule $t$\Rule{label(r)}$t'$.
            Finally, we add the rule $Y$\Rule{Y}$Y$.\\
Let us indicate by $\varphi_{1}$ the following \ALTL formula,
\begin{displaymath}
\textrm{$F(<Y>true)$ $\bigwedge$
  $G\Bigr(\neg(<Y>true)$ $\vee$  $<Y>(G(<Y>true))\Bigr)$}
\end{displaymath}
This formula is satisfied by infinite runs $\pi$ in $\Re'_{S}$
having the form $\pi_{1}\pi_{2}$ where $\pi_{2}$ contains only
occurrences of label $Y$, and $\pi_{1}$ doesn't contain
occurrences of label $Y$.\\
It's easy to deduce that  property 1 is satisfied if, and only if,
there exists a run  $\pi$ in $\Re'_{S}$ with
$\pi\in{}runs_{\Re'_{S}}(X)$ satisfying the following \ALTL
formula:
\begin{displaymath}
\textrm{$\varphi:= \varphi_{1} \wedge$
 $\bigr(\bigwedge_{i\in{}K}\bigvee_{r\in{}\Re^{A}_{S,i}}F(<label(r)>true)\bigr)
\wedge$}\textrm{$\bigr(\bigwedge_{i\notin{}K}\bigwedge_{r\in{}\Re^{A}_{S,i}}G(\neg<label(r)>true)\bigr)$}\nonumber
\footnote{for all $i\in{}K$ if $\Re^{A}_{S,i}=\emptyset$, then
 $\bigvee_{r\in{}\Re^{A}_{S,i}}F(<label(r)>true)$ denotes $false$}
\end{displaymath}
Therefore,  property 1 isn't satisfied if, and only if,
$\forall\pi\in{}runs_{\Re'_{S}}(X)$
$\pi\notin[[\varphi]]_{\Re'_{S}}$, that is if, and only if,
$X\models_{\Re'_{S}}\neg\varphi$.\\

\noindent{}Now, let us consider the second problem. We construct a
  new sequential \PRS
  $\Re'_{S}$ over the alphabet $\zeta$ in the following way. We substitute every rule $r$  in
  $\Re_{S}$ of the form $t$\Rule{a}$t'$ with the rule
  $t$\Rule{label(r)}$t'$.  \\
It's easy to deduce that  property 2 is satisfied if, and only if,
there exists an infinite run  $\pi$ in $\Re'_{S}$ with
$\pi\in{}runs_{\Re'_{S},\infty}(X)$ satisfying the following \ALTL
formula:
\begin{displaymath}
\textrm{$\varphi:=(
\bigwedge_{i\in{}K^{\omega}}\bigvee_{r\in{}\Re^{A}_{S,i}}GF(<label(r)>true))
\wedge(\bigwedge_{i\notin{}K^{\omega}}\bigwedge_{r\in{}\Re^{A}_{S,i}}FG(\neg<label(r)>true))\wedge$}
\footnote{for all $i\in{}K^{\omega}$ if $\Re^{A}_{S,i}=\emptyset$,
then
 $\bigvee_{r\in{}\Re^{A}_{S,i}}GF(<label(r)>true)$ denotes $false$}
\end{displaymath}
\begin{displaymath}
\textrm{$(
\bigwedge_{i\in{}K}\bigvee_{r\in{}\Re^{A}_{S,i}}F(<label(r)>true))
\wedge(\bigwedge_{i\notin{}K}\bigwedge_{r\in{}\Re^{A}_{S,i}}G(\neg<label(r)>true))$}\nonumber
\footnote{for all $i\in{}K$ if $\Re^{A}_{S,i}=\emptyset$, then
 $\bigvee_{r\in{}\Re^{A}_{S,i}}F(<label(r)>true)$ denotes $false$}
\end{displaymath}

Therefore,  property 1 isn't satisfied if, and only if,
$\forall\pi\in{}runs_{\Re'_{S},\infty}(X)$
$\pi\notin[[\varphi]]_{\Re'_{S}}$, that is if, and only if,
$X\models_{\Re'_{S},\infty}\neg\varphi$.
\end{proof}

\begin{Theorem}\label{Theorem:ConditionForProblem2-1} Let us consider two parallel
 \MBRSs
  $M_{P_{1}}=\npla{\Re_{P},\npla{\Re_{P_{1},1}^{A},\ldots,\Re_{P_{1},n}^{A}}}$ and
  $M_{P_{2}}=\npla{\Re_{P},\npla{\Re_{P_{2},1}^{A},\ldots,\Re_{P_{2},n}^{A}}}$
  with the same support $\Re_{P}$, and  a sequential \MBRS
  $M_{S}=\npla{\Re_{S},\npla{\Re_{S,1}^{A},\ldots,\Re_{S,n}^{A}}}$.  Given  a variable $X\in Var$,
  two sets $K,K^{\omega}\in{}P_{n}$, and a subset $\Re^{*}_{P}$ of $\Re_{P}$
  it's decidable if the following condition is satisfied:
\begin{enumerate}
\item There exists a variable  $Y$ reachable from $X$ in
        $\Re_{S}$ through a $(K',\emptyset)$-accepting derivation in $M_{S}$
        with $K'\subseteq{}K$, and there exists a derivation
        $\prsdernorm{Y}{\rho}{\Re_{P}}{}$
        such that \maximalG{M_{P_{1}}}{\rho} = $K$
        and
        \maximalG{M_{P_{2}}}{\rho} $\cup$ \maximalGInf{M_{P_{1}}}{\rho} =
        $K^{\omega}$. Moreover, either $\rho$ is
        infinite or $\rho$ contains some occurrence of rule in
        $\Re_{P}\setminus\Re_{P}^{*}$.
\end{enumerate}
\end{Theorem}
\begin{proof}
Since the sets $\{K'\in{}P_{n}|K'\subseteq{}K\}$ and $Var$ are
finite, the result follows directly from propositions
\ref{Prop:ParMBRS-Inf} and \ref{Prop:SeqMBRS}.
\end{proof}

\begin{Theorem}\label{Theorem:ConditionForProblem2-2} Let us consider two parallel
 \MBRSs
  $M_{P_{1}}=\npla{\Re_{P},\npla{\Re_{P_{1},1}^{A},\ldots,\Re_{P_{1},n}^{A}}}$ and
  $M_{P_{2}}=\npla{\Re_{P},\npla{\Re_{P_{2},1}^{A},\ldots,\Re_{P_{2},n}^{A}}}$
  with the same support $\Re_{P}$, and  a sequential \MBRS
  $M_{S}=\npla{\Re_{S},\npla{\Re_{S,1}^{A},\ldots,\Re_{S,n}^{A}}}$.  Given  a variable $X\in Var$,
  two sets $K,K^{\omega}\in{}P_{n}$, and a subset $\Re^{*}_{P}$ of $\Re_{P}$
  it's decidable if one of the following conditions is satisfied:
\begin{enumerate}
\item There exists a variable  $Y$ reachable from $X$ in
        $\Re_{S}$ through a $(K',\emptyset)$-accepting derivation in $M_{S}$
        with $K'\subseteq{}K$, and there exists a derivation
        $\prsdernorm{Y}{\rho}{\Re_{P}}{}$
        such that \maximalG{M_{P_{1}}}{\rho} = $K$
        and
        \maximalG{M_{P_{2}}}{\rho} $\cup$ \maximalGInf{M_{P_{1}}}{\rho} =
        $K^{\omega}$. Moreover, either $\rho$ is
        infinite or $\rho$ contains some occurrence of rule in
        $\Re_{P}\setminus\Re_{P}^{*}$.
\item There exists a
     $(K,K^{\omega})$-accepting infinite derivation  in $M_{S}$ from
     $X$.
\end{enumerate}
\end{Theorem}
\begin{proof}
Since the sets $\{K'\in{}P_{n}|K'\subseteq{}K\}$ and $Var$ are
finite, the result follows directly from propositions
\ref{Prop:ParMBRS-Inf} and \ref{Prop:SeqMBRS}.
\end{proof}

\subsection{Decidability results on finite derivations of \MBRSs in normal form}\label{sec:finite}

In this section we prove the decidability of Problem 1 stated in
subsection~\ref{sec:Model-checkingPRS},  that for clarity we
recall.
\begin{description}
    \item[Problem 1]\emph{ Given a \MBRS in normal form
    $M=\npla{\Re,\npla{\Re_{1}^{A},\ldots,\Re_{n}^{A}}}$ over Var and
    the alphabet $\Sigma$, given a variable $X\in{}Var$ and a set
    $K\in{}P_{n}$, to decide if there exists a  $(K,\emptyset)$-accepting finite derivation in $M$
    from $X$}.
\end{description}

We show that  problem 1, with input a set $K\in{}P_{n}$, can be
reduced to a similar, but simpler, problem, that is  a
decidability problem on finite derivations restricted to parallel
\MBRSs. In particular, we show that it is possible to construct
effectively a parallel \MBRS
$M^{K}_{PAR}=\npla{\Re^{K}_{PAR},\npla{\Re^{K,A}_{PAR,1},\ldots,\Re_{PAR,n}^{K,A}}}$
in such a way that Pro\-blem 1, with input the set $K$ and a
variable $X$, is reducible to the problem of deciding if  the
following condition is satisfied:
\begin{itemize}
    \item There exists  a $(K,\emptyset)$-accepting finite
    derivation in $M^{K}_{PAR}$ from
    $X$.
\end{itemize}
Since this last problem is decidable (see proposition
\ref{Prop:ParMBRS-F}), decidability of Problem 1 is entailed.

Before illustrating the main idea underlying our approach, we need
few additional definitions and notations, which allows us to look
more in detail at the structure of derivations in \MBRSs in normal
form. The following definition introduces the notion of level of
application of a rule in a derivation:

\begin{Definition}
  Let $t$ \prsoneder{r} $t'$  be a single--step derivation in $\Re$.
  We say that $r$ is \emph{applicable at level 0} in
  $t$ \prsoneder{r} $t'$, if $t = \overline{t}$$\parallel$$s$,
  $t' = \overline{t}$$\parallel$$s'$ $\mathrm{(}$for some $\overline{t}, s, s'\in
  T$$\mathbb{)}$, and $r = s$\Rule{a}$s'$, for some $a\in\Sigma$.

  We say that $r$ is \emph{applicable at level $k > 0$ in}
  $t$ \prsoneder{r} $t'$, if $t = \overline{t}$$\parallel$$X.(s)$,
  $t' = \overline{t}$$\parallel$$X.(s')$ $\mathrm{(}$for some
  $\overline{t}, s,s'\in T$$\mathbb{)}$, $s$ \prsoneder{r} $s'$, and
  $r$ is applicable at level $k-1$ in $s$ \prsoneder{r} $s'$.

  The \emph{level of application}  of $r$ in $t$ \prsoneder{r}
  $t'$ is the greatest level of applicability of $r$ in $t$ \prsoneder{r}
  $t'$.
\end{Definition}

The definition above extends in the obvious way to $n$--step
derivations and to infinite derivations. The next definition
introduces the notion of subderivation starting from a term.

\begin{Definition}[Subderivation]
  Let $\overline{t}$ \multider{\lambda}
  $t$$\parallel$$X.(s)$ \multider{\sigma} be a finite or infinite
  derivation in $\Re$ starting from $\overline{t}$. The set of the
  \emph{subderivations $d'=s$ \multider{\mu} of
    $d=t$$\parallel$$X.(s)$ \multider{\sigma} from $s$} is inductively
  defined as follows:

\begin{enumerate}
\item if $d$ is a null derivation or $s = \varepsilon$, then $d'$
is
  the null derivation from $s$;
\item if $\sigma = r\sigma'$, and $d$ is of the form
\begin{displaymath}
\textrm{
     $t$$\parallel$$X.(Z)$ \prsoneder{r} $t$$\parallel$$Y$
    \multider{\sigma'}  $\quad\mathrm{(}$with $r =X.(Z)$\Rule{a}$Y$ and $s=Z\in{}Var$$\mathrm{)}$}
\end{displaymath}
\noindent  then $d'$ is the null derivation from $s$; \item if
$\sigma = r\sigma'$, and $d$ is of the form
\begin{displaymath}
\textrm{
     $t$$\parallel$$X.(s)$ \prsoneder{r} $t$$\parallel$$X.(s')$
    \multider{\sigma'}   $\quad\mathrm{(}$with $s$ \prsoneder{r} $s'$$\mathrm{)}$}
\end{displaymath}
\noindent then $d' = s$ \prsoneder{r} $s'$ \multider{\mu'}, where
$s'$ \multider{\mu'} is a subderivation of $t$$\parallel$$X.(s')$
\multider{\sigma'} from $s'$;
 \item if
$\sigma = r\sigma'$, and $d$ is of the form
\begin{displaymath}
\textrm{
     $t$$\parallel$$X.(s)$ \prsoneder{r} $t'$$\parallel$$X.(s)$
    \multider{\sigma'}   $\quad\mathrm{(}$with $t$ \prsoneder{r} $t'$$\mathrm{)}$}
\end{displaymath}
\noindent then $d'$ is a subderivation of $t'$$\parallel$$X.(s)$
\multider{\sigma'} from $s$;
\end{enumerate}
Moreover, we say that $d'$ is a subderivation of $\overline{t}$
\multider{\lambda}
  $t$$\parallel$$X.(s)$ \multider{\sigma}.
\end{Definition}

Clearly, in the definition above $\mu$ is a subsequence of
$\sigma$. Moreover, if $k$ is the level of application of a rule
occurrence of $\mu$ in the derivation $d$ then, $k>0$, and the
level of application of this occurrence  in the subderivation
$d'=s$ \multider{\mu} is $k'$ with $k'<k$.

Given a sequence $\sigma=r_{1}r_{2} \ldots{}r_{n} \ldots{}$ of
rules in $\Re$, and a subsequence $\sigma'=r_{k_{1}}r_{k_{2}}
\ldots{} r_{k_{m}} \ldots{}$ of $\sigma$, $\sigma\setminus
\sigma{}'$ denotes the sequence obtained by removing from $\sigma$
all and only the occurrences of rules in $\sigma'$ (namely, those
$r_{i}$ for which there exists a $j=1,\ldots,|\sigma'|$ such that
$k_{j}=i$).\newline

In the following,
$M_{P}=\npla{\Re_{P},\npla{\Re_{P,1}^{A},\ldots,\Re_{P,n}^{A}}}$
denotes the restriction of  $M$  to PAR rules, that is $\Re_{P}$
(resp., $\Re_{P,i}^{A}$ for $i=1,\ldots,n$) is the set $\Re$
(resp., $\Re_{i}^{A}$ for $i=1,\ldots,n$) restricted to the PAR
rules.\\

\setcounter{equation}{0} Let us sketch the main ideas at the basis
of our technique. We show how it is possible to mimic a
$(K,\emptyset)$-accepting finite derivation in $M$ from a variable
by using only PAR rules belonging to an extension of the parallel
\MBRS $M_{P}$, denoted by
$M^{K}_{PAR}=\npla{\Re^{K}_{PAR},\npla{\Re^{K,A}_{PAR,1},\ldots,\Re_{PAR,n}^{K,A}}}$.
More precisely, we show that
\begin{description}
    \item[i.]  if $p$ \multider{\sigma} (with
$p\in{}T_{PAR}$) is a $(\overline{K},\emptyset)$-accepting finite
derivation in $M$ with $\overline{K}\subseteq{}K$ then, there
exists a $(\overline{K},\emptyset)$-accepting finite derivation in
$M^{K}_{PAR}$
 from $p$, and vice versa.
\end{description}
So, let $p$ \multider{\sigma} be a
$(\overline{K},\emptyset)$-accepting finite derivation in $M$ with
$p\in{}T_{PAR}$ and $\overline{K}\subseteq{}K$. Then, all its
possible subderivations contain all, and only, the rule
occurrences in $\sigma$ applied at a level $k$ greater than $0$ in
$p$ \multider{\sigma}. If $\sigma$ contains only PAR rule
occurrences the statement \textbf{i} is evident, since
$M^{K}_{PAR}$ is an extension of $M_{P}$. Otherwise, $p$
\multider{\sigma} can be written in the form:
 \begin{equation}\label{eq:derivation}
 \textrm{$p$ \multider{\lambda} $t$$\parallel$$X$ \prsoneder{r}
 $t$$\parallel$$Y.(Z)$ \multider{\omega}}
 \end{equation}
 where $r = X$\Rule{a}$Y.(Z)$,  $\lambda$ contains only occurrences of rules
 in $\Re_{P}$,  $t\in{}T_{PAR}$ and $X,Y,Z\in{}Var$. Let $Z$ \multider{\rho} be  a subderivation of
 $t$$\parallel$$Y.(Z)$ \multider{\omega} from $Z$. Evidently, \maximal{\rho} $\subseteq{}K$. Moreover,
 only one of the following three cases may occur:
\begin{description}
\item[A] $Z$ \multider{\rho} leads to the term $\varepsilon$, and
  $p$ \multider{\sigma} is of the form
\begin{equation}\label{eq:form1}
 \textrm{$p$ \multider{\lambda} $t$$\parallel$$X$ \prsoneder{r}
 $t$$\parallel$$Y.(Z)$ \multider{\omega_{1}} $\overline{t}$$\parallel$$Y$ \multider{\omega_{2}}}
\end{equation}
  where $\rho$ is a subsequence of $\omega_1$ and $t$
  \multider{\omega_{1}\setminus\rho} $\overline{t}$.  The finite derivation above is
  $(\overline{K},\emptyset)$-accepting if, and only if, the following finite derivation,
  obtained by anticipating (by interleaving) the application of the
  rules in $\rho$ before the application of the rules in $\xi=
  \omega_1\setminus\rho$, is $(\overline{K},\emptyset)$-accepting
  \begin{equation}\label{eq:form1-2}
   \textrm{$p$ \multider{\lambda} $t$$\parallel$$X$ \prsoneder{r}
 $t$$\parallel$$Y.(Z)$ \multider{\rho} $t$$\parallel$$Y$ \multider{\xi}
 $\overline{t}$$\parallel$$Y$ \multider{\omega_{2}}}
  \end{equation}
  Let \maximal{r\rho} =
  $K'\subseteq{}K$. The idea is to collapse the derivation $X$ \prsoneder{r}
 $Y.(Z)$ \multider{\rho} $Y$ into a
  single  PAR rule of the form $r'=X$\Rule{K'}$Y$,
  where  \maximalparK{r'} = $K'$ = \maximal{r\rho}.\\
  Notice that in the step from (2) to (3),
  we exploit the fact that the property on finite derivations we
  are interested in is insensitive to permutations
  of rule applications within a derivation.
  Now, we can apply recursively the same reasoning to the finite
   derivation in $\Re$ from $t$$\parallel$$Y\in{}T_{PAR}$
  \begin{equation}
    \textrm{$t$$\parallel$$Y$ \multider{\xi} $\overline{t}$$\parallel$$Y$
      \multider{\omega_{2}}}
  \end{equation}
   whose finite maximal as to $M$ is
  contained in $K$.
\item[B] $Z$ \multider{\rho} leads to a variable $W$, and
  $p$ \multider{\sigma} can be written as
\begin{equation}\label{eq:form1}
 \textrm{$p$ \multider{\lambda} $t$$\parallel$$X$ \prsoneder{r}
 $t$$\parallel$$Y.(Z)$ \multider{\omega_{1}} $\overline{t}$$\parallel$$Y.(W)$
 \prsoneder{r'} $\overline{t}$$\parallel$$W'$ \multider{\omega_{2}}}
\end{equation}
  where $r'=Y.(W)$\Rule{b}$W'$ (with $W'\in{}Var$),
   $\rho$ is a subsequence of $\omega_1$ and $t$
  \multider{\omega_{1}\setminus\rho} $\overline{t}$.  The finite derivation above is
  $(\overline{K},\emptyset)$-accepting if, and only if, the following finite derivation
  is $(\overline{K},\emptyset)$-accepting
  \begin{equation}\label{eq:form1-2}
   \textrm{$p$ \multider{\lambda} $t$$\parallel$$X$ \prsoneder{r}
 $t$$\parallel$$Y.(Z)$ \multider{\rho} $t$$\parallel$$Y.(W)$  \prsoneder{r'} $t$$\parallel$$W'$
    \multider{\xi}
 $\overline{t}$$\parallel$$W'$ \multider{\omega_{2}}}
  \end{equation}
  where $\xi=\omega_{1}\setminus\rho$.
  Let \maximal{rr'\rho} =
  $K'\subseteq{}K$. The idea is to collapse the derivation $X$ \prsoneder{r}
 $Y.(Z)$ \multider{\rho} $Y.(W)$ \prsoneder{r'} $W'$  into a
  single  PAR rule of the form $r''=X$\Rule{K'}$W'$,
  where  \maximalparK{r''} = $K'$ = \maximal{rr'\rho}.\\
  Now, we can apply recursively the same reasoning to the finite
   derivation in $\Re$ from $t$$\parallel$$W'\in{}T_{PAR}$
  \begin{equation}
    \textrm{$t$$\parallel$$W'$ \multider{\xi} $\overline{t}$$\parallel$$W'$
      \multider{\omega_{2}}}
  \end{equation}
   whose finite maximal as to $M$ is
  contained in $K$.
\item[C] In this case $Z$ \multider{\rho} does not influence the
  applicability of rules in $\omega \setminus \rho$ in the derivation
  $t$$\parallel$$Y.(Z)$ \multider{\omega}
  (i.e. the rule  occurrences in $\rho$ can be arbitrarily
  interleaved with any rule occurrence in $\omega\setminus\rho$).  In other
  words, we have $t$ \multider{\omega\setminus\rho}. Moreover,
  \maximal{r\rho} $=K'$ with $K'\subseteq{}K$.
  Then, we keep track of  the sequence $r\rho$
  by adding a new variable $\hat{Z}_{F}$ (where $\hat{Z}_{F}\notin{}Var$)  and
  a  PAR rule of the form $r'=X$\Rule{K'}$\hat{Z}_{F}$, where
 \maximalparK{r'} = $K'$ =  \maximal{r\rho}.
 Now, we can apply recursively the same reasoning to the finite
 derivation $t$$\parallel$$\hat{Z}_{F}$ \multider{\omega\setminus\rho} in $\Re$ from
  the parallel term $t$$\parallel$$\hat{Z}_{F}$,
   whose finite maximal as to $M$ is
  contained in $K$.
 \end{description}

 The  parallel \MBRS
 $M^{K}_{PAR}=\npla{\Re^{K}_{PAR},\npla{\Re^{K,A}_{PAR,1},\ldots,\Re_{PAR,n}^{K,A}}}$
 is  defined as follows.

\begin{Definition}\label{def:MPAR-K}
 The \MBRS $M^{K}_{PAR}=\npla{\Re^{K}_{PAR},\npla{\Re^{K,A}_{PAR,1},\ldots,\Re_{PAR,n}^{K,A}}}$
  is
 the least parallel \MBRS with $n$ accepting components, over $Var\cup\{\hat{Z}_{F}\}$ and the alphabet $\overline{\Sigma} = \Sigma
  \cup{}P_{n}$\footnote{let us assume that $\Sigma\cap{}P_{n}=\emptyset$}, satisfying the following properties:
\begin{enumerate}
    \item $\Re^{K}_{PAR}\supseteq\{r\in\Re|\textrm{$r$ is a PAR rule}\}$
    and $\forall{}i=1,\ldots,n$
    $\Re^{K,A}_{PAR,i}\supseteq\{r\in\Re_{i}^{A}|\textrm{ $r$ is }$ a PAR rule\}
    \item  Let $r=X$\Rule{a}$Y.(Z)\in{}\Re$ and
    $Z$ \multiderparK{\sigma} $p$ for some term $p$, with \maximal{r} $=K_{1}\subseteq{}K$
            and \maximalparK{\sigma} $=K_{2}\subseteq{}K$. Denoted by $K'$
            the set $K_{1}\cup{}K_{2}$, then we have
             $r'=X$\Rule{K'}$\hat{Z}_{F}\in{}\Re^{K}_{PAR}$ and
            \maximalparK{r'} $=K'$.
    \item  Let $r=X$\Rule{a}$Y.(Z)\in{}\Re$ and
    $Z$ \multiderparK{\sigma} $\varepsilon$ with \maximal{r} $=K_{1}\subseteq{}K$
            and \maximalparK{\sigma} $=K_{2}\subseteq{}K$. Denoted by $K'$
            the set $K_{1}\cup{}K_{2}$, then we have
             $r'=X$\Rule{K'}$Y\in{}\Re^{K}_{PAR}$ and
            \maximalparK{r'} $=K'$.
    \item Let $r=X$\Rule{a}$Y.(Z)\in{}\Re$, $r'=Y.(W)$\Rule{b}$W'\in{}\Re$
         and $Z$ \multiderparK{\sigma} $W$ with \maximal{r} $=K_{1}\subseteq{}K$,
            \maximal{r'} $=K_{2}\subseteq{}K$
             and  \maximalparK{\sigma} $=K_{3}\subseteq{}K$. Denoted by $K'$
            the set $K_{1}\cup{}K_{2}\cup{}K_{3}$, then we have
             $r''=X$\Rule{K'}$W'\in{}\Re^{K}_{PAR}$ and
            \maximalparK{r''} $=K'$.
    \item If $r=X$\Rule{K'}$Y\in{}\Re^{K}_{PAR}\setminus\Re$
          then, $X$ \multider{\sigma} $t$ for some term $t$, with $|\sigma|>0$ and
          \maximal{\sigma} $=K'$. Moreover, if $Y\in{}Var$ then
          $t=Y$.
\end{enumerate}
\end{Definition}

\setcounter{equation}{0}
\begin{Lemma}\label{Lemma:Algo}
The parallel \MBRS
$M^{K}_{PAR}=\npla{\Re^{K}_{PAR},\npla{\Re^{K,A}_{PAR,1},\ldots,\Re_{PAR,n}^{K,A}}}$
 can be effectively constructed.
\end{Lemma}
\begin{proof}
Figure \ref{fig:algo} reports the procedure
BUILD-PARALLEL-MBRS($M$,$K$), which, starting from \MBRS $M$ (in
normal form) and a set $K\in{}P_{n}$, builds a parallel \MBRS
$M_{P}=\npla{\Re_{P},\npla{\Re_{P,1}^{A},\ldots,\Re_{P,n}^{A}}}$.
>From proposition \ref{Prop:ParMBRS-F}, the conditions in each of
the \textbf{if} statements in lines 9, 16 and 27 are decidable,
therefore, the procedure is effective.\\
Now, let us show that the algorithm terminates.  It suffices to
prove that the number of iterations of the \textbf{repeat} loop is
finite. The termination condition of this loop is \emph{flag} =
\emph{false}. On the other hand, at the beginning of every
iteration  the
 \emph{flag} is set to \emph{false}, and can be reset to  \emph{true}
 when a rule of the form
$X$\Rule{K'}$Y$ (with $X\in{}Var$, $Y\in{}Var\cup\{\hat{Z}_{F}\}$
and $K'\in{}P_{n}$) not
    belonging to $\Re_{P}$ is added to $\Re_{P}$
    (lines 11--15, 18--22 and 29--33). Since the set of rules
of the form $X$\Rule{K'}$Y$  with $X\in{}Var$,
$Y\in{}Var\cup\{\hat{Z}_{F}\}$ and $K'\in{}P_{n}$
is finite, termination immediately follows.\\

\begin{figure}[htbp]
\ \\\noindent \textbf{Algorithm}
  B{\small{}UILD}--{\small{}PARALLEL}--MBRS($M$,$K$)\vspace{4pt}
\\\small
1 $\Re_{P}:=\{r\in\Re|$ \emph{$r$ is a PAR rule}\};\\
2 \textbf{for} $i=1,\ldots,n$ \textbf{do}\\
3 $\quad\quad\Re_{P,i}^{A}:=\{r\in\Re_{i}^{A}|$ \emph{$r$ is a PAR rule}\};\\
4 \textbf{repeat}\\
5 $\quad\quad{}$\emph{flag:=false;}\\
6 $\quad\quad{}$\textbf{for each} $r=X$\Rule{a}$Y.(Z)\in\Re$ \emph{such that} \maximal{r} $\subseteq{}K$  \textbf{do}\\
7 $\quad\quad\quad\quad{}$\emph{Set} $K_{1}=$ \maximal{r}\\
8 $\quad\quad\quad\quad{}$\textbf{for each} $K_{2}\subseteq{}K$ \textbf{do}\\
9 $\quad\quad\quad\quad\quad\quad{}$\textbf{if}
$\prsdernorm{Z}{\sigma}{\Re_{P}}{p}$ for some $p$
            \emph{such that} \maximalG{M_{P}}{\sigma} $=K_{2}$ \textbf{then}\\
10 $\quad\quad\quad\quad\quad\quad\quad\quad{}$\emph{Set}
            $K'=K_{1}\cup{}K_{2}=\{i_{1},\ldots,i_{|K'|}\}$;\\
11 $\quad\quad\quad\quad\quad\quad\quad\quad{}$\textbf{if}
        $X$\Rule{K'}$\hat{Z}_{F}\notin{}\Re_{P}$ \textbf{then}\\
12
$\quad\quad\quad\quad\quad\quad\quad\quad\quad\quad{}\Re_{P}:=\Re_{P}\cup\{X$\Rule{K'}$\hat{Z}_{F}\};$\\
13 $\quad\quad\quad\quad\quad\quad\quad\quad\quad\quad{}$\textbf{for} $j=1,\ldots,|K'|$ \textbf{do}\\
14
$\quad\quad\quad\quad\quad\quad\quad\quad\quad\quad\quad\quad{}\Re_{P,i_{j}}^{A}:=\Re_{P,i_{j}}^{A}\cup\{X$\Rule{K'}$\hat{Z}_{F}\};$\\
15 $\quad\quad\quad\quad\quad\quad\quad\quad\quad\quad{}$\emph{flag:=true;}\\
16 $\quad\quad\quad\quad\quad\quad{}$\textbf{if}
$\prsdernorm{Z}{\sigma}{\Re_{P}}{\varepsilon}$
            \emph{such that} \maximalG{M_{P}}{\sigma} $=K_{2}$ \textbf{then}\\
17 $\quad\quad\quad\quad\quad\quad\quad\quad{}$\emph{Set}
            $K'=K_{1}\cup{}K_{2}=\{i_{1},\ldots,i_{|K'|}\}$;\\
18 $\quad\quad\quad\quad\quad\quad\quad\quad{}$\textbf{if}
        $X$\Rule{K'}$Y\notin{}\Re_{P}$ \textbf{then}\\
19
$\quad\quad\quad\quad\quad\quad\quad\quad\quad\quad{}\Re_{P}:=\Re_{P}\cup\{X$\Rule{K'}$Y\};$\\
20 $\quad\quad\quad\quad\quad\quad\quad\quad\quad\quad{}$\textbf{for} $j=1,\ldots,|K'|$ \textbf{do}\\
21
$\quad\quad\quad\quad\quad\quad\quad\quad\quad\quad\quad\quad{}\Re_{P,i_{j}}^{A}:=\Re_{P,i_{j}}^{A}\cup\{X$\Rule{K'}$Y\};$\\
22 $\quad\quad\quad\quad\quad\quad\quad\quad\quad\quad{}$\emph{flag:=true;}\\
23 $\quad\quad\quad\quad{}$\textbf{done} $\rhd$ for\\
24 $\quad\quad\quad\quad{}$\textbf{for each} $r'=Y.(W)$\Rule{b}$W'\in\Re$  \emph{such that} \maximal{r'} $\subseteq{}K$\textbf{do}\\
25 $\quad\quad\quad\quad\quad\quad{}$\emph{Set} $K_{2}=$ \maximal{r'}\\
26 $\quad\quad\quad\quad\quad\quad{}$\textbf{for each} $K_{3}\subseteq{}K$ \textbf{do}\\
27 $\quad\quad\quad\quad\quad\quad\quad\quad{}$\textbf{if}
$\prsdernorm{Z}{\sigma}{\Re_{P}}{W}$
            \emph{such that} \maximalG{M_{P}}{\sigma} $=K_{3}$ \textbf{then}\\
28
$\quad\quad\quad\quad\quad\quad\quad\quad\quad\quad{}$\emph{Set}
            $K'=K_{1}\cup{}K_{2}\cup{}K_{3}=\{i_{1},\ldots,i_{|K'|}\}$;\\
29
$\quad\quad\quad\quad\quad\quad\quad\quad\quad\quad{}$\textbf{if}
        $X$\Rule{K'}$W'\notin{}\Re_{P}$ \textbf{then}\\
30
$\quad\quad\quad\quad\quad\quad\quad\quad\quad\quad\quad\quad{}\Re_{P}:=\Re_{P}\cup\{X$\Rule{K'}$W'\};$\\
31 $\quad\quad\quad\quad\quad\quad\quad\quad\quad\quad\quad\quad{}$\textbf{for} $j=1,\ldots,|K'|$ \textbf{do}\\
32
$\quad\quad\quad\quad\quad\quad\quad\quad\quad\quad\quad\quad\quad\quad{}\Re_{P,i_{j}}^{A}:=\Re_{P,i_{j}}^{A}\cup\{X$\Rule{K'}$W'\};$\\
33 $\quad\quad\quad\quad\quad\quad\quad\quad\quad\quad\quad\quad{}$\emph{flag:=true;}\\
34 $\quad\quad\quad\quad\quad\quad{}$\textbf{done} $\rhd$ for\\
35 $\quad\quad\quad\quad{}$\textbf{done} $\rhd$ for\\
36 $\quad\quad{}$\textbf{done} $\rhd$ for\\
37 \textbf{until} \emph{flag = false} \\
\caption{Algorithm to turn a \MBRS into a parallel
\MBRS.}\label{fig:algo}
\end{figure}

Now, let us denote by $\Re^{K}_{PAR}$, (resp., $\Re^{K,A}_{PAR,i}$
for $i=1,\ldots,n$)  the set $\Re_{P}$ (resp., $\Re_{P,i}^{A}$ for
$i=1,\ldots,n$) at termination of the algorithm. Let $M^{K}_{PAR}$
the parallel \MBRS
$\npla{\Re^{K}_{PAR},\npla{\Re^{K,A}_{PAR,1},\ldots,\Re_{PAR,n}^{K,A}}}$.
 Since a new rule is
added to $\Re_{P}$ only once, it follows that the following
property holds:
\begin{description}
\item[a.] Each rule $r$ in
$\Re^{K}_{PAR}\setminus\{r\in\Re|\textrm{$r$ is a PAR rule}\}$ has
the form $X$\Rule{K'}$Y$, with $X\in{}Var$,
$Y\in{}Var\cup\{\hat{Z}_{F}\}$ and
    $K'\in{}P_{n}$. Moreover, \maximalparK{r} $=K'$.
\end{description}
Now, let us show that $M^{K}_{PAR}$ satisfies  properties 1-5 of
definition \ref{def:MPAR-K}. Evidently, property  1 is
satisfied.\\
Let us prove  property 2. Let $r=X$\Rule{a}$Y.(Z)\in{}\Re$ and $Z$
\multiderparK{\sigma} $p$ for some term $p$, with \maximal{r}
$=K_{1}\subseteq{}K$ and \maximalparK{\sigma}
$=K_{2}\subseteq{}K$. Denoted by $K'$ the set $K_{1}\cup{}K_{2}$,
then we have to prove that
$r'=X$\Rule{K'}$\hat{Z}_{F}\in{}\Re^{K}_{PAR}$, and
\maximalparK{r'} $=K'$. From property \textbf{a} (notice that
$r'\notin\Re$) it suffices to prove that $r'$ is added to
$\Re_{P}$ during the computation. Let us consider the last
iteration of the \textbf{repeat} loop. Since any update of the
sets $\Re_{P}$,$\Re_{P,1}^{A},\ldots,\Re_{P,n}^{A}$ (the
\emph{flag} is set to $true$) involves a new iteration of this
loop, we deduce that in this computation phase
\begin{equation}
   \textrm{$\Re_{P}=\Re^{K}_{PAR}$  and
$\forall{}i=1,\ldots,n$ $\Re_{P,i}^{A}=\Re^{K,A}_{PAR,i}$}
\end{equation}
and they will not be updated anymore. Now, the rule
$r=X$\Rule{a}$Y.(Z)$ will be examined during an iteration of the
\textbf{for} loop in lines 6--36. From (1) it follows that during
the inner \textbf{for} loop (lines 8--23) iteration associated to
$K_{2}$, the condition of the \textbf{if} statement in line 9 is
 satisfied.
Since $\Re_{P}$  cannot be updated anymore, we deduce that the
condition of the  \textbf{if} statement in line 11 cannot be
satisfied. Therefore, $X$\Rule{K'}$\hat{Z}_{F}\in\Re_{P}$,
and the assertion is proved.\\
Following a similar reasoning we can easily prove that also
properties 3 and 4 in definition \ref{def:MPAR-K} are satisfied. \\
\noindent Now, let us prove  property 5 of definition
\ref{def:MPAR-K}. Let
$\overline{r}=X$\Rule{K'}$p'\in\Re^{K}_{PAR}\setminus\Re$.
Therefore, $X\in{}Var$, $p'\in{}Var\cup\{\hat{Z}_{F}\}$ and
$K'\in{}P_{n}$. We have to prove that $X$ \multider{\sigma} $t$
for some term $t$, with $|\sigma|>0$, \maximal{\sigma} $=K'$
 and $t=p'$ if $p'\in{}Var$.
 Let us assume  that $\overline{r}$ is
 the \emph{n}-th rule added to $\Re_{P}$
during the computation. Then, $\overline{r}$ is added to $\Re_{P}$
during an iteration of the \textbf{for} loop in lines 6-36, in
which a rule $r$ of the form $X$\Rule{a}$Y.(Z)\in\Re$ is examined.
Let $K_{1}=$ \maximal{r}.
The proof is by induction on $n$.\\
 \textbf{Base Step}: $n=1$. In this phase
\begin{enumerate}
    \item[1.] $\Re_{P}=\{r\in\Re|\textrm{ $r$ is a PAR rule}\}$
    and $\forall{}i=1,\ldots,n$
     $\Re_{P,i}^{A}=\{r\in\Re_{i}^{A}|\textrm{ $r$ is}$ a PAR
     rule\}.
     \item[2.] $\forall{}r'\in{}\Re_{P}$ we have \maximal{r'} =
     \maximalG{M_{p}}{r'}.
\end{enumerate}
 There are three cases:
\begin{itemize}
    \item $\overline{r}$ is added to
    $\Re_{P}$ in line 12.
    Then, $p'=\hat{Z}_{F}\notin{}Var$, and  the condition of the
    \textbf{if} statement in line 9 is satisfied:
    $\prsdernorm{Z}{\sigma}{\Re_{P}}{p}$ for some $p$, with
    \maximalG{M_{P}}{\sigma} $=K_{2}$ and
    $K'=K_{1}\cup{}K_{2}$. From 1 we deduce that $\sigma$ is a sequence
    of PAR rules in $\Re$. From 2 it follows that \maximal{\sigma} =
    \maximalG{M_{P}}{\sigma}. Therefore,
    $X$ \prsoneder{r} $Y.(Z)$ \multider{\sigma} $Y.(p)$ with
    \maximal{r\sigma} $=K_{1}\cup{}K_{2}=K'$, and the assertion is proved.
    \item $\overline{r}$ is added to
    $\Re_{P}$ in line 19.
    Then, $p'=Y\in{}Var$, and  the condition of the
    \textbf{if} statement in line 16 is satisfied:
    $\prsdernorm{Z}{\sigma}{\Re_{P}}{\varepsilon}$ with
    \maximalG{M_{P}}{\sigma} $=K_{2}$ and
    $K'=K_{1}\cup{}K_{2}$. From 1 we deduce that $\sigma$ is a sequence
    of PAR rules in $\Re$. From 2 it follows that \maximal{\sigma} =
    \maximalG{M_{P}}{\sigma}. Therefore,
    $X$ \prsoneder{r} $Y.(Z)$ \multider{\sigma} $Y$ with
    \maximal{r\sigma} $=K_{1}\cup{}K_{2}=K'$, and the assertion is proved.
    \item $\overline{r}$ is added to
    $\Re_{P}$ by the inner \textbf{for} loop in lines 24-35, when
     a rule  $r'$ of the form $Y.(W)$\Rule{b}$W'\in\Re$ is examined.
    Then, $p'=W'\in{}Var$, and $\overline{r}$ is added to
      $\Re_{P}$ in line 30. So,
    the condition of the \textbf{if} statement  in linea 27 is satisfied:
    $\prsdernorm{Z}{\sigma}{\Re_{P}}{W}$ with
    \maximalG{M_{P}}{\sigma} $=K_{3}$ and
    $K'=K_{1}\cup{}K_{2}\cup{}K_{3}$, where $K_{2}=$ \maximal{r'}.
    From 1 we deduce that $\sigma$ is a sequence
    of PAR rules in $\Re$. From 2 it follows that \maximal{\sigma} =
    \maximalG{M_{P}}{\sigma}. Therefore,
    $X$ \prsoneder{r'} $Y.(Z)$ \multider{\sigma} $Y.(W)$ \prsoneder{r''} $W'$
    with
    \maximal{r\sigma{}r'} $=K_{1}\cup{}K_{2}\cup{}K_{3}=K'$, and the assertion is proved.
\end{itemize}
 \textbf{Induction Step}: $n>1$.  Let $\overline{\Re}_{P}$
  (resp., $\overline{\Re}_{P,i}^{A}$  for $i=1,\ldots,n$) be
  the set of the rules in $\Re^{K}_{PAR}\setminus\Re$
  (resp., $\Re^{K,A}_{PAR,i}\setminus\Re$ for $i=1,\ldots,n$) after $n-1$ rules have been
  added to $\Re_{P}$. Then,
  in this phase we have
\begin{itemize}
    \item $\Re_{P}$ $=\{r\in\Re|$ $r$ is a PAR
    rule$\}\cup\overline{\Re}_{P}$, and $\forall{}i=1,\ldots,n$ $\Re_{P,i}^{A}$
$=\{r\in\Re_{i}^{A}|$ $r$ is a PAR
rule$\}\cup\overline{\Re}_{P,i}^{A}$.
    \item $\forall{}\hat{r}=\hat{t}$\Rule{\hat{K}}$\hat{t}'\in\overline{\Re}_{P}$
    we have \maximalG{M_{P}}{\hat{r}} $=\hat{K}$.
    \item $\forall{}r'\in{}\{r\in\Re|$ $r$ is a PAR
    rule$\}$ we have \maximal{r'} =
     \maximalG{M_{p}}{r'}.
\end{itemize}
>From the inductive hypothesis, we deduce easily that the following
property is satisfied
\begin{enumerate}
    \item[3.] If $\prsdernorm{\hat{t}}{\sigma}{\Re_{P}}{\hat{t}'}$, where
     $\hat{t}$ is a parallel term,
    then $\hat{t}$ \multider{\rho} $\hat{t}''$ for some term $\hat{t}''$,  with \maximal{\rho}
    = \maximalG{M_{P}}{\sigma}. Moreover, if $\hat{t}'$ doesn't
    contain occurrences of $\hat{Z}_{F}$ then
    $\hat{t}''=\hat{t}'$.
\end{enumerate}
As before, there are three cases:
\begin{itemize}
    \item $\overline{r}$ is added to
    $\Re_{P}$ in line 12. Then, $p'=\hat{Z}_{F}\notin{}Var$,
     and  the condition of the
    \textbf{if} statement in line 9 is satisfied:
    $\prsdernorm{Z}{\sigma}{\Re_{P}}{p}$ for some term $p$, with
    \maximalG{M_{P}}{\sigma} $=K_{2}$ and
    $K'=K_{1}\cup{}K_{2}$. From 3 it follows that
    $Z$ \multider{\rho} $t$ for some term $t$, with \maximal{\rho}
    = \maximalG{M_{P}}{\sigma}. Therefore,
    $X$ \prsoneder{r} $Y.(Z)$ \multider{\rho} $Y.(t)$ with
    \maximal{r\rho} $=K_{1}\cup{}K_{2}=K'$, and the assertion is proved.
    \item $\overline{r}$ is added to
    $\Re_{P}$ in line 19.
    Then, $p'=Y\in{}Var$, and  the condition of the
    \textbf{if} statement in line 16 is satisfied:
    $\prsdernorm{Z}{\sigma}{\Re_{P}}{\varepsilon}$ with
    \maximalG{M_{P}}{\sigma} $=K_{2}$ and
    $K'=K_{1}\cup{}K_{2}$. From 3 it follows that
    $Z$ \multider{\rho} $\varepsilon$, with \maximal{\rho}
    = \maximalG{M_{P}}{\sigma}. Therefore,
    $X$ \prsoneder{r} $Y.(Z)$ \multider{\rho} $Y$ with
    \maximal{r\rho} $=K_{1}\cup{}K_{2}=K'$, and the assertion is proved.
    \item $\overline{r}$ is added to
    $\Re_{P}$ by the inner \textbf{for} loop in lines 24-35, when
     a rule  $r'$ of the form $Y.(W)$\Rule{b}$W'\in\Re$ is examined.
    Then, $p'=W'\in{}Var$, and $\overline{r}$ is added to
      $\Re_{P}$ in line 30. So,
    the condition of the \textbf{if} statement in line 27 is satisfied:
    $\prsdernorm{Z}{\sigma}{\Re_{P}}{W}$ with
    \maximalG{M_{P}}{\sigma} $=K_{3}$ and
    $K'=K_{1}\cup{}K_{2}\cup{}K_{3}$, where $K_{2}=$ \maximal{r'}.
    From 3 it follows that
    $Z$ \multider{\rho} $W$ with \maximal{\rho}
    = \maximalG{M_{P}}{\sigma}. Therefore,
    $X$ \prsoneder{r'} $Y.(Z)$ \multider{\rho} $Y.(W)$ \prsoneder{r''} $W'$
    with
    \maximal{r\rho{}r'} $=K_{1}\cup{}K_{2}\cup{}K_{3}=K'$, and the assertion is proved.
\end{itemize}

\noindent{} Finally, it's easy to show that $M^{K}_{PAR}$ is the
least parallel \MBRS over $Var$ and the alphabet
$\overline{\Sigma}$ satisfying  properties 1-5 of definition
\ref{def:MPAR-K}.
\end{proof}

\begin{Remark}\label{Remark:MKPAR}By construction the following
properties hold:
\begin{itemize}
    \item $\forall{}r\in\Re\cap\Re^{K}_{PAR}$ we have \maximalparK{r}
    = \maximal{r}.
    \item $\forall{}r=X$\Rule{K'}$p\in\Re_{PAR}^{K}\setminus\Re$ we have \maximalparK{r}
        = $K'$.
\end{itemize}
\end{Remark}

Soundness and completeness of the procedure described above is
stated by the following theorem, whose proof is reported in
appendix (section B).

\begin{Theorem}\label{Theorem:Problem1}
For all $X\in{}Var$ there exists  a   $(K,\emptyset)$-accepting
finite derivation in $M$ from $X$ if, and only if, there exists a
$(K,\emptyset)$-accepting finite derivation in
    $M^{K}_{PAR}$ from $X$.
\end{Theorem}

This result, together  with proposition \ref{Prop:ParMBRS-F},
allow us to conclude that
 Problem 1, stated at the beginning of this section, is decidable.

\subsection{Decidability results on infinite derivations of \MBRSs in normal form}\label{sec:infinite}

\setcounter{equation}{0}

In this section we prove the decidability of Problem 2 stated in
subsection~\ref{sec:Model-checkingPRS},  that for clarity we
recall.
\begin{description}
    \item[Problem 2]\emph{ Given a \MBRS in normal form
    $M=\npla{\Re,\npla{\Re_{1}^{A},\ldots,\Re_{n}^{A}}}$ over Var and
    the alphabet $\Sigma$, given a variable $X\in{}Var$ and two
    sets
    $K,K^{\omega}\in{}P_{n}$, to decide if there exists a  $(K,K^{\omega})$-accepting infinite derivation in $M$
    from $X$}.
\end{description}

Let us observe that a necessary condition for the existence of a
$(K,K^{\omega})$-accepting infinite derivation
in $M$ is that $K\supseteq{}K^{\omega}$.\\
The proof of decidability is by induction on $|K|+|K^{\omega}|$.\\
 \textbf{Base Step}: $|K|=0$ and $|K^{\omega}|=0$. Let $M_{F}=\npla{\Re,\Re_{F}}$
 be the \BRS with $\Re_F=\bigcup_{i=1}^{n}\Re_{i}^{A}$. Given an
 infinite derivation $X$ \multider{\sigma} in $\Re$ from a variable $X$ then,
this derivation is $(\emptyset,\emptyset)$-accepting in $M$ if,
and only if, it doesn't contain occurrences of accepting rules in
$M_{F}$. So, the decidability result follows from theorem
\ref{Theorem:OldResults}.\\
  \textbf{Inductive Step}:  $|K|+|K^{\omega}|>0$. From the
inductive hypothesis, for each $K'\subseteq{}K$ and
$K'^{\omega}\subseteq{}K^{\omega}$ with
 $|K'|+|K'^{\omega}|<|K|+|K^{\omega}|$ the
result holds. In other words,   it is decidable if there exists a
$(K',K'^{\omega})$-accepting infinite derivation in $M$ from a
variable $X$.
 Starting from this assumption
we show that  problem 2, with input the sets $K$ and $K^{\omega}$,
can be reduced to (a combination of) two similar, but simpler,
problems: the first is a decidability problem on infinite
derivations restricted to parallel \MBRSs; the second is a
decidability problem on infinite derivations restricted to
sequential \MBRSs. More precisely, we show that it is possible to
 construct effectively two parallel \MBRSs
$M^{K,K^{\omega}}_{PAR}=\npla{\Re^{K,K^{\omega}}_{PAR},\npla{\Re^{K,K^{\omega},A}_{PAR,1},\ldots,\Re_{PAR,n}^{K,K^{\omega},A}}}$
and
$M^{K,K^{\omega}}_{PAR,\infty}=\npla{\Re^{K,K^{\omega}}_{PAR},\npla{\Re^{K,K^{\omega},A}_{PAR,\infty,1},\ldots,\Re_{PAR,\infty,n}^{K,K^{\omega},A}}}$
 with the same support,  and a sequential \MBRS
$M_{SEQ}^{K}=\npla{\Re_{SEQ}^{K},\npla{\Re^{K,A}_{SEQ,1},\ldots,\Re_{SEQ,n}^{K,A}}}$,
in such a way that Problem 2, with input the sets $K$ and
$K^{\omega}$ and a variable $X\in{}Var$, is reducible to one of
two following problems depending if $K\supset{}K^{\omega}$ or
$K=K^{\omega}$.\\

\noindent\textbf{Problem 3} $(K\supset{}K^{\omega})$. To decide if
the following condition is satisfied:
\begin{itemize}
        \item There exists a variable  $Y\in{}Var$ reachable from $X$ in
        $\Re_{SEQ}^{K}$ through a $(K',\emptyset)$-accepting derivation in $M_{SEQ}^{K}$
        with $K'\subseteq{}K$, and there exists a derivation
        $Y$ \multiderparKomega{\rho}
        such that \maximalparKomega{\rho} = $K$
        and
        \maximalparKomegaInf{\rho} $\cup$ \maximalparKomegaPlus{\rho} =
        $K^{\omega}$. Moreover, either $\rho$ is
        infinite or $\rho$ contains some occurrence of rule in
        $\Re^{K,K^{\omega}}_{PAR}\setminus\Re_{PAR}^{K}$ (where $\Re_{PAR}^{K}$ is the
        support of the parallel \MBRS $M_{PAR}^{K}$ defined in the previous subsection).
\end{itemize}

\noindent\textbf{Problem 4} $(K=K^{\omega})$. To decide if one of
the following conditions is satisfied:
\begin{itemize}
        \item There exists a variable  $Y\in{}Var$ reachable from $X$ in
        $\Re_{SEQ}^{K}$ through a $(K',\emptyset)$-accepting derivation in $M_{SEQ}^{K}$
        with $K'\subseteq{}K$, and there exists a derivation
        $Y$ \multiderparKomega{\rho}
        such that \maximalparKomega{\rho} = $K$
        and
        \maximalparKomegaInf{\rho} $\cup$ \maximalparKomegaPlus{\rho} =
        $K^{\omega}$. Moreover, either $\rho$ is
        infinite or $\rho$ contains some occurrence of rule in
        $\Re^{K,K^{\omega}}_{PAR}\setminus\Re_{PAR}^{K}$.
        \item There exists a
     $(K,K^{\omega})$-accepting infinite derivation  in $M_{SEQ}^{K}$ from
     $X$.
\end{itemize}

 Since these last problems are decidable (see theorems
\ref{Theorem:ConditionForProblem2-1} and
\ref{Theorem:ConditionForProblem2-2}), decidability of Problem 2
is entailed.

Before illustrating the main idea underlying our approach, we need
the following definition.

\begin{Definition}
Let us denote by $\Pi^{K,K^{\omega}}_{PAR,\infty}$ the set of
derivations $t$ \multider{\sigma} in $\Re$ \emph{\textbf{not}}
satisfying the following property:
\begin{itemize}
    \item There exists a subderivation of $t$ \multider{\sigma}
    that is  a $(K,K^{\omega})$-accepting infinite derivation in
    $M$.
\end{itemize}
\end{Definition}

 In the following we use a new variable  $\hat{Z}_{\infty}$, and denote
by  $T$ (resp., $T_{PAR}$, $T_{SEQ}$)  the set of process terms
(resp., the set of terms in which no sequential composition
occurs, the set of terms in which no  parallel composition occurs)
over
$Var\cup\{\hat{Z}_{F},\hat{Z}_{\infty}\}$\footnote{$\hat{Z}_{F}$
is the variable used in the previous subsection for the
construction of the parallel \MBRS $M_{PAR}^{K}$}.\newline

\noindent{}Let us sketch the main ideas at the basis of our
technique. At first, let us focus  on the class of derivations
$\Pi^{K,K^{\omega}}_{PAR,\infty}$, showing how it is possible to
mimic a $(K,K^{\omega})$-accepting infinite derivation in $M$ from
a variable, belonging to this class, by using only PAR rules
belonging to  extensions of the parallel \MBRS $M_{PAR}^{K}$
computed by the algorithm of lemma \ref{Lemma:Algo} (with input
$M$ and $K$). More precisely, we'll show, as anticipated, that it
is possible construct two parallel extensions of $M_{PAR}^{K}$
with the same support, denoted by
$M^{K,K^{\omega}}_{PAR}=\npla{\Re^{K,K^{\omega}}_{PAR},\npla{\Re^{K,K^{\omega},A}_{PAR,1},\ldots,\Re_{PAR,n}^{K,K^{\omega},A}}}$
and
$M^{K,K^{\omega}}_{PAR,\infty}=\npla{\Re^{K,K^{\omega}}_{PAR},\npla{\Re^{K,K^{\omega},A}_{PAR,\infty,1},\ldots,\Re_{PAR,\infty,n}^{K,K^{\omega},A}}}$
(with $\Re^{K,K^{\omega},A}_{PAR,i}\supseteq{}\Re^{K,A}_{PAR,i}$
and
$\Re^{K,K^{\omega},A}_{PAR,\infty,i}\cap\Re^{K,A}_{PAR,i}=\emptyset$
for $i=1,\ldots,n$), in such way that the following condition
holds:
\begin{description}
    \item[i.]  There exists a $(\overline{K},\overline{K}^{\omega})$-accepting
derivation  in $M$ belonging to $\Pi^{K,K^{\omega}}_{PAR,\infty}$
of the form $p$ \multider{\sigma}, with $p\in{}T_{PAR}$,
$\overline{K}\subseteq{}K$ and
$\overline{K}^{\omega}\subseteq{}K^{\omega}$ if, and only if,
there exists a derivation  $p$ \multiderparKomega{\rho} in
$\Re^{K,K^{\omega}}_{PAR}$
 from $p$ such that \maximalparKomega{\rho} = $\overline{K}$ and
    \maximalparKomegaInf{\rho} $\cup$ \maximalparKomegaPlus{\rho} =
    $\overline{K}^{\omega}$. Moreover, if $\sigma$ is infinite then, either $\rho$ is
    infinite or $\rho$ contains some occurrence of rule in
    $\Re^{K,K^{\omega}}_{PAR}\setminus\Re_{PAR}^{K}$, and vice versa.
\end{description}

So, let $p$ \multider{\sigma} be a
$(\overline{K},\overline{K}^{\omega})$-accepting derivation  in
$M$ belonging to $\Pi^{K,K^{\omega}}_{PAR,\infty}$ with
$p\in{}T_{PAR}$, $\overline{K}\subseteq{}K$ and
$\overline{K}^{\omega}\subseteq{}K^{\omega}$.  Then, all its
possible subderivations contain all, and only, the rule
occurrences in $\sigma$ applied at a level $k$ greater than $0$ in
$p$ \multider{\sigma}. If $\sigma$ contains only PAR rule
occurrences the statement \textbf{i} is evident, since by
construction (remember that $\Re_{PAR}^{K}$ contains all PAR rules
of $\Re$) we have $p$ \multiderparKomega{\sigma} with
\maximalparKomega{\sigma} = $\overline{K}$,
    \maximalparKomegaInf{\sigma}  =
    $\overline{K}^{\omega}$ and \maximalparKomegaPlus{\sigma} = $\emptyset$. Otherwise,
$p$ \multider{\sigma} can be written in the form:
 \begin{equation}\label{eq:derivation}
 \textrm{$p$ \multider{\lambda} $t$$\parallel$$X$ \prsoneder{r}
 $t$$\parallel$$Y.(Z)$ \multider{\omega}}
 \end{equation}
 where $r = X$\Rule{a}$Y.(Z)$,  $\lambda$ contains only occurrences of PAR rules
 in $\Re$,  $t\in{}T_{PAR}$ and $X,Y,Z\in{}Var$. Let $Z$ \multider{\rho} be a subderivation of
 $t$$\parallel$$Y.(Z)$ \multider{\omega} from $Z$.
Since $p$ \multider{\sigma} is in
$\Pi^{K,K^{\omega}}_{PAR,\infty}$, $Z$ \multider{\rho} is
\emph{not} a $(K,K^{\omega})$-accepting infinite derivation in
$M$. More precisely, \maximal{\rho} $\subseteq{}K$,
\maximalInf{\rho} $\subseteq{}K^{\omega}$ (since $\rho$ is a
subsequence of $\sigma$) and
$|$\maximal{\rho}$|+|$\maximalInf{\rho}$|<|K|+|K^{\omega}|$. Thus,
 only one of the following four cases may occur:
\begin{description}
\item[A] $Z$ \multider{\rho} leads to the term $\varepsilon$, and
  $p$ \multider{\sigma} is of the form
\begin{equation}\label{eq:form1}
 \textrm{$p$ \multider{\lambda} $t$$\parallel$$X$ \prsoneder{r}
 $t$$\parallel$$Y.(Z)$ \multider{\omega_{1}} $\overline{t}$$\parallel$$Y$ \multider{\omega_{2}}}
\end{equation}
  where $\rho$ is a subsequence of $\omega_1$ and $t$
  \multider{\omega_{1}\setminus\rho} $\overline{t}$.  The  derivation above is
  $(\overline{K},\overline{K}^{\omega})$-accepting in $M$ if, and only if, the following
   derivation
   is $(\overline{K},\overline{K}^{\omega})$-accepting in $M$
  \begin{equation}\label{eq:form1-2}
   \textrm{$p$ \multider{\lambda} $t$$\parallel$$X$ \prsoneder{r}
 $t$$\parallel$$Y.(Z)$ \multider{\rho} $t$$\parallel$$Y$ \multider{\xi}
 $\overline{t}$$\parallel$$Y$ \multider{\omega_{2}}}
  \end{equation}
 where $\xi=\omega_{1}\setminus\rho$.
 Let us consider the derivation $X$ \prsoneder{r} $Y.(Z)$
 \multider{\rho} $Y$ where \maximal{r\rho} $\subseteq{}K$. From properties of $M^{K}_{PAR}$ (see lemma \ref{Lemma:From-M-To-MKPAR} in appendix)
 there exists a derivation of the form $X$ \multiderparK{\eta} $Y$
 such that \maximalparK{\eta} = \maximal{r\rho}.
  Now, we can apply recursively the same reasoning to the
  derivation in $\Re$ from $t$$\parallel$$Y\in{}T_{PAR}$
  \begin{equation}
    \textrm{$t$$\parallel$$Y$ \multider{\xi} $\overline{t}$$\parallel$$Y$
      \multider{\omega_{2}}}
  \end{equation}
  which belongs to $\Pi^{K,K^{\omega}}_{PAR,\infty}$ and whose finite (resp., infinite)
  maximal as to $M$ is
  contained in $K$ (resp., $K^{\omega}$).
\item[B] $Z$ \multider{\rho} leads to a variable $W$, and
  $p$ \multider{\sigma} can be written as
\begin{equation}\label{eq:form1}
 \textrm{$p$ \multider{\lambda} $t$$\parallel$$X$ \prsoneder{r}
 $t$$\parallel$$Y.(Z)$ \multider{\omega_{1}} $\overline{t}$$\parallel$$Y.(W)$
 \prsoneder{r'} $\overline{t}$$\parallel$$W'$ \multider{\omega_{2}}}
\end{equation}
  where $r'=Y.(W)$\Rule{b}$W'$ (with $W'\in{}Var$),
   $\rho$ is a subsequence of $\omega_1$ and $t$
  \multider{\omega_{1}\setminus\rho} $\overline{t}$.  The derivation above is
  $(\overline{K},\overline{K}^{\omega})$-accepting if, and only if, the following derivation
  is $(\overline{K},\overline{K}^{\omega})$-accepting
  \begin{equation}\label{eq:form1-2}
   \textrm{$p$ \multider{\lambda} $t$$\parallel$$X$ \prsoneder{r}
 $t$$\parallel$$Y.(Z)$ \multider{\rho} $t$$\parallel$$Y.(W)$  \prsoneder{r'} $t$$\parallel$$W'$
    \multider{\xi}
 $\overline{t}$$\parallel$$W'$ \multider{\omega_{2}}}
  \end{equation}
  where $\xi=\omega_{1}\setminus\rho$.
  Let us consider the derivation $X$ \prsoneder{r} $Y.(Z)$
 \multider{\rho} $Y.(W)$ \prsoneder{r'} $W'$ where \maximal{rr'\rho} $\subseteq{}K$.
 From properties of $M^{K}_{PAR}$ (see lemma \ref{Lemma:From-M-To-MKPAR} in appendix)
 there exists a derivation of the form $X$ \multiderparK{\eta} $W'$
 such that \maximalparK{\eta} = \maximal{rr'\rho}.
  Now, we can apply recursively the same reasoning to the
   derivation in $\Re$ from $t$$\parallel$$W'\in{}T_{PAR}$
  \begin{equation}
    \textrm{$t$$\parallel$$W'$ \multider{\xi} $\overline{t}$$\parallel$$W'$
      \multider{\omega_{2}}}
  \end{equation}
  which belongs to $\Pi^{K,K^{\omega}}_{PAR,\infty}$ and whose finite (resp., infinite)
  maximal as to $M$ is
  contained in $K$ (resp., $K^{\omega}$).
\item[C]  $Z$ \multider{\rho} is finite and does not influence
 the
  applicability of rule occurrences in $\omega \setminus \rho$ in the derivation
  $t$$\parallel$$Y.(Z)$ \multider{\omega}.  In other
  words, we have $t$ \multider{\omega\setminus\rho}. Moreover,
  \maximal{r\rho} $\subseteq{}K$.
  Let us consider the finite derivation $X$ \prsoneder{r} $Y.(Z)$
 \multider{\rho}.
 From properties of $M^{K}_{PAR}$ (see lemma \ref{Lemma:From-M-To-MKPAR} in appendix)
 there exists a finite derivation of the form $X$
 \multiderparK{\eta} $\overline{p}$ with $\overline{p}\in{}T_{PAR}$
 and \maximalparK{\eta} = \maximal{r\rho}.
  Now, we can apply recursively the same reasoning to the
   derivation $t$$\parallel$$\overline{p}$ \multider{\omega\setminus\rho} in $\Re$ from
   $t$$\parallel$$\overline{p}$ (where $t$$\parallel$$\overline{p}\in{}T_{PAR}$),
  which belongs to $\Pi^{K,K^{\omega}}_{PAR,\infty}$ and whose finite (resp., infinite)
  maximal as to $M$ is
  contained in $K$ (resp., $K^{\omega}$).
\item[D]  $Z$ \multider{\rho} is infinite. From the definition of
   subderivation we have $t$ \multider{\omega\setminus\rho}.
Moreover,
  \maximal{\rho} $=K_1\subseteq{}K$, \maximalInf{\rho} $=K_{1}^{\omega}\subseteq{}K^{\omega}$,
\maximal{r} $=K_2\subseteq{}K$, \maximalInf{r} $=\emptyset$
  and  $|K_1|+|K_{1}^{\omega}|<|K|+|K^{\omega}|$.
  From our assumptions (inductive hypothesis)
   it is decidable if there exists a
  $(K_1,K_{1}^{\omega})$-accepting infinite derivation in $M$ from
  variable $Z$.
  Then, we keep track of  the infinite sequence $r\rho$
  by adding the new variable $\hat{Z}_{\infty}$ (where $\hat{Z}_{\infty}\notin{}Var$)  and
  a  PAR rule of the form $r'=X$\Rule{K',K_{1}^{\omega}}$\hat{Z}_{F}$  with
  $K'=K_1\cup{}K_2$,
  \maximalparKomega{r'} = $K'$ and \maximalparKomegaPlus{r'} = $K_{1}^{\omega}$.
   Now, we can apply recursively the same reasoning to the
   derivation $t$$\parallel$$\hat{Z}_{\infty}$ \multider{\omega\setminus\rho} in $\Re$ from
   $t$$\parallel$$\hat{Z}_{\infty}$ (where $t$$\parallel$$\hat{Z}_{\infty}\in{}T_{PAR}$),
  which belongs to $\Pi^{K,K^{\omega}}_{PAR,\infty}$ and whose finite (resp., infinite)
  maximal as to $M$ is
  contained in $K$ (resp., $K^{\omega}$).
 \end{description}

In other words,  all  subderivations in $p$ \multider{\sigma} are
abstracted away by PAR rules not belonging to $\Re$,
 according to the intuitions given above.\newline

 By the parallel \MBRS $M_{PAR}^{K}$  we  keep track of subderivations of the forms \textbf{A},
 \textbf{B} and \textbf{C}.
In  order to simulate subderivations of the form \textbf{D}, we
need to add additional PAR rules in $M_{PAR}^{K}$.
 The following definition provides an extension of
$M_{PAR}^{K}$ suitable for our purposes.

\begin{Definition}\label{Def:M-PARs-Infinite}
By
    $M^{K,K^{\omega}}_{PAR}=
   \npla{\Re^{K,K^{\omega}}_{PAR},\npla{\Re^{K,K^{\omega},A}_{PAR,1},\ldots,
   \Re^{K,K^{\omega},A}_{PAR,n}}}$
   and \newline
   $M^{K,K^{\omega}}_{PAR,\infty}=
   \npla{\Re^{K,K^{\omega}}_{PAR},\npla{\Re^{K,K^{\omega},A}_{PAR,\infty,1},\ldots,
   \Re^{K,K^{\omega},A}_{PAR,\infty,n}}}$ we denote the parallel \MBRSs
   over $Var\cup\{\hat{Z}_{F},\hat{Z}_{\infty}\}$ and the alphabet
   $\Sigma\cup{}P_n\cup{}P_n\times{}P_n$,
  defined by $M$ and $M_{PAR}^{K}$
   in the following way:
\begin{itemize}
\item $\Re^{K,K^{\omega}}_{PAR}=\begin{array}[t]{l}%
    \Re_{PAR}^{K} \cup \\
    \{\prsrule{X}{\overline{K},\overline{K}^{\omega}}{\hat{Z}_{\infty}} \mid \begin{array}[t]{l}
         \overline{K}\subseteq{}K,
         \overline{K}^{\omega}\subseteq{}K^{\omega},
          \text{ there exists a
         rule } r=\prsrule{X}{a}{Y.(Z)}\in\Re \\[4pt]
       \text{and
             an infinite  derivation } \prsder{Z}{\sigma}{} \text{
         such that } \\[4pt] |\text{\maximal{\sigma}}| +
         |\text{\maximalInf{\sigma}}|
         < |K|+|K^{\omega}| \text{ and} \\[4pt]
        \text{ \maximal{\sigma} $\cup$ \maximal{r} } =\overline{K} \text{ and}
           \text{ \maximalInf{\sigma} } =\overline{K}^{\omega}
         \}\
     \end{array}
  \end{array}$
\item
   $\Re^{K,K^{\omega},A}_{PAR,i}=\Re_{PAR,i}^{K,A}\cup\{
  X$\Rule{\overline{K},\overline{K}^{\omega}}$\hat{Z}_{\infty}\in\Re^{K,K^{\omega}}_{PAR}|$
  $i\in{}\overline{K}$\} for all $i=1,\ldots,n$
\item
  $\Re^{K,K^{\omega},A}_{PAR,i,\infty}=\{
  X$\Rule{\overline{K},\overline{K}^{\omega}}$\hat{Z}_{\infty}\in\Re^{K,K^{\omega}}_{PAR}|$
  $i\in{}\overline{K}^{\omega}$\} for all $i=1,\ldots,n$
\end{itemize}
\end{Definition}

\noindent{}From the inductive hypothesis on the decidability of
problem 2 for sets $\overline{K},\overline{K}^{\omega}\in{}P_{n}$
such that $\overline{K}\subseteq{}K$,
$\overline{K}^{\omega}\subseteq{}K^{\omega}$ and
$|\overline{K}|+|\overline{K}^{\omega}|<|K|+|K^{\omega}|$,   it
follows that $M^{K,K^{\omega}}_{PAR}$ and
$M^{K,K^{\omega}}_{PAR,\infty}$ can be built effectively. Thus,
the following result holds.

\begin{Lemma}
$M^{K,K^{\omega}}_{PAR}$ and $M^{K,K^{\omega}}_{PAR,\infty}$ can
be built effectively.
\end{Lemma}

\begin{Remark}\label{Remark:MPAR-Inf}By construction, the
following properties hold:
\begin{itemize}
    \item for all $r\in\Re_{PAR}^{K}$  we have \maximalparKomega{r} =
          \maximalparK{r} and \maximalparKomegaPlus{r}
          $=\emptyset$.
    \item for all $r\in\Re^{K,K^{\omega}}_{PAR}\cap\Re$  we have \maximalparKomega{r} =
          \maximal{r} and \maximalparKomegaPlus{r}
          $=\emptyset$.
    \item for all $r=X$\Rule{\overline{K},\overline{K}^{\omega}}$\hat{Z}_{\infty}\in
    \Re^{K,K^{\omega}}_{PAR}$
    we have \maximalparKomega{r} = $\overline{K}$ and
          \maximalparKomegaPlus{r} = $\overline{K}^{\omega}$.
\end{itemize}
\end{Remark}

Now, let us go back to Problem 2 and consider a
$(K,K^{\omega})$-accepting infinite derivation in $M$ from a
variable $X$ of the form $X$ \multider{\sigma}, and non belonging
to $\Pi^{K,K^{\omega}}_{PAR,\infty}$.  In this case, the
derivation $X$ \multider{\sigma} can be written in the form $X$
\multider{} $t$$\parallel$$Y.(Z)$ \multider{\rho}, with
$Z\in{}Var$, and such that there exists a subderivation of
$t$$\parallel$$Y.(Z)$ \multider{\rho} from $Z$ that is a
$(K,K^{\omega})$-accepting infinite derivation in $M$. To manage
this kind of derivation, we build, starting from the \MBRSs $M$
and $M_{PAR}^{K}$, a sequential \MBRS $M_{SEQ}^{K}$ according to
the following definition:

\begin{Definition}\label{Def:M-SEQ}
  By $M_{SEQ}^{K}=\npla{\Re_{SEQ}^{K},\npla{\Re_{SEQ,1}^{K,A},\ldots,\Re_{SEQ,n}^{K,A}}}$ we denote
  the sequential \MBRS
  over $Var$ and the alphabet $\overline{\Sigma}=\Sigma\cup{}P_{n}$ so defined:
\begin{itemize}
\item $\Re_{SEQ}^{K}=\begin{array}[t]{l}%
    \{\prsrule{X}{a}{Y.(Z)}\in\Re\}\ \cup \\
    \{\prsrule{X}{K'}{Y} \mid \begin{array}[t]{l}
         X,Y\in{}Var, K'\subseteq{}K \text{ and there exists a
         derivation } \prsderpar{X}{\sigma}{\parcomp{p}{Y}} \\[4pt]
       \text{ in } \Re_{PAR}^{K} \text{
         for some } p\in{}T_{PAR}
    \text{, with } |\sigma|>0 \text{ and  \maximalparK{\sigma} = $K'$} \}\
     \end{array}
  \end{array}$
\item
$\Re_{SEQ,i}^{K,A}=\{X$\Rule{a}$Y.(Z)\in\Re_{i}^{A}\}\cup\{X$\Rule{K'}$Y\in\Re_{SEQ}^{K}|
i\in{}K'$\} for all $i=1,\ldots,n$
\end{itemize}
\end{Definition}

\begin{Remark}\label{Remark:MKSEQ} By construction the following
properties hold
\begin{itemize}
    \item  for all $r\in\Re\cap\Re_{SEQ}^{K}$ we have
        \maximal{r} = \maximalseqK{r}.
    \item for all $r=X$\Rule{K'}$Y\in\Re_{SEQ}^{K}\setminus\Re$  we have \maximalseqK{r}
            = $K'$.
\end{itemize}
\end{Remark}

\begin{Lemma}
$M_{SEQ}^{K}$ can be built effectively.
\end{Lemma}

\begin{proof}
 The result   follows directly from the definition of $M_{SEQ}^{K}$
 and proposition \ref{Prop:ParMBRS-F}.
\end{proof}

\setcounter{equation}{0}

Soundness and completeness of the procedure described above is
stated by the following two theorems, whose proof is reported  in
 appendix (section C).

\begin{Theorem}\label{Theorem:Problem2.1}
Let us assume that $K\neq{}K^{\omega}$. Given $X\in{}Var$, there
exists a
 $(K,K^{\omega})$-accepting infinite derivation  in $M$ from
 $X$ if, and only if,  the following property is satisfied:
\begin{itemize}
    \item There exists a variable  $Y\in{}Var$ reachable from $X$ in
    $\Re_{SEQ}^{K}$ through a $(K',\emptyset)$-accepting derivation in $M_{SEQ}^{K}$
    with $K'\subseteq{}K$, and there exists a derivation
    $Y$ \multiderparKomega{\rho}
    such that \maximalparKomega{\rho} = $K$
    and
    \maximalparKomegaInf{\rho} $\cup$ \maximalparKomegaPlus{\rho} =
    $K^{\omega}$. Moreover, either $\rho$ is
    infinite or $\rho$ contains some occurrence of rule in
    $\Re^{K,K^{\omega}}_{PAR}\setminus\Re_{PAR}^{K}$.
\end{itemize}
\end{Theorem}

\begin{Theorem}\label{Theorem:Problem2.2}
Let us assume that $K=K^{\omega}$. Given $X\in{}Var$, there exists
a
 $(K,K^{\omega})$-accepting infinite derivation  in $M$ from
 $X$ if, and only if, one of  the following properties is satisfied:
\begin{enumerate}
    \item There exists a variable  $Y\in{}Var$ reachable from $X$ in
    $\Re_{SEQ}^{K}$ through a $(K',\emptyset)$-accepting derivation in $M_{SEQ}^{K}$
    with $K'\subseteq{}K$, and there exists a derivation
    $Y$ \multiderparKomega{\rho}
    such that \maximalparKomega{\rho} = $K$
    and
    \maximalparKomegaInf{\rho} $\cup$ \maximalparKomegaPlus{\rho} =
    $K^{\omega}$. Moreover, either $\rho$ is
    infinite or $\rho$ contains some occurrence of rule in
    $\Re^{K,K^{\omega}}_{PAR}\setminus\Re_{PAR}^{K}$.
    \item There exists a
     $(K,K^{\omega})$-accepting infinite derivation  in $M_{SEQ}^{K}$ from
     $X$.
\end{enumerate}
\end{Theorem}

These two results, together  with theorems
\ref{Theorem:ConditionForProblem2-1} and
\ref{Theorem:ConditionForProblem2-2}, allow us to conclude that
 Pro\-blem 2, stated at the beginning of this subsection, is decidable.

\bibliographystyle{plain}


\newpage

\appendix
\begin{LARGE}
\textbf{APPENDIX}
\end{LARGE}

\section{Definitions and simple properties}


In this section we give some definitions and deduce simple
properties that will be used in  sections B and C for the proof of
theorems 4.1-4.3.

In the following $\hat{Var}$ denotes the set of variables
$Var\cup\{\hat{Z}_{F},\hat{Z}_{\infty}\}$, $T$ denotes the set of
terms over $\hat{Var}$, and $T_{PAR}$ (resp., $T_{SEQ}$) the set
of terms in $T$ not containing sequential (resp., parallel)
composition.

\begin{Definition}
The set of {\em subterms\/} of a term $t\in{}T$, denoted by
$SubTerms(t)$, is defined inductively as follows:
\begin{itemize}
\item $SubTerms(\varepsilon)=\{\varepsilon\}$ \item
$SubTerms(X)=\{X\}$, for all  $X\in{}\hat{Var}$ \item
$SubTerms(X.(t))=SubTerms(t)\cup\{X.(t)\}$, for all $X.(t)\in{}T$
with  $t\neq\varepsilon$ \item
$SubTerms($$t_{1}$$\parallel$$t_{2})=\bigcup_{(t_{1}',t_{2}')\in{}S}
(SubTerms(t_{1}')\cup{} SubTerms(t_{2}'))$ $\cup$
$\{t_{1}$$\parallel$$t_{2}\}$,
\newline
with $S=\{(t_{1}',t_{2}')\in{}T\times{}T \mid
t_{1}',t_{2}'\neq\varepsilon$
     and $t_{1}$$\parallel$$t_{2}={}t_{1}'$$\parallel$$t_{2}'\}$ and
$t_{1},t_{2}\in{}T\setminus \{\varepsilon\}$.
\end{itemize}
\end{Definition}

\begin{Definition} The set of terms obtained from a term $t\in T$
{\em substituting\/} an occurrence of a subterm $st$ of $t$ with a
term $t'\in T$, denoted by $t[st\rightarrow{}t']$, is defined
inductively as follows:
\begin{itemize}
    \item $t[t\rightarrow{}t']=\{t'\}$
    \item $X.(t)[st\rightarrow{}t']=
     \{X.(s) \mid s\in{}t[st\rightarrow{}t']\}$, for all terms
$X.(t)\in{}T$ with $t\neq\varepsilon$ and
     $st\in{}SubTerms(X.(t))\setminus \{X.(t)\}$
    \item $t_{1}$$\parallel$$t_{2}[st\rightarrow{}t']=$
$\{\parcomp{t''}{t'_2} \mid (t'_1,t'_2) \in T \times T, t'_1,t'_2
\neq \varepsilon$, $\parcomp{t'_1}{t'_2} = \parcomp{t_1}{t_2}$,
$st \in SubTerms(t'_1)$, $t''\in{}t_{1}'[st\rightarrow{}t']\}$,
for all $t_1,t_2\in T\setminus\{\varepsilon\}$ and $st \in
SubTerms(\parcomp{t_1}{t_2})\setminus \{\parcomp{t_1}{t_2}\}$.
\end{itemize}
\end{Definition}

\begin{Definition}
  For a term $t\in{}T$, the set of terms $SEQ(t)$ is the subset of
  $T_{SEQ}\setminus\{\varepsilon\}$ defined inductively as follows:
\begin{itemize}
    \item $SEQ(\varepsilon)=\emptyset$
    \item $SEQ(X)=\{X\}$, for all $X\in{}\hat{Var}$
    \item $SEQ(X.(t))=\{X.(t') \mid t'\in{}SEQ(t)\}$, for all
      $X\in{}\hat{Var}$ and $t\in{}T\setminus \{\varepsilon\}$
\item $SEQ(t_{1}$$\parallel$$t_{2})=SEQ(t_{1})\cup{}SEQ(t_{2})$.
    \end{itemize}
\end{Definition}

For a term $t\in{}T_{SEQ}\setminus \{\varepsilon\}$ having the
form $t=X_{1}.(X_{2}.(\ldots{}X_{n}.(Y)\ldots))$, with $n\geq0$,
we denote the variable $Y$ by $last(t)$. Given two terms
 $t,t'\in{}T_{SEQ}\setminus\{\varepsilon\}$, with
 $t=X_{1}.(X_{2}.(\ldots{}X_{n}.(Y)\ldots))$ and
 $t'=X_{1}'.(X_{2}'.(\ldots{}X_{k}'.(Y')\ldots))$, we denote by
 $t\circ{}t'$ the term
 $X_{1}.(X_{2}.(\ldots{}X_{n}.(X_{1}'.(X_{2}'$
 $.(\ldots{}X_{k}'.(Y')\ldots)))\ldots))$. Notice that $t\circ{}t'$  is
the only term  in $t[Y\rightarrow{}t']$, and that the operation
$\circ$ on terms in $T_{SEQ} \setminus \{\varepsilon\}$ is
associative.


\begin{Proposition}[see~\cite{Bozz03}]
\label{Prop:Subterms1} The following properties hold:
\begin{enumerate}
    \item If $t$ \multider{\sigma} $t'$ and $t\in{}SubTerms(s)$,
for some $s\in{}T$, then it holds $s$ \multider{\sigma} $s'$ for
all $s'\in{}s[t\rightarrow{}t']$; \item If $t$ \multider{\sigma}
is an infinite derivation in $\Re$ and $t\in{}SubTerms(s)$, for
some $s\in{}T$, then it holds $s$ \multider{\sigma}.
\end{enumerate}
\end{Proposition}


\begin{Proposition}[see~\cite{Bozz03}]\label{Prop:Subterms2}
If
 $t,t'\in{}T_{SEQ}\setminus \{\varepsilon\}$ such that
$last(t)$ \multider{\rho}
 $t'$, then it holds that
 \begin{enumerate}
    \item $t$ \multider{\rho} $t\circ{}t'$;
    \item $t''\circ{}t$ \multider{\rho} $t''\circ{}t\circ{}t'$
 for all $t''\in{}T_{SEQ}\setminus\{\varepsilon\}$.
 \end{enumerate}
\end{Proposition}


\begin{Lemma}\label{Lemma:Subderivations1}
 Let $t$$\parallel$$X.(s)$ \multider{\sigma} be a
derivation in $\Re$, and let $s$ \multider{\sigma{}'} be a
subderivation of  $t$$\parallel$$X.(s)$ \multider{\sigma} from
$s$. Then, the following properties are satisfied:
\begin{enumerate}
    \item If $s$ \multider{\sigma{}'} is infinite, then  it holds
    that
    $t$ \multider{\sigma\setminus\sigma{}'}.
     Moreover, if $t$$\parallel$$X.(s)$ \multider{\sigma}
    is in $\Pi^{K,K^{\omega}}_{PAR,\infty}$,  then also  $t$ \multider{\sigma\setminus\sigma{}'} is in
    $\Pi^{K,K^{\omega}}_{PAR,\infty}$.
    \item If $s$ \multider{\sigma{}'} leads
    to $\varepsilon$, then the derivation
    $t$$\parallel$$X.(s)$ \multider{\sigma} can be written in the form
    \begin{displaymath}
     \textrm{$t$$\parallel$$X.(s)$ \multider{\sigma_{1}} $t'$$\parallel$$X$ \multider{\sigma_{2}}}
    \end{displaymath}
    where  $t$ \multider{\lambda} $t'$ with $\sigma_{1}\in{}Interleaving(\lambda,\sigma{}')$.
    Moreover, if $t$$\parallel$$X.(s)$ \multider{\sigma}
    is in $\Pi^{K,K^{\omega}}_{PAR,\infty}$, there is a derivation
    of the form $t$$\parallel$$X$ \multider{\lambda} $t'$$\parallel$$X$ \multider{\sigma_{2}}
    belonging to $\Pi^{K,K^{\omega}}_{PAR,\infty}$.
    \item If $s$ \multider{\sigma{}'} leads
    to a term $s'\neq\varepsilon$ one of the following conditions is satisfied:
    \begin{itemize}
        \item $t$ \multider{\sigma\setminus\sigma{}'}.
            If  $t$$\parallel$$X.(s)$ \multider{\sigma}
            is in $\Pi^{K,K^{\omega}}_{PAR,\infty}$, then also $t$ \multider{\sigma\setminus\sigma{}'}
            is in
            $\Pi^{K,K^{\omega}}_{PAR,\infty}$. Moreover, if
            $t$$\parallel$$X.(s)$ \multider{\sigma} is finite
            and leads to $\overline{t}$, then
            $\overline{t}=X.(s')$$\parallel$$t'$ where $t$
            \multider{\sigma\setminus\sigma{}'} $t'$.
        \item $s'=W\in{}Var$ and the derivation
        $t$$\parallel$$X.(s)$ \multider{\sigma} can be written in the
        form
        \begin{displaymath}
        \textrm{$t$$\parallel$$X.(s)$ \multider{\sigma_{1}}
        $t'$$\parallel$$X.(W)$ \prsoneder{r}
        $t'$$\parallel$$W'$ \multider{\sigma_{2}}}
        \end{displaymath}
        where  $r=X.(W)$\Rule{a}$W'\in{}\Re$, and
         $t$ \multider{\lambda} $t'$
          with $\sigma_{1}\in{}Interleaving(\lambda,\sigma{}')$.
          Moreover, if $t$$\parallel$$X.(s)$ \multider{\sigma}
        is in $\Pi^{K,K^{\omega}}_{PAR,\infty}$, there is a derivation
        of the form $t$$\parallel$$W'$ \multider{\lambda} $t'$$\parallel$$W'$ \multider{\sigma_{2}}
    belonging to $\Pi^{K,K^{\omega}}_{PAR,\infty}$.
    \end{itemize}
\end{enumerate}
\end{Lemma}

\begin{proof}
 The assertion  follows easily from the definition of subderivation.
\end{proof}


\begin{Lemma}\label{Lemma:Base}
\setcounter{equation}{0}
 Let $p$ \multider{\sigma}
$t$$\parallel$$Y.(s)$ \multider{\omega}  with $s\neq\varepsilon$
and $p\in{}T_{PAR}$. Then,  $p$ \multider{\sigma}
$t$$\parallel$$Y.(s)$ can be written in the form
\begin{equation}
\textrm{$p$ \multider{\sigma_{1}} $t'$$\parallel$$Z$ \prsoneder{r}
$t'$$\parallel$$Y.(Z')$ \multider{\sigma_{2}}
$t$$\parallel$$Y.(s)$}
\end{equation}
with $r=Z$\Rule{a}$Y.(Z')$, and
\begin{equation}
\textrm{$Z'$ \multider{\sigma_{2}'} $s\quad$ and $t'$
\multider{\sigma_{2}''} $t$}
\end{equation}
with $\sigma_{2}\in{}Interleaving(\sigma_{2}',\sigma_{2}'')$.
Moreover, the following property is satisfied:
\begin{description}
    \item[A] Let $s$ \multider{\omega'} be a subderivation of $t$$\parallel$$Y.(s)$
    \multider{\omega}  from $s$. Then, the derivation
    \begin{displaymath}
        \textrm{$Z'$ \multider{\sigma_{2}'} $s$
        \multider{\omega'}}
    \end{displaymath}
    is a  subderivation of $t'$$\parallel$$Y.(Z')$ \multider{\sigma_{2}}
    $t$$\parallel$$Y.(s)$ \multider{\omega}  from $Z'$.
\end{description}
\end{Lemma}

\begin{proof}
 The assertion  follows easily  by induction on the length of $\sigma$.
\end{proof}


\section{Proof of Theorem \ref{Theorem:Problem1}}

In order to prove theorem \ref{Theorem:Problem1} we need the
following two lemmata
\ref{Lemma:From-MKPAR-To-M}--\ref{Lemma:From-M-To-MKPAR}

\begin{Lemma}\label{Lemma:From-MKPAR-To-M} Let $p$
\multiderparK{\sigma} $p'$$\parallel$$p''$ with
$p,p',p''\in{}T_{PAR}$, $p'$ not containing occurrences of
$\hat{Z}_{F}$ and $\hat{Z}_{\infty}$, and $p''$ not containing
occurrences of variables in $Var$. Then, there exists a $t\in{}T$
such that $p$ \multider{\rho} $p'$$\parallel$$t$ with
\maximal{\rho} = \maximalparK{\sigma}, and $|\rho|>0$ if
$|\sigma|>0$.
\end{Lemma}
\begin{proof}
The proof is by induction on $|\sigma|$.\\
\textbf{Base Step}: $|\sigma|=0$. In this case the assertion is
obvious.\\
\textbf{Induction Step}: $|\sigma|>0$. In this case the derivation
$p$ \multiderparK{\sigma} $p'$$\parallel$$p''$ can be written in
the  form
\begin{displaymath}
\textrm{$p$ \multiderparK{\sigma'}
$\overline{p}'$$\parallel$$\overline{p}''$ \onederparK{r}
$p'$$\parallel$$p''\quad$ with $|\sigma'|<|\sigma|$,
$r\in\Re^{K}_{PAR}$ and
$\overline{p}',\overline{p}''\in{}T_{PAR}$}
\end{displaymath}
Moreover, $\overline{p}'$ doesn't contain occurrences of
$\hat{Z}_{F}$ and $\hat{Z}_{\infty}$, and $\overline{p}''$ doesn't
contain occurrences of variables in $Var$. From the inductive
hypothesis, there exists a $\overline{t}\in{}T$ such that $p$
\multider{\rho'} $\overline{p}'$$\parallel$$\overline{t}$ with
\maximal{\rho'} = \maximalparK{\sigma'}.\\
There are two cases:
\begin{enumerate}
    \item $r$ is a PAR rule of $\Re$. From remark \ref{Remark:MKPAR}
    \maximal{r} = \maximalparK{r}. Moreover, $\overline{p}''=p''$
    and $\overline{p}'$ \onederparK{r} $p'$. Then,  we deduce that
    $p$ \multider{\rho'}
    $\overline{p}'$$\parallel$$\overline{t}$ \prsoneder{r}
    $p'$$\parallel$$\overline{t}$ with
    \maximal{\rho'r} = \maximalparK{\sigma'r}, and the assertion is
    proved.
    \item $r\in\Re^{K}_{PAR}\setminus\Re$. There are two subcases:
    \begin{itemize}
        \item $r=X$\Rule{K'}$Y$
        with $X,Y\in{}Var$ and $K'\in{}P_{n}$.
        From remark \ref{Remark:MKPAR} \maximalparK{r} = $K'$.
        From property 5 in the definition of $M_{PAR}^{K}$
        we have $X$ \multider{\rho''} $Y$ with
        \maximal{\rho''} $=K'$ and $|\rho''|>0$. Moreover,
        $\overline{p}''=p''$ and $\overline{p}'$ \onederparK{r} $p'$. Then,
        we deduce that
        $p$ \multider{\rho'}
        $\overline{p}'$$\parallel$$\overline{t}$ \multider{\rho''}
        $p'$$\parallel$$\overline{t}$ with
        \maximal{\rho'\rho''} = \maximalparK{\sigma'r}, and the assertion is
        proved.
        \item $r=X$\Rule{K'}$\hat{Z}_{F}$ with $X\in{}Var$, and
        \maximalparK{r} = $K'$. From property 5 in the definition of $M_{PAR}^{K}$, we
        deduce that $X$ \multider{\rho''} $t$, for some term $t$,
        with
        \maximal{\rho''} = \maximalparK{r} and $|\rho''|>0$. Evidently,
        $p''=\overline{p}''$$\parallel$$\hat{Z}_{F}$ and
        $\overline{p}'=p'$$\parallel$$X$. Then,
        we deduce that
        $p$ \multider{\rho'}
    $\overline{p}'$$\parallel$$\overline{t}=p'$$\parallel$$X$$\parallel$$\overline{t}$
    \multider{\rho''} $p'$$\parallel$$t$$\parallel$$\overline{t}$
    with
        \maximal{\rho'\rho''} = \maximalparK{\sigma'r}, and the assertion is
        proved.
    \end{itemize}
\end{enumerate}
\end{proof}

 \setcounter{equation}{0}
$\newline$\begin{Lemma}\label{Lemma:From-M-To-MKPAR}
 Let $p$ \multider{\sigma} $t$$\parallel$$p'$
 with $p,p'\in{}T_{PAR}$ and \maximal{\sigma} $\subseteq{}K$. Then,
 there exists a $s\in{}T_{PAR}$ such that $p$ \multiderparK{\rho} $s$$\parallel$$p'$
 with
\maximal{\sigma} = \maximalparK{\rho}, and $s=\varepsilon$ if
$t=\varepsilon$.
\end{Lemma}
\begin{proof}
The proof is by induction on the length of finite derivations $p$
\multider{\sigma} in $\Re$ from terms in $T_{PAR}$ with
\maximal{\sigma} $\subseteq{}K$.\\
\\
\textbf{Base Step}: $|\sigma|=0$. In this case the assertion is
obvious.\\
\\
\textbf{Induction Step}: $|\sigma|>0$. The derivation $p$
\multider{\sigma}  can be written in the form
\begin{equation}
\textrm{$p$ \prsoneder{r} $\overline{t}$ \multider{\sigma'}
$t$$\parallel$$p'$ $\quad$ }
\end{equation}
with $r\in\Re$,  $|\sigma'|<|\sigma|$ and \maximal{\sigma'}
$\subseteq{}K$. There are two cases:
\begin{enumerate}
    \item r is a PAR rule. Then, we have $\overline{t}\in{}T_{PAR}$. >From the
    inductive hypothesis, there exists a $s\in{}T_{PAR}$ such that
    $\overline{t}$ \multiderparK{\rho'} $s$$\parallel$$p'$
    with \maximal{\sigma'} = \maximalparK{\rho'}, and
    $s=\varepsilon$ if $t=\varepsilon$.
    By construction, $r\in\Re^{K}_{PAR}$. From remark \ref{Remark:MKPAR}, it follows  that
    \maximal{r} = \maximalparK{r}. By proposition
    \ref{Prop:Maximal} we obtain
    $p$ \onederparK{r} $\overline{t}$ \multiderparK{\rho'} $s$$\parallel$$p'$ with
    \maximal{r\sigma'} = \maximalparK{r\rho'}, and
    $s=\varepsilon$ if $t=\varepsilon$. Hence, the assertion is proved.
    \item $r=Z$\Rule{a}$Y.(Z')$ with \maximal{r} $\subseteq{}K$. Then, we have
    $p=p''$$\parallel$$Z$ and $\overline{t}=p''$$\parallel$$Y.(Z')$, with  $p''\in{}T_{PAR}$.
    From (1), let $Z'$ \multider{\lambda} $t_{1}$ a subderivation of $\overline{t}=p''$$\parallel$$Y.(Z')$
    \multider{\sigma'} from $Z'$. Evidently,
    \maximal{\lambda} $\subseteq{}K$. From lemma \ref{Lemma:Subderivations1} we can
    distinguish three subcases:
    \begin{itemize}
        \item $t_{1}\neq\varepsilon$ and $p''$ \multider{\sigma'\setminus\lambda} $t'$. Moreover,
              $t$$\parallel$$p'=t'$$\parallel$$Y.(t_{1})$,   $t'=p'$$\parallel$$t''$ for some
              term $t''$, and
              $t=t''$$\parallel$$Y.(t_{1})$. In particular,
              $t\neq\varepsilon$. Since
              \maximal{\sigma'\setminus\lambda} $\subseteq{}K$, from the
              inductive hypothesis there exists an $s\in{}T_{PAR}$
              such that $p''$ \multiderparK{\rho'}
              $s$$\parallel$$p'$ with
              \maximal{\sigma'\setminus\lambda} = \maximalparK{\rho'}.
                Moreover, since \maximal{\lambda} $\subseteq{}K$,
                from the inductive hypothesis we have that
            $Z'$ \multiderparK{\rho''} $\overline{p}$ for some
            $\overline{p}\in{}T_{PAR}$, with
            \maximalparK{\rho''} = \maximal{\lambda}.
              Since \maximalparK{\rho''} = \maximal{\lambda} $\subseteq{}K$ and
              \maximal{r} $\subseteq{}K$,  from property 2 in the
              definition of $M^{K}_{PAR}$ we deduce that
              $r'=Z$\Rule{K'}$\hat{Z}_{F}\in\Re^{K}_{PAR}$ with
              $K'=$ \maximal{\lambda} $\cup{}$ \maximal{r} and
              \maximalparK{r} = $K'$. Then, by proposition \ref{Prop:Maximal}
              we have $p=p''$$\parallel$$Z$ \onederparK{r'}
              $p''$$\parallel$$\hat{Z}_{F}$ \multiderparK{\rho'}
              $s$$\parallel$$p'$$\parallel$$\hat{Z}_{F}$ with
              \maximalparK{r'\rho'} = \maximal{r\lambda(\sigma'\setminus\lambda)} =
              \maximal{\sigma}, and the assertion is proved.
        \item $t_{1}=\varepsilon$ and the derivation $p''$$\parallel$$Y.(Z')$
                \multider{\sigma'} $t$$\parallel$$p'$ can be written in the
                 form
              \begin{equation}
                \textrm{$p''$$\parallel$$Y.(Z')$
                \multider{\sigma_{1}} $t'$$\parallel$$Y$ \multider{\sigma_{2}} $t$$\parallel$$p'\quad$
                with $p''$ \multider{\sigma'_{1}} $t'$, and
                $\sigma_{1}\in{}Interleaving(\lambda,\sigma'_{1})$}
              \end{equation}
              Now,  $Z'$ \multider{\lambda} $\varepsilon$ with
              $|\lambda|<|\sigma|$ and
              \maximal{\lambda} $\subseteq{}K$. From the inductive
              hypothesis, we have $Z'$ \multiderparK{\rho}
              $\varepsilon$
              such that
              \maximal{\lambda} = \maximalparK{\rho}.
              From property 3 in the definition of $M^{K}_{PAR}$
              it follows that $r'=Z$\Rule{K'}$Y\in\Re^{K}_{PAR}$
              where
              $K'=$ \maximal{r} $\cup{}$ \maximalparK{\rho} and
              \maximalparK{r'} = $K'$.
              Now, we have $p''$$\parallel$$Y$
                \multider{\sigma'_{1}} $t'$$\parallel$$Y$ \multider{\sigma_{2}}
                $t$$\parallel$$p'$ with
              \maximal{\sigma'_{1}\sigma_{2}} $\subseteq{}K$ and
               $|\sigma'_{1}\sigma_{2}|<|\sigma|$. From the
              inductive hypothesis, there exists a $s\in{}T_{PAR}$
              such that $p''$$\parallel$$Y$
                \multiderparK{\rho'}
                $s$$\parallel$$p'$, with
                \maximalparK{\rho'} = \maximal{\sigma'_{1}\sigma_{2}},
                and $s=\varepsilon$ if $t=\varepsilon$. After all,
                considering  proposition \ref{Prop:Maximal},
                we have $p=p''$$\parallel$$Z$ \onederparK{r'}
              $p''$$\parallel$$Y$ \multiderparK{\rho'}
              $s$$\parallel$$p'$ with
              \maximal{r'\rho'} = \maximal{r\lambda\sigma'_{1}\sigma_{2}} =
              \maximal{\sigma}, and $s=\varepsilon$ if $t=\varepsilon$. So, the assertion is proved.
        \item
            $t_{1}=W\in{}Var$ and the derivation $p''$$\parallel$$Y.(Z')$
                \multider{\sigma'} $t$$\parallel$$p'$ can be written in the
               form
              \begin{eqnarray}
                \textrm{$p''$$\parallel$$Y.(Z')$
                \multider{\sigma_{1}} $t'$$\parallel$$Y.(W)$
                \prsoneder{r'} $t'$$\parallel$$W'$ \multider{\sigma_{2}}
                $t$$\parallel$$p'$}\\
                \textrm{with $p''$ \multider{\sigma'_{1}} $t'$,
                $r'=Y.(W)$\Rule{b}$W'$ and
                $\sigma_{1}\in{}Interleaving(\lambda,\sigma'_{1})$}
              \end{eqnarray}
            Now,  $Z'$ \multider{\lambda} $W$ with
              $|\lambda|<|\sigma|$ and
              \maximal{\lambda} $\subseteq{}K$. From the inductive
              hypothesis, we have $Z'$ \multiderparK{\rho}
              $W$
              such that
              \maximal{\lambda} = \maximalparK{\rho}.
              From property 4 in the definition of $M^{K}_{PAR}$,
              considering that $r=Z$\Rule{a}$Y.(Z')\in\Re$ and
              $r'=Y.(W)$\Rule{b}$W'\in\Re$ with \maximal{r} $\subseteq{}K$
              and \maximal{r'} $\subseteq{}K$, it follows that $r''=Z$\Rule{K'}$W'\in\Re^{K}_{PAR}$
              where
              $K'=$ \maximal{rr'} $\cup{}$ \maximalparK{\rho}
              and \maximalparK{r''} = $K'$.
              Now, we have $p''$$\parallel$$W'$
                \multider{\sigma'_{1}} $t'$$\parallel$$W'$ \multider{\sigma_{2}}
                $t$$\parallel$$p'$ with
              \maximal{\sigma'_{1}\sigma_{2}} $\subseteq{}K$ and $|\sigma'_{1}\sigma_{2}|<|\sigma|$.
               From the
              inductive hypothesis there exists a $s\in{}T_{PAR}$
              such that $p''$$\parallel$$W'$
                \multiderparK{\rho'}
                $s$$\parallel$$p'$, with
                \maximalparK{\rho'} = \maximal{\sigma'_{1}\sigma_{2}},
                and $s=\varepsilon$ if $t=\varepsilon$. After all,
                considering  proposition \ref{Prop:Maximal},
                we obtain $p=p''$$\parallel$$Z$ \onederparK{r''}
              $p''$$\parallel$$W'$ \multiderparK{\rho'}
              $s$$\parallel$$p'$ with
              \maximalparK{r''\rho'} = \maximal{rr'\lambda\sigma'_{1}\sigma_{2}} =
              \maximal{\sigma}, and $s=\varepsilon$ if $t=\varepsilon$. So, the assertion is proved.
    \end{itemize}
\end{enumerate}
\end{proof}

At this point, theorem \ref{Theorem:Problem1} follows directly
from lemmata
\ref{Lemma:From-MKPAR-To-M}--\ref{Lemma:From-M-To-MKPAR}.

\section{Proof of Theorems \ref{Theorem:Problem2.1} and \ref{Theorem:Problem2.2}}

\setcounter{equation}{0}

In order to prove theorems \ref{Theorem:Problem2.1} and
\ref{Theorem:Problem2.2} we need the following lemmata
\ref{Lemma:next}--\ref{Lemma:From-M-To-MSEQ3}\newline

To prove lemma \ref{Lemma:From-MPAR-To-M-Inf} we use  a mapping
for coding pairs of  integers by single  integers. In particular,
we consider the following bijective mapping from $N\times{}N$ to
$N$, that is a primitive recursive function  \cite{davis83}
\begin{displaymath}
\textrm{$<$ $>$$: (x,y)\in{}N\times{}N\rightarrow{}2^{x}(2y+1)-1$}
\end{displaymath}
Let $\ell$ (resp. $\wp$) be the first component (resp. the second
component) of $<$ $>^{-1}$.  The following properties are
satisfied:
\begin{enumerate}
    \item $\forall{}x,y\in{}N$ $\ell(<$$x,y$$>)=x$ and $\wp(<$$x,y$$>)=y$.
    \item $\forall{}z\in{}N$ $<$$\ell(z),\wp(z)$$>$ $=z$.
    \item $\forall{}z\in{}N$ $\ell(z),\wp(z)\leq{}z$.
    \item  $\forall{}z,z'\in{}N$  if $z>z'$ and $\ell(z)=\ell(z')$ then $\wp(z)>\wp(z')$.
\end{enumerate}
Now, we introduce a new function
$next:N\times{}N\rightarrow{}N\times{}N$ defined by primitive
recursion in the following way
\begin{displaymath}
 next(x,0)=(x,0)
\end{displaymath}
\begin{displaymath}
 next(x,y+1)= \left\{ \begin{array}{ll}
             (\ell(y),\wp(y)+1) & \textrm{if $next(x,y)=(\ell(y),\wp(y))$}\\
             next(x,y) & \textrm{otherwise}
             \end{array}
             \right.
\end{displaymath}

For all $x,y\in{}N$ let us denote by $next_{x}(y)$ the second
component of $next(x,y)$. The following lemma establishes some
properties of $next$.

\begin{Lemma}\label{Lemma:next}
The function $next$ satisfies the following properties:
\begin{enumerate}
    \item $\forall{}x,y\in{}N$ if $y\leq{}x$ then
              $next(x,y)=(x,0)$.
    \item $\forall{}x,y\in{}N$ $next(x,y)=(x,z_{x,y})$ for some $z_{x,y}\in{}N$.
    \item  $\forall{}x,y\in{}N$
          $next_{x}(y)\leq{}next_{x}(y+1)$.
    \item Let $x,y_{1},y_{2}\in{}N$ with
    $next_{x}(y_{1})<next_{x}(y_{2})$. Then, there exists a
    $k\in{}N$ such that $next(x,k)=(\ell(k),\wp(k))$, $\wp(k)=next_{x}(y_{2})-1$
    and $y_{1}\leq{}k<y_{2}$
    \item $\forall{}x,n\in{}N$  there exists a $y\in{}N$
    such that $next(x,y)=(x,n)$.
    \item $\forall{}x\in{}N$  $next(\ell(x),x)=(\ell(x),\wp(x))$.
     \item $\forall{}x,i\in{}N$ if $i\neq\ell(x)$  then $next(i,x+1)=next(i,x)$.
\end{enumerate}
\end{Lemma}

\begin{proof}
At first, let us consider property 1. We prove it by induction
on $y$. By construction, for $y=0$  we have $next(x,0)=(x,0)$.\\
Now, let $0<y\leq{}x$. From the inductive hypothesis
$next(x,y-1)=(x,0)$. So, it suffices to prove that
$next(x,y)=next(x,y-1)$. By absurd, let us assume that
$next(x,y)\neq{}next(x,y-1)$. Then, by construction we have
$next(x,y-1)=(\ell(y-1),\wp(y-1))$ and
$next(x,y)=(\ell(y-1),\wp(y-1)+1)$. Therefore, $x=\ell(y-1)$. But,
$x>y-1\geq{}\ell(y-1)$. So, we obtain an
absurd.\\
Now, let us consider property 2. We prove it by induction
on $y$. By construction, for $y=0$  we have $next(x,0)=(x,0)$.\\
Now, let $y>0$. From the inductive hypothesis $next(x,y-1)=(x,m)$
for some $m\in{}N$. Now, by construction either
$next(x,y)=next(x,y-1)=(x,m)$ or $next(x,y)=(x,m+1)$. In both
cases the assertion is satisfied.\\
Now, let us consider  property 3. By construction,
$\forall{}x,y\in{}N$ either $next_{x}(y+1)=next_{x}(y)$ or
$next_{x}(y+1)=next_{x}(y)+1$. So, the assertion is satisfied.\\
\\
Now, let us consider property 4. From property 3 $y_{1}<y_{2}$.
The proof is by induction on
$y_{2}-y_{1}$.\\
\textbf{Base Step}: $y_{2}-y_{1}=1$. So, $y_{2}=y_{1}+1$. From
hypothesis $next(x,y_{1})\neq{}next(x,y_{2})$. Therefore, by
construction we deduce that
$next(x,y_{1})=(\ell(y_{1}),\wp(y_{1}))$ and
$next(x,y_{2})=(\ell(y_{1}),\wp(y_{1})+1)$. So, setting $k=y_{1}$
we have $next(x,k)=(\ell(k),\wp(k))$, $\wp(k)=next_{x}(y_{2})-1$
and $y_{1}\leq{}k<y_{2}$. The
assertion is proved.\\
\textbf{Induction Step}: $y_{2}-y_{1}>1$. From property 3 there
are two cases:
\begin{itemize}
    \item $next_{x}(y_{2}-1)=next_{x}(y_{2})$. So,
    $next_{x}(y_{1})<next_{x}(y_{2}-1)$. Since
    $(y_{2}-1)-y_{1}<y_{2}-y_{1}$, from the inductive hypothesis
    there exists a
    $k\in{}N$ such that $next(x,k)=(\ell(k),\wp(k))$,
    $\wp(k)=next_{x}(y_{2}-1)-1$ and $y_{1}\leq{}k<y_{2}-1$. Since
    $next_{x}(y_{2}-1)=next_{x}(y_{2})$ we obtain the assertion.
    \item $next_{x}(y_{2}-1)<next_{x}(y_{2})$. We reason as in the
    base step.
\end{itemize}

Now, let us consider  property 5. The proof is by induction on
$n$. By construction, for $n=0$ we have $next(x,0)=(x,0)$. So, in
this case the assertion is satisfied. Now, let $n>0$. From the
inductive hypothesis there exists a $y\in{}N$ such that
$next(x,y)=(x,n-1)$. From property 4, we deduce that it suffices
to prove that there exists a $m>n-1$ such that $next(x,z)=(x,m)$
for some $z\in{}N$. By absurd we assume that this property isn't
satisfied. From property 2 and 3, we deduce that
\begin{equation}
\forall{}z\geq{}y{}  \textrm{ $next(x,z)=(x,n-1)$}
\end{equation}
Let $k=$ $<$$x,n-1$$>$. Then, $x=\ell(k)$ and $n-1=\wp(k)$. There
are two cases:
\begin{itemize}
    \item $k\geq{}y$. From (1) we have $next(x,k)=(\ell(k),\wp(k))$. By
    construction, we obtain $next(x,k+1)=(\ell(k),\wp(k)+1)=(x,n)$ in
    contrast with (1).
    \item $k<y$. Now, $next(x,k)=(\ell(k),\overline{n})$. From
    property 3, it follows that $\overline{n}\leq{}\wp(k)$. There
    are two subcases:
    \begin{itemize}
        \item[-] $\overline{n}=\wp(k)$. So,
        $next(x,k)=(\ell(k),\wp(k))$. By construction, we obtain
        $next(x,k+1)=(\ell(k),\wp(k)+1)=(x,n)$. So, we have
        $k+1\leq{}y$ and $next_{x}(k+1)>next_{x}(y)$ in contrast with
        property 3.
        \item[-] $\overline{n}<\wp(k)$. In other words,
        we have $k<y$ and
         $next_{x}(k)<next_{x}(y)=\wp(k)$. From property 4, there exists
         a $k'\geq{}k$ such that
          $next(x,k')=(\ell(k'),\wp(k'))$ and
          $\wp(k')=next_{x}(y)-1=\wp(k)-1$. From property 2,
          $\ell(k')=x$. So, we have $\ell(k)=\ell(k')$, $k'\geq{}k$ and
           $\wp(k')<\wp(k)$. This is in contradiction with properties
           of $\wp$ and $\ell$.
    \end{itemize}
\end{itemize}

Now, let us consider property 6.  Let $x\in{}N$. So, $x=$
$<$$\ell(x),\wp(x)$$>$. From property 5 there exist two integers
$y,z\in{}N$ such that $next(\ell(x),y)=(\ell(x),\wp(x)+1)$ and
$next(\ell(x),z)=(\ell(x),\wp(x))$ where $y>z$. Since
$next_{\ell(x)}$ is crescent there exists the greatest
$\overline{z}$ such that
$next(\ell(x),\overline{z})=(\ell(x),\wp(x))$. In particular,
$next(\ell(x),\overline{z}+1)\neq{}next(\ell(x),\overline{z})$.
>From definition of $next$ it follows that
$next(\ell(x),\overline{z})=(\ell(\overline{z}),\wp(\overline{z}))$.
Therefore,
$(\ell(x),\wp(x))=(\ell(\overline{z}),\wp(\overline{z}))$. From
this we deduce that $\overline{z}=x$, and the assertion is proved.

Finally, let us consider property 7.  Let $x,i\in{}N$ with
$i\neq\ell(x)$. By absurd let us assume that
$next(i,x+1)\neq{}next(i,x)$. Then, by construction
$next(i,x)=(\ell(x),\wp(x))$. From property 2 we obtain
$i=\ell(x)$, an absurd.
\end{proof}

Now, we give the notion of \emph{Interleaving} of a succession of
rule sequences in a \PRS $\Re$. To formalize this concept and
facilitate the proof of some connected results, we redefine the
notion of sequence rule. Precisely, a sequence rule in $\Re$ can
be seen as a mapping $\sigma:N'\rightarrow{}\Re$ where $N'$ can be
a generic subset of $N$. In particular, this facilitates the
formalization of the  notion of subsequence. A rule sequence
$\sigma':N''\rightarrow{}\Re$ is a subsequence of
$\sigma:N'\rightarrow{}\Re$ iff $N''\subseteq{}N'$ and
$\sigma'=\sigma|_{N''}$, that is $\sigma'$
 is the restriction of $\sigma$ to set $N''$.\\
Given a rule sequence $\sigma:N'\rightarrow{}\Re$, we denote by
$pr(\sigma)$ the set $N'$.\\
Given a set $N'$, subset of $N$, we denote by $min(N')$ the
smallest element of $N'$.\\
Finally, given two rule sequences $\sigma$ and $\sigma'$, we say
that they are disjoint if $pr(\sigma)\cap{}pr(\sigma')=\emptyset$.

\begin{Definition}\label{Def:Interleaving}
Let $\{\rho_{h}\}_{h\in{}N}$ be a succession of rule sequences in
a \PRS $\Re$. The \emph{Interleaving} of $\{\rho_{h}\}_{h\in{}N}$,
denoted by $Interleaving(\{\rho_{h}\})$, is the set of rule
sequences $\sigma$ in $\Re$ such that there exists an
\emph{injective} mapping
$M_{\sigma}:\bigcup_{h\in{}N}(\{h\}\times{}pr(\rho_{h}))\rightarrow{}N$
$\mathrm{(}$depending on $\sigma$$\mathrm{)}$ satisfying the
following properties $\mathrm{(}$where $\Delta$ is the set
$\bigcup_{h\in{}N}(\{h\}\times{}pr(\rho_{h})$$\mathrm{))}$
\begin{itemize}
    \item $\forall{}h\in{}N$ $\forall{}n,n'\in{}pr(\rho_{h})$ with
     $n<n'$ then $M_{\sigma}(h,n)<M_{\sigma}(h,n')$.
    \item $pr(\sigma)=M_{\sigma}(\Delta)$.
    \item $\forall{}(h,n)\in{}\Delta$ we have
    $\sigma(M_{\sigma}(h,n))=\rho_{h}(n)$.
\end{itemize}
\end{Definition}

\begin{Proposition}\label{Prop:MaximalInterleaving}
Let
  $M=\npla{\Re,\npla{\Re_{1}^{A},\ldots,\Re_{n}^{A}}}$ be
  a  \MBRS and let $\{\sigma_{h}\}_{h\in{}N}$ be a succession of rule
    sequences in $\Re$. Then,
    $\forall{}\pi\in{}Interleaving(\{\sigma_{h}\})$ we have
\begin{enumerate}
    \item \maximal{\pi} = $\bigcup_{h\in{}N}$\maximal{\sigma_{h}}.
    \item \maximalInf{\pi} =
    $\bigcup_{h\in{}N}$\maximalInf{\sigma_{h}} $\cup$
                           $\bigoplus_{h\in{}N}$\maximal{\sigma_{h}}.
\end{enumerate}
\end{Proposition}
\begin{proof}
We prove property 2. In similar way it's possible to prove
property 1. Let
$\Delta=\bigcup_{h\in{}N}(\{h\}\times{}pr(\sigma_{h})$$\mathrm{)}$.
>From hypothesis
 there exists an \emph{injective} mapping
$M_{\pi}:\Delta\rightarrow{}N$ such that
$pr(\pi)=M_{\pi}(\Delta)$, and
 for all $(h,k)\in{}\Delta$
    $\pi(M_{\pi}(h,k))=\sigma_{h}(k)$.\\
Let $i\in{}\{1,\ldots,n\}$. We have to prove that
\begin{displaymath}
\textrm{$i\in$\maximalInf{\pi} $\Leftrightarrow$ $i\in$
 $\bigcup_{h\in{}N}$\maximalInf{\sigma_{h}} $\cup$
                           $\bigoplus_{h\in{}N}$\maximal{\sigma_{h}}.}
\end{displaymath}
$(\Rightarrow)$. Let $i\in$\maximalInf{\pi}. So, $\pi$ contains
infinite occurrences of a rule $r\in\Re^{A}_{i}$. Therefore, the
set $\{k\in{}M_{\pi}(\Delta)|$ $\pi(k)=r\}$ is infinite. Then, the
set $\{(h,k)\in{}\Delta|$ $\sigma_{h}(k)=r\}$ is infinite. There
are two cases:
\begin{itemize}
    \item $\exists{}h\in{}N$ such that the set $\{j\in{}pr(\sigma_{h})|$ $\sigma_{h}(j)=r\}$
    is infinite. Therefore, $i\in$ \maximalInf{\sigma_{h}}
    \item The set $\{h\in{}N|$ $\sigma_{h}$ contains some occurrence of $r\}$ is infinite.
    Therefore, $i\in$ $\bigoplus_{h\in{}N}$\maximal{\sigma_{h}}.
\end{itemize}
In both cases the result holds.\\
 $(\Leftarrow)$. Let
$i\in$ $\bigcup_{h\in{}N}$\maximalInf{\sigma_{h}} $\cup$
                           $\bigoplus_{h\in{}N}$\maximal{\sigma_{h}}.
There are two cases:
\begin{itemize}
    \item $\exists{}h\in{}N$ such that $i\in$
    \maximalInf{\sigma_{h}}. Since $M_{\pi}$ is injective, the set
     $\{k\in{}M_{\pi}(\Delta)|$ $\pi(k)\in\Re^{A}_{i}\}$ is infinite. So,
    $i\in$\maximalInf{\pi}.
    \item The set $\{h\in{}N|$ $\sigma_{h}$ contains some occurrence of a rule in
    $\Re_{i}^{A}\}$ is infinite. Since $M_{\pi}$ is injective, it follows
    that
    $i\in$\maximalInf{\pi}.
\end{itemize}
\end{proof}

 \setcounter{equation}{0}
\begin{Lemma}\label{Lemma:From-MPAR-To-M-Inf}
Let $p$ \multiderparKomega{\sigma}  with $p\in{}T_{PAR}$. Then,
there exists in $\Re$ a derivation from $p$ of the form $p$
\multider{\delta} such that \maximal{\delta} =
\maximalparKomega{\sigma} and \maximalInf{\delta} =
\maximalparKomegaInf{\sigma} $\cup$ \maximalparKomegaPlus{\sigma}.
Moreover, if $\sigma$ is infinite or contains some occurrence of
rule in $\Re^{K,K^{\omega}}_{PAR}\setminus\Re_{PAR}^{K}$ then,
$\delta$ is infinite.
\end{Lemma}

\begin{proof}
For the proof we use the following property
\begin{description}
    \item[A.] Let $p'$$\parallel$$p''$\multiderparKomega{\sigma} with $p',p''\in{}T_{PAR}$ and $p''$
    not
    containing variables in $Var$. Then
    $p'$ \multiderparKomega{\sigma}.
\end{description}
Property \textbf{A} follows easily from the observation that the
left-hand side of each rule in $\Re^{K,K^{\omega}}_{PAR}$ doesn't
contain occurrences of $\hat{Z}_{F}$ and
$\hat{Z}_{\infty}$.\newline Let $\lambda$ be the subsequence of
$\sigma$ containing all, and only, the occurrences of rules in
$\Re^{K,K^{\omega}}_{PAR}\setminus\Re_{PAR}^{K}$. Let us assume
that $\lambda$ is infinite. In similar way we reason if $\lambda$
is finite. Now, $\lambda=r_{0}r_{1}r_{2}\ldots$, where
$\forall{}h\in{}N$
$r_{h}\in\Re^{K,K^{\omega}}_{PAR}\setminus\Re_{PAR}^{K}$.
Moreover, $\sigma$ can be written in the form
$\rho_{0}r_{0}\rho_{1}r_{1}\rho_{2}r_{2}\ldots$, where
$\sigma\setminus\lambda=\rho_{0}\rho_{1}\rho_{2}\ldots$ and
$\forall{}h\in{}N$ $\rho_{h}$ is a rule sequence (possibly empty)
in $\Re_{PAR}^{K}$. For all $h\in{}N$ we denote by $\sigma^{h}$
the suffix of
$\sigma$ $\rho_{h}r_{h}\rho_{h+1}r_{h+1}\ldots$.\\
 Now,
we prove  that there exists a succession of terms in $T_{PAR}$,
 $(p_{h})_{h\in{N}}$, a succession of variables
$(X_{h})_{h\in{}N}$ and a succession of terms $(t_{h})_{h\in{}N}$
such that:
\begin{description}
    \item[i.] $p_{0}=p$.
    \item[ii.] $\forall{}h\in{}N$   $p_{h}$ \multiderparKomega{\sigma^{h}}.
    \item[iii.] $\forall{}h\in{}N$ $p_{h}$ \multider{\eta_{h}}
    $p_{h+1}$$\parallel$$t_{h}$$\parallel$$X_{h}$ with \maximal{\eta_{h}} =
     \maximalparKomega{\rho_{h}}.
    \item[iv.] $\forall{}h\in{}N$ $X_{h}$ \multider{\pi_{h}}
     with $\pi_{h}$ infinite, \maximal{\pi_{h}} =
     \maximalparKomega{r_{h}} and \maximalInf{\pi_{h}} =
    \maximalparKomegaPlus{r_{h}}.
\end{description}

Setting $p_{0}=p$, property ii is satisfied for $h=0$. So, let us
assume that the statement is true $\forall{}h=0,\ldots,k$. Then,
it suffices to prove that
\begin{description}
    \item[B.] there exists a $p_{k+1}\in{}T_{PAR}$, a term $t_{k}$ and a variable
    $X_{k}$ such that $p_{k}$ \multider{\eta_{k}}
    $p_{k+1}$$\parallel$$t_{k}$$\parallel$$X_{k}$,   $p_{k+1}$ \multiderparKomega{\sigma^{k+1}},
    and $X_{k}$ \multider{\pi_{k}}
     with $\pi_{k}$ infinite. Moreover, \maximal{\eta_{k}} =
     \maximalparKomega{\rho_{k}}, \maximal{\pi_{k}} =
     \maximalparKomega{r_{k}} and \maximalInf{\pi_{k}} =
    \maximalparKomegaPlus{r_{k}}.
\end{description}
>From the inductive hypothesis we have $p_{k}$
\multiderparKomega{\sigma^{k}}. The  derivation  $p_{k}$
\multiderparKomega{\sigma^{k}} can be written in the form
\begin{displaymath}
\textrm{$p_{k}$ \multiderparKomega{\rho_{k}}
$p'$$\parallel$$p''$$\parallel$$X$
        \onederparKomega{r_{k}} $p'$$\parallel$$p''$$\parallel$$\hat{Z}_{\infty}$
     \multiderparKomega{\sigma^{k+1}}}
\end{displaymath}
where $r_{k}=X$\Rule{K',K'^{\omega}}$\hat{Z}_{\infty}$ with
$X\in{}Var$ and $K',K'^{\omega}\in{}P_{n}$. Moreover, $p'$ doesn't
contain occurrences of $\hat{Z}_{F}$ and $\hat{Z}_{\infty}$,  and
$p''$ doesn't contain occurrences of variables in $Var$. From the
definition of $\Re^{K,K^{\omega}}_{PAR}$ we have $X$
\multider{\pi_{k}}
     with $\pi_{k}$ infinite, \maximal{\pi_{k}} $=K'$ and \maximalInf{\pi_{k}}
     $=K'^{\omega}$. From remark \ref{Remark:MPAR-Inf} we have
     \maximalparKomega{r_{k}} $=K'$ and
     \maximalparKomegaPlus{r_{k}} $=K'^{\omega}$.
      From property \textbf{A}  it follows that $p'$
\multiderparK{\sigma^{k+1}}. Since $\rho_{k}$ is a rule sequence
in $\Re_{PAR}^{K}$, from lemma \ref{Lemma:From-MKPAR-To-M} it
follows that $p_{k}$ \multider{\eta_{k}}
    $p'$$\parallel$$t$$\parallel$$X$ for some term $t$ and
    \maximal{\eta_{k}} = \maximalparK{\rho_{k}}. From remark \ref{Remark:MPAR-Inf}
    we deduce that \maximal{\eta_{k}} =
    \maximalparKomega{\rho_{k}}.
Setting  $p_{k+1}=p'$, $t_{k}=t$ and
$X_{k}=t$ we obtain that property \textbf{B} is satisfied.\\
Thus, properties i-iv are satisfied. Now, let us consider
$\forall{}h\in{}N$ the infinite derivation $X_{h}$
\multider{\pi_{h}}. It can be written in the form
\begin{equation}
\textrm{$s_{(h,0)}$ \prsoneder{r_{(h,0)}} $s_{(h,1)}$
\prsoneder{r_{(h,1)}} $s_{(h,2)}\ldots$}
\end{equation}
where $s_{(h,0)}=X_{h}$ and $\forall{}k\in{}N$ $r_{(h,k)}\in\Re$.
For all $k\in{}N$ we denote by $\overline{r}_{k}$ the rule
$r_{(\ell(k),\wp(k))}$. For all $h,k\in{}N$ we denote by
$s_{h}(k)$ the term $s_{next(h,k)}$. Now, we show that
$\forall{}k\in{}N$ the following result holds
\begin{eqnarray}
\textrm{$p_{k+1}$$\parallel$$t_{0}$$\parallel$$\ldots$$\parallel$$t_{k}$$\parallel$$s_{0}(k)$$\parallel$$s_{1}(k)$$\parallel$$\ldots$$\parallel$$s_{k}(k)$
        \multider{\eta_{k+1}\overline{r}_{k}}}\nonumber\\
\textrm{
$p_{k+2}$$\parallel$$t_{0}$$\parallel$$\ldots$$\parallel$$t_{k}$$\parallel$$t_{k+1}$$\parallel$$s_{0}(k+1)$$\parallel$$s_{1}(k+1)$$\parallel$$\ldots$$\parallel$$s_{k+1}(k+1)$
        }
\end{eqnarray}
Since $\forall{}k\in{}N$ $s_{k}(k)=s_{next(k,k)}$, from property 1
of lemma \ref{Lemma:next} it follows that
$s_{k}(k)=s_{(k,0)}=X_{k}$. From iii we deduce that
\begin{eqnarray}
\textrm{$p_{k+1}$$\parallel$$t_{0}$$\parallel$$\ldots$$\parallel$$t_{k}$$\parallel$$s_{0}(k)$$\parallel$$s_{1}(k)$$\parallel$$\ldots$$\parallel$$s_{k}(k)$
        \multider{\eta_{k+1}}}\nonumber\\
\textrm{
$p_{k+2}$$\parallel$$t_{0}$$\parallel$$\ldots$$\parallel$$t_{k}$$\parallel$$t_{k+1}$$\parallel$$s_{0}(k)$$\parallel$$s_{1}(k)$$\parallel$$\ldots$$\parallel$$s_{k}(k)$$\parallel$$s_{k+1}(k+1)$
        }
\end{eqnarray}
So, to obtain (2) it suffices to prove that
\begin{equation}
\textrm{$s_{0}(k)$$\parallel$$s_{1}(k)$$\parallel$$\ldots$$\parallel$$s_{k}(k)$
        \prsoneder{\overline{r}_{k}}
        $s_{0}(k+1)$$\parallel$$s_{1}(k+1)$$\parallel$$\ldots$$\parallel$$s_{k}(k+1)$
        }
\end{equation}
>From property 6 of lemma \ref{Lemma:next} $\forall{}k\in{}N$
$next(\ell(k),k)=(\ell(k),\wp(k))$. Moreover,
$next(\ell(k),k+1)=(\ell(k),\wp(k)+1)$. Therefore, we have
$s_{\ell(k)}(k)=s_{(\ell(k),\wp(k))}$ \prsoneder{\overline{r}_{k}}
$s_{(\ell(k),\wp(k)+1)}=s_{\ell(k)}(k+1)$. From property 7 of
lemma \ref{Lemma:next} $\forall{}i\neq{}\ell(k)$
$next(i,k+1)=next(i,k)$. So, $\forall{}i\neq{}\ell(k)$
$s_{i}(k+1)=s_{i}(k)$. Since $\ell(k)\leq{}k$,  we obtain
evidently (4). So,  (2) is satisfied $\forall{}k\in{}N$. Moreover,
since $s_{0}(0)=X_{0}$, we have
\begin{equation}
\textrm{$p=p_{0}$ \multider{\eta_{0}}
        $p_{1}$$\parallel$$t_{0}$$\parallel$$s_{0}(0)$
        }
\end{equation}
Setting
$\delta=\eta_{0}\eta_{1}\overline{r}_{0}\eta_{2}\overline{r}_{1}\eta_{3}\overline{r}_{2}\ldots$,
from (2) and (5) we obtain that $p$ \multider{\delta} with
$\delta$ infinite. So, to obtain the assertion it remains to prove
that \maximal{\delta} = \maximalparKomega{\sigma} and
\maximalInf{\delta} = \maximalparKomegaInf{\sigma} $\cup$
\maximalparKomegaPlus{\sigma}. Let
$\mu=\overline{r}_{0}\overline{r}_{1}\overline{r}_{2}\ldots$. Let
us observe that $\mu\in{}Interleaving(\{\pi_{h}\})$.
 From iii-iv, propositions \ref{Prop:Maximal} and \ref{Prop:MaximalInterleaving},
 and remembering that
$\sigma=\rho_{0}r_{0}\rho_{1}r_{1}\ldots$, we obtain
\begin{displaymath}
\textrm{\maximal{\delta} = $\bigcup_{h\in{}N}$\maximal{\eta_{h}}
$\cup$ \maximal{\mu} =
$\bigcup_{h\in{}N}$\maximalparKomega{\rho_{h}} $\cup$
$\bigcup_{h\in{}N}$\maximal{\pi_{h}}
        =}
\end{displaymath}
\begin{displaymath}
\textrm{$\bigcup_{h\in{}N}$\maximalparKomega{\rho_{h}} $\cup$
$\bigcup_{h\in{}N}$\maximalparKomega{r_{h}} =
\maximalparKomega{\sigma}.}
\end{displaymath}
>From remark \ref{Remark:MPAR-Inf}, $\forall{}r\in\Re_{PAR}^{K}$
\maximalparKomegaPlus{r} = $\emptyset$. Remembering that
$\lambda=r_{0}r_{1}r_{2}\ldots$, from iii-iv and propositions
\ref{Prop:Maximal}, \ref{Prop:MaximalInterleaving} we obtain
\begin{displaymath}
\textrm{\maximalInf{\delta} =
$\bigoplus_{h\in{}N}$\maximal{\eta_{h}} $\cup$ \maximalInf{\mu} =
$\bigoplus_{h\in{}N}$\maximalparKomega{\rho_{h}} $\cup$
$\bigcup_{h\in{}N}$\maximalInf{\pi_{h}} $\cup$
$\bigoplus_{h\in{}N}$\maximal{\pi_{h}}
        =}
\end{displaymath}
\begin{displaymath}
\textrm{\maximalparKomegaInf{\sigma\setminus\lambda} $\cup$
$\bigcup_{h\in{}N}$\maximalparKomegaPlus{r_{h}} $\cup$
$\bigoplus_{h\in{}N}$\maximalparKomega{r_{h}} = }
\end{displaymath}
\begin{displaymath}
\textrm{\maximalparKomegaInf{\sigma\setminus\lambda} $\cup$
\maximalparKomegaPlus{\sigma} $\cup$ \maximalparKomegaInf{\lambda}
= \maximalparKomegaInf{\sigma} $\cup$
\maximalparKomegaPlus{\sigma}.}
\end{displaymath}
This concludes the proof.
\end{proof}

\setcounter{equation}{0}
\begin{Lemma}\label{Lemma:From-MKSEQ-To-M}
Let $t,t'\in{}T_{SEQ}$ and $s$ be any term in $T$ such that
$t\in{}SEQ(s)$. The following results hold
\begin{enumerate}
    \item  If $t$ \onederseqK{r} $t'$, then
     there exists a $s'\in{}T$ with $t'\in{}SEQ(s')$ such that $s$ \multider{\sigma} $s'$,
     with \maximal{\sigma} = \maximalseqK{r} and $|\sigma|>0$.
      \item If $t$ \multiderseqK{\sigma} $t'$ with $t\neq\varepsilon$,
    then there exists a $s'\in{}T$ with $t'\in{}SEQ(s')$ such that $s$
    \multider{\rho} $s'$, with \maximal{\rho} =
    \maximalseqK{\sigma}, and $|\rho|>0$ if $|\sigma|>0$.
    \item If $t$ \multiderseqK{\sigma} is a $(K,K^{\omega})$-accepting infinite derivation  in
    $M_{SEQ}^{K}$ from $t\in{}T_{SEQ}$, then  there exists a $(K,K^{\omega})$-accepting infinite derivation
     in $M$ from s.
\end{enumerate}
\end{Lemma}
\begin{proof}
At first, we prove property 1. We use the following properties,
whose proof is immediate. Let $t\in{}SEQ(s)$,
    $s\in{}T$ and $t=X_{1}.(X_{2}.(\ldots{}X_{n}.(Y)\ldots))$, with
    $n\geq0$. Then
\begin{description}
    \item[A.] if
    $st\in{}T_{SEQ}\setminus\{\varepsilon\}$ and
     $t'=X_{1}.(X_{2}.(\ldots{}X_{n}.(st)\ldots))$, then there
     exists a
     $s'\in{}s[Y\rightarrow{}st]$
     (notice that $Y$ is a subterm of $s$)  such that $t'\in{}SEQ(s')$.
    \item[B.] if  $Z\in{}Var$, $st'\in{}T$ and
    $st=st'$$\parallel$$Z$, then
     there exists a $s'\in{}s[Y\rightarrow{}st]$
       such that
        $X_{1}.(X_{2}.(\ldots{}X_{n}.(Z)\ldots))$ $\in{}SEQ(s')$.
\end{description}
We can, now, distinguish the following two cases:
\begin{itemize}
    \item  $r=Y$\Rule{a}$Z_{1}.(Z_{2})\in{}\Re$. From remark \ref{Remark:MKSEQ}
     \maximal{r} = \maximalseqK{r}. Moreover,
    $t=X_{1}.(X_{2}.(\ldots{}X_{n}.(Y)\ldots))$ and $t'=X_{1}.(X_{2}.(\ldots{}X_{n}$
    $.(Z_{1}.(Z_{2}))\ldots))$.
    Let  $s\in{}T$ be such that $t\in{}SEQ(s)$. From \textbf{A} above,
     there exists a $s'\in{}s[Y\rightarrow{}Z_{1}.(Z_{2})]$
      such that $t'\in{}SEQ(s')$. Since $Y$ \prsoneder{r}
      $Z_{1}.(Z_{2})$,
      by proposition \ref{Prop:Subterms1} it follows that $s$ \prsoneder{r} $s'$, and the
      thesis is proved.
    \item $r=Y$\Rule{K'}$Z$ with $Y,Z\in{}Var$ and
    \maximalseqK{r} = $K'$. Moreover, $t=X_{1}.(X_{2}.(\ldots{}$
      $X_{n}.(Y)\ldots))$
     and $t'=X_{1}.(X_{2}.(\ldots{}X_{n}.(Z)\ldots))$.
     From the definition of $\Re^{K}_{SEQ}$ there exists a derivation
      in
     $\Re^{K}_{PAR}$ of the form $Y$ \multiderparK{\sigma} $p$$\parallel$$Z$
     for some
     $p\in{}T_{PAR}$, with \maximalparK{\sigma} = \maximalseqK{r} and
     $|\sigma|>0$.
      From lemma \ref{Lemma:From-MKPAR-To-M} there exists a term
      $st$ such that $Y$ \multider{\rho} $st$$\parallel$$Z$ with
      \maximal{\rho} = \maximalparK{\sigma} and $|\rho|>0$. So, \maximal{\rho} = \maximalseqK{r}.
       Let $s\in{}T$ be
      such that
    $t\in{}SEQ(s)$. From property \textbf{B} above, there exists a
    $s'\in{}s[Y\rightarrow{}st$$\parallel$$Z]$
    such that
    $t'\in{}SEQ(s')$.
    Since $Y$ \multider{\rho} $st$$\parallel$$Z$,
    by proposition \ref{Prop:Subterms1} we conclude that $s$ \multider{\rho} $s'$
    with $|\rho|>0$. Hence, the thesis.
\end{itemize}
Property 2 can be easily proved by induction on the length of
$\sigma$, and using  property 1.  It remains to prove property 3.
 The infinite  derivation $t$
\multiderseqK{\sigma} can be written in the form
\begin{displaymath}
\textrm{$t_{0}$ \onederseqK{r_{0}} $t_{1}$ \onederseqK{r_{1}}
$t_{2}$ \onederseqK{r_{2}} $\ldots$}
\end{displaymath}
where $t_{0}=t$ and $\forall{}i\in{}N$ $r_{i}\in{}\Re_{SEQ}^{K}$.
Let $s\in{}T$ such that $t\in{}SEQ(s)$. From property 1 there
exists a $s_{1}\in{}T$ with $t_{1}\in{}SEQ(s_{1})$ such that $s$
\multider{\lambda_{0}} $s_{1}$ with \maximal{\lambda_{0}} =
\maximalseq{r_{0}} and $|\lambda_{0}|>0$. Iterating such reasoning
we deduce that there exists a succession of terms,
$(s_{h})_{h\in{}N}$, such that for all $h\in{}N$
\begin{displaymath}
\textrm{ $s_{h}$ \multider{\lambda_{h}} $s_{h+1}$ with
\maximal{\lambda_{h}} = \maximalseqK{r_{h}}, $|\lambda_{h}|>0$ and
$s_{0}=s$.}
\end{displaymath}
Therefore,
\begin{displaymath}
\textrm{$s=s_{0}$  \multider{\lambda_{0}} $s_{1}$
\multider{\lambda_{1}}$\ldots$ $s_{h}$ \multider{\lambda_{h}}
$s_{h+1}\ldots$}
\end{displaymath}
Let $\rho=\lambda_{0}\lambda_{1}\lambda_{2}\ldots$. So, $s$
\multider{\rho}
 is an  infinite derivation  in $\Re$ from $s$ with $t\in{}SEQ(s)$. Moreover, by
 proposition \ref{Prop:Maximal} we
 obtain
\begin{displaymath}
\textrm{\maximal{\rho} = $\bigcup_{h\in{}N}$ \maximal{\lambda_{h}}
= $\bigcup_{h\in{}N}$ \maximalseqK{r_{h}} = \maximalseqK{\sigma} =
$K$. }
\end{displaymath}
\begin{displaymath}
\textrm{\maximalInf{\rho} = $\bigoplus_{h\in{}N}$
\maximal{\lambda_{h}} = $\bigoplus_{h\in{}N}$ \maximalseqK{r_{h}}
= \maximalseqInfK{\sigma} = $K^{\omega}$.  }
\end{displaymath}
This proves the thesis.
\end{proof}

 \setcounter{equation}{0}

\begin{Proposition}\label{Prop:Interleaving}
Let $\sigma$ be a rule sequence in $\Re$ and
$\{\rho_{h}\}_{h\in{}N}$ be a succession of subsequences of
$\sigma$ two by two disjoints and such that
$\bigcup_{h\in{}N}pr(\rho_{h})=pr(\sigma)$. Then,
$\sigma\in{}Interleaving(\{\rho_{h}\})$.
\end{Proposition}
\begin{proof}
Setting $\Delta=\bigcup_{h\in{}N}(\{h\}\times{}pr(\rho_{h}))$, let
us consider the following mapping
\begin{displaymath}
 M_{\sigma}: (h,n)\in{}\Delta\rightarrow{}n
\end{displaymath}
Since for all $h,h'\in{}N$ with $h\neq{}h'$ we have
$pr(\rho_{h})\cap{}pr(\rho_{h'})=\emptyset$, it follows that
$M_{\sigma}$ is an injective mapping. Let us observe that
$M_{\sigma}(h,n_{1})<M_{\sigma}(h,n_{2})$ if $n_{1}<n_{2}$.
Moreover, $pr(\sigma)=M_{\sigma}(\Delta)$, and
$\forall{}(h,n)\in\Delta$ we have
$\sigma(M_{\sigma}(h,n))=\sigma(n)=\rho_{h}(n)$. From definition
\ref{Def:Interleaving} we obtain the assertion.
\end{proof}

\begin{Lemma}\label{Lemma:From-M-To-MPAROMEGA}
 Let $p$ \multider{\sigma}$ $ be a $(\overline{K},\overline{K}^{\omega})$-accepting
 non--null derivation  in $M$ belonging to
 $\Pi^{K,K^{\omega}}_{PAR,\infty}$,
 with $p\in{}T_{PAR}$, $\overline{K}\subseteq{}K$ and
 $\overline{K}^{\omega}\subseteq{}K^{\omega}$.  Then,
 there exists a derivation  $p$ \multiderparKomega{\rho} in $\Re^{K,K^{\omega}}_{PAR}$
 from $p$ such that
 \begin{description}
    \item[a] \maximalparKomega{\rho} = $\overline{K}$
    \item[b]
    \maximalparKomegaInf{\rho} $\cup$ \maximalparKomegaPlus{\rho} = $\overline{K}^{\omega}$
    \item[c] If $\sigma$ is infinite then, either $\rho$ is
    infinite or $\rho$ contains some occurrence of rule in
    $\Re^{K,K^{\omega}}_{PAR}\setminus\Re_{PAR}^{K}$.
 \end{description}
\end{Lemma}
\begin{proof}

\noindent{}At first, let us prove the following property
\begin{description}
    \item[d] There exists a $p'\in{}T_{PAR}$, a non empty finite rule
    sequence $\lambda$ in $\Re^{K,K^{\omega}}_{PAR}$, and a non
    empty subsequence $\eta$ (possibly infinite) of $\sigma$ such that:
    \begin{enumerate}
        \item $min(pr(\eta))=min(pr(\sigma))$ (i.e. the first rule
        occurrence in $\eta$ is the first rule occurrence in
        $\sigma$).
        \item $p$ \multiderparKomega{\lambda} $p'$
        \item  \maximalparKomega{\lambda} = \maximal{\eta}
        \item \maximalparKomegaPlus{\lambda} = \maximalInf{\eta}
        \item $p'$ \multider{\sigma\setminus\eta}, and this derivation
        is in $\Pi^{K,K^{\omega}}_{PAR,\infty}$.
        \item If $\sigma$ is infinite then, either $\sigma\setminus\eta$
        is infinite or $\lambda$ is a rule in $\Re^{K,K^{\omega}}_{PAR}\setminus\Re_{PAR}^{K}$.
    \end{enumerate}
\end{description}

Now, let us show that from property \textbf{d} the thesis follows.
So, let us assume that  property \textbf{d} is satisfied. Since
$\sigma\setminus\eta$ is a subsequence of $\sigma$, we have
\maximal{\sigma\setminus\eta} $\subseteq{}K$ and
\maximalInf{\sigma\setminus\eta} $\subseteq{}K^{\omega}$. Thus, if
$\sigma\neq\eta$ we can apply  property \textbf{d} to the
derivation $p'$ \multider{\sigma\setminus\eta}. Iterating this
reasoning it follows that there exists a $m\in{}N\cup\{\infty\}$,
a succession $\{p_{h}\}_{h=0}^{m+1}$ of terms in $T_{PAR}$, a
succession $\{\lambda_{h}\}_{h=0}^{m}$ of non empty finite rule
sequences in $\Re^{K,K^{\omega}}_{PAR}$, two successions
$\{\sigma_{h}\}_{h=0}^{m}$ and $\{\eta_{h}\}_{h=0}^{m}$ of non
empty rule sequences in $\Re$ such that
\begin{enumerate}
    \item[7.] $p=p_{0}$ and $\sigma=\sigma_{0}$.
    \item[8.] for all $h=0,\ldots,m\quad$ $\eta_{h}$ is a subsequence of
    $\sigma_{h}$ and $min(pr(\eta_{h}))=min(pr(\sigma_{h}))$.
    \item[9.] for all $h=0,\ldots,m-1\quad$
    $\sigma_{h+1}=\sigma_{h}\setminus\eta_{h}$.
    \item[10.] for all $h=0,\ldots,m\quad$
        $p_{h}$ \multiderparKomega{\lambda_{h}} $p_{h+1}$.
    \item[11.] for all $h=0,\ldots,m\quad$
    \maximalparKomega{\lambda_{h}} = \maximal{\eta_{h}} and
        \maximalparKomegaPlus{\lambda_{h}} = \maximalInf{\eta_{h}}
    \item[12.] for all $h=1,\ldots,m\quad$ $p_{h}$
    \multider{\sigma_{h}}.
    \item[13.] If $m$ is finite then, $\sigma_{m}=\eta_{m}$.
    \item[14.] If $\sigma$ is infinite then, either  $m$ is infinite or
    there exists an $h$ such that $\lambda_{h}$ is a rule in
    $\Re^{K,K^{\omega}}_{PAR}\setminus\Re_{PAR}^{K}$.
\end{enumerate}
\noindent{}By setting $\rho=\lambda_{0}\lambda_{1}\ldots$ we have
that $p$ \multiderparKomega{\rho}. From property 14 it follows
that if $\sigma$ is infinite then, either $\rho$ is infinite or
$\rho$ contains some occurrence of rule in
$\Re^{K,K^{\omega}}_{PAR}\setminus\Re_{PAR}^{K}$. So, property
\textbf{c} is satisfied. It remains to prove  properties
\textbf{a} and \textbf{b}. Let us assume that $m=\infty$. The
proof in the case where $m$ is finite is simpler.  From properties
7-9 $\eta_{0},\eta_{1},\ldots$ are non empty subsequences of
$\sigma$ two by two disjoints. Since $\sigma$ is infinite,  we can
assume that $pr(\sigma)=N$. Now, let us show that
\begin{enumerate}
    \item[15.] $\sigma\in{}Interleaving(\{\eta_{h}\})$
\end{enumerate}
>From proposition \ref{Prop:Interleaving} it suffices to prove that
$\forall{}h\in{}N$ there exists a $i\in{}N$ such that
$h\in{}pr(\eta_{i})$. From properties 8-9 it follows  that
$\forall{}h\in{}N$ $min(pr(\sigma_{h}))< min(pr(\sigma_{h+1}))$.
Let $h\in{}N$, then there exists the smallest $i\in{}N$ such that
$h\notin{}pr(\sigma_{i})$. Since $\sigma_0=\sigma$, $i>0$ and
$h\in{}pr(\sigma_{i-1})$. Since
$\sigma_{i}=\sigma_{i-1}\setminus\eta_{i-1}$,
$h\notin{}pr(\sigma_{i})$ and $h\in{}pr(\sigma_{i-1})$, it follow
that
$h\in{}pr(\eta_{i-1})$. Thus, property 15 holds.\\
>From properties 11, 15, and propositions \ref{Prop:Maximal} and
\ref{Prop:MaximalInterleaving} it follows that
\begin{displaymath}
\textrm{\maximalparKomega{\rho} $=$
$\bigcup_{h\in{}N}$\maximalparKomega{\lambda_{h}} $=$
$\bigcup_{h\in{}N}$\maximal{\eta_{h}} $=$ \maximal{\sigma} $=$
$\overline{K}$.}
\end{displaymath}
Therefore, property \textbf{a} holds. Moreover,
 \begin{displaymath}
\textrm{\maximalparKomegaInf{\rho} $\cup$
\maximalparKomegaPlus{\rho} $=$
$\bigoplus_{h\in{}N}$\maximalparKomega{\lambda_{h}} $\cup$
$\bigcup_{h\in{}N}$\maximalparKomegaPlus{\lambda_{h}} =}
\end{displaymath}
\begin{displaymath}
\textrm{$\bigoplus_{h\in{}N}$\maximal{\eta_{h}} $\cup$
$\bigcup_{h\in{}N}$\maximalInf{\eta_{h}} $=$ \maximalInf{\sigma}
$=$ $\overline{K}^{\omega}$.}
\end{displaymath}
Therefore,
property \textbf{b} is satisfied.\\
 \noindent{}At this point, it remains to prove  property \textbf{d}. The derivation
$p$ \multider{\sigma} can be written in the form
\begin{equation}
\textrm{$p$ \prsoneder{r} $t$ \multider{\sigma'}}
\end{equation}
Since each subderivation of $t$ \multider{\sigma'} is also a
subderivation of $p$ \multider{\sigma}, it follows that $t$
\multider{\sigma'} is in $\Pi^{K,K^{\omega}}_{PAR,\infty}$.
 There are two cases:
\begin{enumerate}
    \item r is a PAR rule. Then, we have that $t\in{}T_{PAR}$ and
     $r\in\Re^{K,K^{\omega}}_{PAR}$. From remark \ref{Remark:MPAR-Inf}
    \maximalparKomega{r} = \maximal{r}, and
    \maximalparKomegaPlus{r} = $\emptyset$ = \maximalInf{r}.
    Moreover,
    $t$ \multider{\sigma'}
        is in $\Pi^{K,K^{\omega}}_{PAR,\infty}$ with
        $\sigma'=\sigma\setminus{}r$.
    Thus, since $\sigma'$ is infinite if $\sigma$ is infinite,
    property \textbf{d} follows,  setting $p'=t$, $\lambda=r$ and
    $\eta=r$.
    \item $r=Z$\Rule{a}$Y.(Z')$. So, $p=p''$$\parallel$$Z$ and
    $t=p''$$\parallel$$Y.(Z')$ with $p''\in{}T_{PAR}$.
    From (1), let $Z'$ \multider{\nu}  a subderivation of $t=p''$$\parallel$$Y.(Z')$
    \multider{\sigma'} from $Z'$.  From lemma \ref{Lemma:Subderivations1} we can
    distinguish four subcases:
    \begin{itemize}
        \item $Z'$ \multider{\nu} is infinite,
               and $p''$ \multider{\sigma'\setminus\nu}.
        Moreover, $p''$ \multider{\sigma'\setminus\nu}
                is in $\Pi^{K,K^{\omega}}_{PAR,\infty}$.
        Clearly, for every $\overline{p}\in{}T_{PAR}$
        we have that $p''$$\parallel$$\overline{p}$
        \multider{\sigma'\setminus\nu}, and this derivation
        belongs to $\Pi^{K,K^{\omega}}_{PAR,\infty}$.
                From hypothesis, we have that
                $($\maximal{\nu},\maximalInf{\nu}$)\neq(K,K^{\omega})$,
                \maximal{\nu} $\subseteq{}K$ and
                \maximalInf{\nu} $\subseteq{}K^{\omega}$.
                Hence,   $|$\maximal{\nu}$|+|$\maximalInf{\nu}$|<|K|+|K^{\omega}|$.
                Moreover, $r=Z$\Rule{a}$Y.(Z')$ with \maximal{r} $\subseteq{}K$.
                From the definition of $\Re^{K,K^{\omega}}_{PAR}$, it follows that
                $r'=Z$\Rule{K_{1},K^{\omega}_{1}}$\hat{Z}_{\infty}\in\Re^{K,K^{\omega}}_{PAR}\setminus\Re_{PAR}^{K}$
                where $K_{1}$ = \maximal{\nu} $\cup$ \maximal{r} and
                $K^{\omega}_{1}$ = \maximalInf{\nu}.
                From remark \ref{Remark:MPAR-Inf}, we have that
                \maximalparKomega{r'} = $K_{1}$ and \maximalparKomegaPlus{r'} = $K^{\omega}_{1}$.
                Therefore, we deduce that $p=p''$$\parallel$$Z$
                \onederparKomega{r'}
                $p''$$\parallel$$\hat{Z}_{\infty}$.
                Moreover, $p''$$\parallel$$\hat{Z}_{\infty}$
                \multider{\sigma'\setminus\nu} and this derivation is in $\Pi^{K,K^{\omega}}_{PAR,\infty}$.
                Since
                $\sigma'\setminus\nu=\sigma\setminus{}r\nu$ and
                \maximalparKomegaPlus{r'} = \maximalInf{\nu} = \maximalInf{r\nu},
                property \textbf{d} follows, setting
                $p'=p''$$\parallel$$\hat{Z}_{\infty}$, $\lambda=r'$ and
                $\eta=r\nu$.
        \item $Z'$ \multider{\nu} leads to a term $t_{1}\neq\varepsilon$,
               and $p''$ \multider{\sigma'\setminus\nu}. Moreover, $p''$ \multider{\sigma'\setminus\nu}
                is in $\Pi^{K,K^{\omega}}_{PAR,\infty}$.
                Since \maximal{\nu} $\subseteq$ $K$, from lemma
                \ref{Lemma:From-M-To-MKPAR} there exists a
                $\overline{p}\in{}T_{PAR}$ such that
                $Z'$ \multiderparK{\gamma} $\overline{p}$, where
                \maximalparK{\gamma} = \maximal{\nu}.
                Since \maximal{r} $\subseteq$ $K$,  from the
                definition of $\Re_{PAR}^{K}$ it follows that
                $r'=Z$\Rule{K'}$\hat{Z}_{F}\in\Re_{PAR}^{K}$
                with \maximalparK{r'} = $K'$, where
                $K'=$ \maximal{r\nu}. By construction
                $r'\in\Re^{K,K^{\omega}}_{PAR}$, and
                from remark \ref{Remark:MPAR-Inf} \maximalparKomega{r'} = $K'$ and
                \maximalparKomegaPlus{r'} = $\emptyset$.  Since
                $\sigma'\setminus\nu=\sigma\setminus{}r\nu$,
                \maximalparKomegaPlus{r'}  =
                $\emptyset$ = \maximalInf{r\nu}, and $\sigma'\setminus\nu$ is infinite if
                $\sigma$ is infinite,
                property \textbf{d} follows,  setting
                $p'=p''$$\parallel$$\hat{Z}_{F}$, $\lambda=r'$ and
                $\eta=r\nu$.
        \item $Z'$ \multider{\nu} leads to $\varepsilon$ and the derivation $p''$$\parallel$$Y.(Z')$
                \multider{\sigma'}  can be written in the
              following form
              \begin{equation}
                \textrm{$p''$$\parallel$$Y.(Z')$
                \multider{\sigma_{1}} $t'$$\parallel$$Y$
                \multider{\sigma_{2}} $\quad$
                with $p''$ \multider{\sigma'_{1}} $t'$ and
                $\sigma_{1}\in{}Interleaving(\nu,\sigma'_{1})$ }
              \end{equation}
              Moreover, $p''$$\parallel$$Y$
                \multider{\sigma'_{1}} $t'$$\parallel$$Y$
                \multider{\sigma_{2}} and this derivation is in $\Pi^{K,K^{\omega}}_{PAR,\infty}$.
                Since  $Z'$ \multider{\nu} $\varepsilon$ and
                \maximal{\nu} $\subseteq$ $K$, from lemma
                \ref{Lemma:From-M-To-MKPAR}
                $Z'$ \multiderparK{\chi}
                $\varepsilon$ with
                \maximalparK{\chi} = \maximal{\nu}.
                Since $r=Z$\Rule{a}$Y.(Z')$  with \maximal{r} $\subseteq$
                $K$, from the definition of $\Re_{PAR}^{K}$ it
                follows that  $r'=Z$\Rule{K'}$Y\in\Re_{PAR}^{K}$
                where
                $K'=$ \maximal{r\nu}  and
                \maximalparK{r'} = $K'$.
                By construction
                $r'\in\Re^{K,K^{\omega}}_{PAR}$, and
                from remark \ref{Remark:MPAR-Inf} \maximalparKomega{r'} = $K'$ and
                \maximalparKomegaPlus{r'} = $\emptyset$.
                Since
                $\sigma\setminus{}r\nu=\sigma'_{1}\sigma_{2}$,
                \maximalparKomegaPlus{r'} =
                $\emptyset$ = \maximalInf{r\nu} , and $\sigma'_{1}\sigma_{2}$ is
                infinite if $\sigma$ is infinite,
                property \textbf{d} follows, setting
                $p'=p''$$\parallel$$Y$, $\lambda=r'$ and
                $\eta=r\nu$.
        \item
             $Z'$ \multider{\nu} leads to a variable $W\in{}Var$ and the derivation
             $p''$$\parallel$$Y.(Z')$
                \multider{\sigma'}  can be written in the
              following form
              \begin{eqnarray}
                \textrm{$p''$$\parallel$$Y.(Z')$
                \multider{\sigma_{1}} $t'$$\parallel$$Y.(W)$
                \prsoneder{r'} $t'$$\parallel$$W'$ \multider{\sigma_{2}}}\\
                \textrm{with $p''$ \multider{\sigma'_{1}} $t'$,
                $r'=Y.(W)$\Rule{b}$W'$ and
                $\sigma_{1}\in{}Interleaving(\lambda,\sigma'_{1})$}
              \end{eqnarray}
              Moreover, $p''$$\parallel$$W'$
                \multider{\sigma'_{1}} $t'$$\parallel$$W'$
                \multider{\sigma_{2}} and this derivation  is in $\Pi^{K,K^{\omega}}_{PAR,\infty}$.
                Since  $Z'$ \multider{\nu} $W$
                and
                \maximal{\nu} $\subseteq$ $K$, from lemma
                \ref{Lemma:From-M-To-MKPAR} we have that $Z'$ \multiderparK{\chi}
                $W$  with
                \maximalparK{\chi} = \maximal{\nu}.
              Since $r=Z$\Rule{a}$Y.(Z')\in\Re$ and
              $r'=Y.(W)$\Rule{a}$W'\in\Re$, where \maximal{r} $\subseteq$
              $K$ and
              \maximal{r'} $\subseteq$ $K$, from the definition of
               $\Re_{PAR}^{K}$ it follows that
              $r''=Z$\Rule{K'}$W'\in\Re_{PAR}^{K}$
                where
                $K'=$ \maximal{rr'} $\cup{}$ \maximalpar{\chi} = \maximal{r\nu{}r'}  and
                \maximalparK{r''} = $K'$. By construction,
                $r''\in\Re^{K,K^{\omega}}_{PAR}$, and
                from remark \ref{Remark:MPAR-Inf} \maximalparKomega{r''} = $K'$ and
                \maximalparKomegaPlus{r''} = $\emptyset$.
               Since
                $\sigma\setminus{}r\nu{}r'=\sigma'_{1}\sigma_{2}$,
                \maximalparKomegaPlus{r''} = $\emptyset$ = \maximalInf{r\nu{}r'},
                and $\sigma'_{1}\sigma_{2}$ is
                infinite if $\sigma$ is infinite,
                property \textbf{d} follows setting
                $p'=p''$$\parallel$$W'$, $\lambda=r''$ and
                $\eta=r\nu{}r'$.
    \end{itemize}
\end{enumerate}
\end{proof}

\noindent{}Now, let us assume that $K\neq{}K^{\omega}$. Then, the
following result holds.

\begin{Lemma}\label{Lemma:From-M-To-MSEQ1}
Let us assume that $K\neq{}K^{\omega}$. Given a variable
$X\in{}Var$ and  a
 $(K,K^{\omega})$-accepting infinite derivation  in $M$ from
 $X$, the following property is satisfied:
\begin{enumerate}
    \item There exists a variable  $Y\in{}Var$ reachable from $X$ in
    $\Re_{SEQ}^{K}$ through a $(K',\emptyset)$-accepting derivation in $M_{SEQ}^{K}$
    with $K'\subseteq{}K$, and there exists a derivation
    $Y$ \multiderparKomega{\rho}
    such that \maximalparKomega{\rho} = $K$
    and
    \maximalparKomegaInf{\rho} $\cup$ \maximalparKomegaPlus{\rho} =
    $K^{\omega}$. Moreover, either $\rho$ is
    infinite or $\rho$ contains some occurrence of rule in
    $\Re^{K,K^{\omega}}_{PAR}\setminus\Re_{PAR}^{K}$.
\end{enumerate}
\end{Lemma}
\begin{proof}
Since $K\neq{}K^{\omega}$ and $K\supseteq{}K^{\omega}$, it follows
that $K\supset{}K^{\omega}$.
 Let  $d=X$ \multider{\sigma} be
 a $(K,K^{\omega})$-accepting infinite derivation  in $M$ from
 $X$.  Evidently,
$K\setminus{}K^{\omega}=\{i\in\{1,\ldots,n\}|$ $\sigma$ contains a
finite non--null number of occurrences of rules in
$\Re_{i}^{A}\}$. Then, for all $i\in{}K\setminus{}K^{\omega}$ it's
defined the greatest application level, denoted by $h_{i}(d)$, of
occurrences of rules of $\Re_{i}^{A}$ in the derivation $d$.  The
proof is by induction on
$max_{i\in{}K\setminus{}K^{\omega}}\{h_{i}(d)\}$.\\
 \textbf{Base Step}:
$max_{i\in{}K\setminus{}K^{\omega}}\{h_{i}(d)\}=0$. In this case
it follows that  each subderivation  of  $d=X$ \multider{\sigma}
does not contain occurrences of rules in
$\bigcup_{i\in{}K\setminus{}K^{\omega}}\Re_{i}^{A}$. So, $d$ is
belonging to
 $\Pi^{K,K^{\omega}}_{PAR,\infty}$. Then, from lemma
 \ref{Lemma:From-M-To-MPAROMEGA} we obtain the assertion
 setting $Y=X$.\\
 \textbf{Induction Step}: $max_{i\in{}K\setminus{}K^{\omega}}\{h_{i}(d)\}>0$. If
 $d=X$ \multider{\sigma} is in
$\Pi^{K,K^{\omega}}_{PAR,\infty}$, from lemma
\ref{Lemma:From-M-To-MPAROMEGA} we obtain the assertion setting
$Y=X$. Otherwise, from  lemma \ref{Lemma:Base} it follows that the
derivation $X$ \multider{\sigma} can be written in the form
\begin{displaymath}
\textrm{$X$ \multider{\sigma_{1}} $t$$\parallel$$Z$ \prsoneder{r}
$t$$\parallel$$Y.(Z')$ \multider{\sigma_{2}}}
\end{displaymath}
where $r=Z$\Rule{a}$Y.(Z')$, and there exists a subderivation of
$t$$\parallel$$Y.(Z')$ \multider{\sigma_{2}} from $Z'$, namely
$d'=Z'$ \multider{\sigma_{2}'}, that  is a
$(K,K^{\omega})$-accepting infinite derivation in $M$. Evidently,
$max_{i\in{}K\setminus{}K^{\omega}}\{h_{i}(d')\}<max_{i\in{}K\setminus{}K^{\omega}}\{h_{i}(d)\}$.
By inductive hypothesis, the thesis holds for the derivation $d'$.
Therefore, it suffices to prove that $Z'$ is reachable from $X$ in
    $\Re_{SEQ}^{K}$ through a $(K',\emptyset)$-accepting derivation in $M_{SEQ}^{K}$
    with $K'\subseteq{}K$. From lemma \ref{Lemma:From-M-To-MKPAR}, applied to
    the derivation    $X$ \multider{\sigma_{1}} $t$$\parallel$$Z$
    where \maximal{\sigma_{1}} $\subseteq{}K$, there exists a $p\in{}T_{PAR}$
    such that $X$ \multiderparK{\rho_{1}} $p$$\parallel$$Z$
    with \maximalparK{\rho_{1}} = \maximal{\sigma_{1}}. From the
    definition of $\Re_{SEQ}^{K}$ we obtain that  $X$
    \multiderseqK{\gamma} $Z$ \onederseqK{r} $Y.(Z')$,
    with \maximalseqK{\gamma} = \maximalparK{\rho_{1}} and
    \maximalseqK{r} = \maximal{r} $\subseteq{}K$. So,
    \maximalseqK{\gamma{}r} $\subseteq{}K$. This concludes the proof.
\end{proof}

\noindent{}Now, let us assume that $K=K^{\omega}$. The next two
lemmata manage this case.

\begin{Lemma}\label{Lemma:From-M-To-MSEQ2}
 Let $i\in{}K$, $X\in{}Var$ and
 $X$ \multider{\sigma} be a
 $(K,K^{\omega})$-accepting infinite derivation  in $M$ from
 $X$.
Then, one of the following conditions is satisfied:
\begin{enumerate}
    \item There exists a variable  $Y\in{}Var$ reachable from $X$ in
    $\Re_{SEQ}^{K}$ through a $(K',\emptyset)$-accepting derivation in $M_{SEQ}^{K}$
    with $K'\subseteq{}K$, and there exists a derivation
    $Y$ \multiderparKomega{\rho} such that \maximalparKomega{\rho} = $K$
    and
    \maximalparKomegaInf{\rho} $\cup$ \maximalparKomegaPlus{\rho} =
    $K^{\omega}$. Moreover, either $\rho$ is
    infinite or $\rho$ contains some occurrence of rule in
    $\Re^{K,K^{\omega}}_{PAR}\setminus\Re_{PAR}^{K}$.
    \item There exists a variable  $Y\in{}Var$ reachable from $X$ in
    $\Re_{SEQ}^{K}$ through a $(K_{i},\emptyset)$-accepting derivation in $M_{SEQ}^{K}$
    with $\{i\}\subseteq{}K_{i}\subseteq{}K$, and
    there exists a $(K,K^{\omega})$-accepting infinite derivation  in $M$ from
    $Y$.
\end{enumerate}
\end{Lemma}
\begin{proof}
The proof is by induction on the level $k$ of application of the
first occurrence of a rule $r$ of $\Re_{i}^{A}$  in a
$(K,K^{\omega})$-accepting infinite derivation  in $M$ from
 a variable.\\
\textbf{Base Step}: $k=0$. If $X$ \multider{\sigma} is in
$\Pi^{K,K^{\omega}}_{PAR,\infty}$, from lemma
\ref{Lemma:From-M-To-MPAROMEGA} property 1 follows, setting $Y=X$.
Otherwise, from  lemma \ref{Lemma:Base} it follows that the
derivation $X$ \multider{\sigma} can be written in the form
\begin{displaymath}
\textrm{$X$ \multider{\sigma_{1}} $t$$\parallel$$Z$ \prsoneder{r'}
$t$$\parallel$$Y.(Z')$ \multider{\sigma_{2}}}
\end{displaymath}
where $r'=Z$\Rule{a}$Y.(Z')$, and there exists a subderivation of
$t$$\parallel$$Y.(Z')$ \multider{\sigma_{2}} from $Z'$, namely
$Z'$ \multider{\sigma_{2}'}, that is a $(K,K^{\omega})$-accepting
infinite derivation in $M$. By noticing that every rule occurrence
in $\sigma_{2}'$ is applied to a  level greater than zero in $X$
\multider{\sigma}, and that we are considering the case where
$k=0$, it follows that $r$ must occur in the rule sequence
$\sigma_{1}r'(\sigma_{2}\setminus\sigma_{2}')$.
 From lemma \ref{Lemma:Subderivations1}, we have $t$
\multider{\sigma_{2}\setminus\sigma_{2}'}. Therefore,  there
exists a derivation of the form $X$ \multider{\lambda}
$t'$$\parallel$$Z$ \prsoneder{r'} $t'$$\parallel$$Y.(Z')$ with
$\{i\}\subseteq$ \maximal{\lambda{}r'} $\subseteq{}K$. From lemma
 \ref{Lemma:From-M-To-MKPAR}, applied to the derivation $X$ \multider{\lambda}
$t'$$\parallel$$Z$, there exists a $p\in{}T_{PAR}$ such that $X$
\multiderparK{\rho} $p$$\parallel$$Z$, with \maximalparK{\rho} =
\maximal{\lambda}.  From the definition of $\Re_{SEQ}^{K}$ we have
 that $X$ \multiderseqK{\mu} $Z$ \onederseqK{r'}
$Y.(Z')$, with \maximalseqK{\mu} = \maximalparK{\rho} and
\maximalseqK{r'} = \maximal{r'}. Therefore, \maximalseqK{\mu{}r'}
= \maximal{\lambda{}r'}.
 Thus,  variable $Z'$ is reachable from $X$ in
    $\Re_{SEQ}^{K}$ through a $(K_{i},\emptyset)$-accepting derivation in $M_{SEQ}^{K}$
    with $\{i\}\subseteq{}K_{i}\subseteq{}K$, and
    there exists a $(K,K^{\omega})$-accepting infinite derivation  in $M$ from
    $Z'$. This is exactly what property 2 states.\\
 \textbf{Induction Step}: $k>0$. If $X$ \multider{\sigma} is in
$\Pi^{K,K^{\omega}}_{PAR,\infty}$, from lemma
\ref{Lemma:From-M-To-MPAROMEGA} property 1 follows, setting $Y=X$.
Otherwise, from  lemma \ref{Lemma:Base} it follows that the
derivation $X$ \multider{\sigma} can be written in the form
\begin{displaymath}
\textrm{$X$ \multider{\sigma_{1}} $t$$\parallel$$Z$ \prsoneder{r'}
$t$$\parallel$$Y.(Z')$ \multider{\sigma_{2}}}
\end{displaymath}
where $r'=Z$\Rule{a}$Y.(Z')$, and there exists a subderivation of
$t$$\parallel$$Y.(Z')$ \multider{\sigma_{2}} from $Z'$, namely
$Z'$ \multider{\sigma_{2}'}, that is a $(K,K^{\omega})$-accepting
infinite derivation in $M$.
 There can be two
cases:
\begin{itemize}
    \item The rule sequence
    $\sigma_{1}r'( \sigma_{2}\setminus\sigma_{2}')$ contains the first occurrence of $r$
    in $\sigma$. In this case, the thesis follows by reasoning as
    in the base step.
    \item  $\sigma_{2}'$ contains the first occurrence of $r$ in $\sigma$.
    Clearly, this occurrence is the first occurrence of a  rule of $\Re_{i}^{A}$
    in the $(K,K^{\omega})$-accepting
    infinite derivation
     $Z'$ \multider{\sigma_{2}'}, and it is applied to level $k'$ in
    $Z'$ \multider{\sigma_{2}'} with $k'<k$. By inductive hypothesis, the thesis holds for
    the derivation $Z'$ \multider{\sigma_{2}'}.
    Therefore, it suffices to prove that  $Z'$
    is reachable from  $X$ in $\Re_{SEQ}^{K}$
    through a $(K',\emptyset)$-accepting derivation in $M_{SEQ}^{K}$
    with $K'\subseteq{}K$. From lemma \ref{Lemma:From-M-To-MKPAR}, applied
    to the
    derivation   $X$ \multider{\sigma_{1}} $t$$\parallel$$Z$, there exists a $p\in{}T_{PAR}$
    such that $X$ \multiderparK{\rho} $p$$\parallel$$Z$
    with \maximalparK{\rho} = \maximal{\sigma_{1}} $\subseteq{}K$. >From the
    definition of $\Re_{SEQ}^{K}$ we obtain that  $X$
    \multiderseqK{\mu} $Z$ \onederseqK{r'} $Y.(Z')$
    with  \maximalseqK{\mu} = \maximalparK{\rho} and
    \maximalseqK{r'} = \maximal{r'} $\subseteq{}K$.
    So, \maximalseqK{\mu{}r'} $\subseteq{}K$. This concludes the proof.
\end{itemize}
\end{proof}

\begin{Lemma}\label{Lemma:From-M-To-MSEQ3}
 Let  $X\in{}Var$ and
 $X$ \multider{\sigma} be a
 $(K,K^{\omega})$-accepting infinite derivation  in $M$ from
 $X$.
Then, one of the following conditions is satisfied:
\begin{enumerate}
    \item There exists a variable  $Y\in{}Var$ reachable from $X$ in
    $\Re_{SEQ}^{K}$ through a $(K',\emptyset)$-accepting derivation in $M_{SEQ}^{K}$
    with $K'\subseteq{}K$, and there exists a derivation
    $Y$ \multiderparKomega{\rho} such that \maximalparKomega{\rho} = $K$
    and
    \maximalparKomegaInf{\rho} $\cup$ \maximalparKomegaPlus{\rho} =
    $K^{\omega}$.
    Moreover, either $\rho$ is
    infinite or $\rho$ contains some occurrence of rule in
    $\Re^{K,K^{\omega}}_{PAR}\setminus\Re_{PAR}^{K}$.
    \item There exists a variable  $Y\in{}Var$ reachable from $X$ in
    $\Re_{SEQ}^{K}$ through a $(K,\emptyset)$-accepting derivation in $M_{SEQ}^{K}$, and
    there exists a $(K,K^{\omega})$-accepting infinite derivation  in $M$ from
    $Y$.
\end{enumerate}
\end{Lemma}
\begin{proof}
It suffices to prove that, assuming that property 1 is not
satisfied, property 2 must hold. If $|K|=0$, property 2  is
obviously satisfied. So, let us assume that $|K|>0$. Let
$K=\{j_{1},\ldots,j_{|K|}\}$, and for all $p=1,\ldots,|K|$ let
$K_{p}=\{j_{1},\ldots,j_{p}\}$. Let us prove by induction on $p$
that for all $p=1,\ldots,|K|$ the following property is satisfied
(assuming that  property 1 isn't satisfied):
\begin{description}
    \item[a] There exists a variable  $Y$ reachable from $X$ in
    $\Re_{SEQ}^{K}$ through a $(K',\emptyset)$-accepting derivation in $M_{SEQ}^{K}$
    with $K_{p}\subseteq{}K'\subseteq{}K$, and
    there exists a $(K,K^{\omega})$-accepting infinite derivation  in $M$ from
    $Y$.
\end{description}
 \textbf{Base Step}: $p=1$. Considering that property 1 isn't satisfied,
 the result follows from lemma \ref{Lemma:From-M-To-MSEQ2},
 setting $i=j_{1}$.\\
 \textbf{Induction Step}: $1<p\leq{}|K|$.
>From the inductive hypothesis there exists a
$t\in{}T_{SEQ}\setminus\{\varepsilon\}$ such that $X$
\multiderseqK{\rho} $t$ with $K_{p-1}\subseteq{}$
\maximalseqK{\rho} $\subseteq{}K$, and there exists a
$(K,K^{\omega})$-accepting infinite derivation in $M$ of the form
$last(t)$ \multider{\eta}. From lemma \ref{Lemma:From-M-To-MSEQ2},
applied to the derivation $last(t)$ \multider{\eta}, and
considering that  property 1 isn't satisfied, it follows that
there exists a $\overline{t}\in{}T_{SEQ}\setminus\{\varepsilon\}$
such that $last(t)$ \multiderseqK{\overline{\rho}} $\overline{t}$
with $\{j_{p}\}\subseteq{}$\maximalseqK{\overline{\rho}}
$\subseteq{}K$, and there exists a $(K,K^{\omega})$-accepting
infinite derivation in $M$ from $last(\overline{t})$. So, we have
$X$ \multiderseqK{\rho\overline{\rho}} $t\circ{}\overline{t}$ with
$K_{p}\subseteq{}$\maximalseqK{\rho\overline{\rho}}
$\subseteq{}K$.
Therefore, setting $Y=last(\overline{t})$, we obtain the assertion.\\
>From property \textbf{a}, since $K_{|K|}=K$, the thesis follows.
\end{proof}

\subsection{Proof of Theorem \ref{Theorem:Problem2.1}}

Let us assume that $K\neq{}K^{\omega}$. Given $X\in{}Var$, we have
to prove that there exists a
 $(K,K^{\omega})$-accepting infinite derivation  in $M$ from
 $X$ if, and only if,  the following property is satisfied:
\begin{itemize}
    \item There exists a variable  $Y\in{}Var$ reachable from $X$ in
    $\Re_{SEQ}^{K}$ through a $(K',\emptyset)$-accepting derivation in $M_{SEQ}^{K}$
    with $K'\subseteq{}K$, and there exists a derivation
    $Y$ \multiderparKomega{\rho}
    such that \maximalparKomega{\rho} = $K$
    and
    \maximalparKomegaInf{\rho} $\cup$ \maximalparKomegaPlus{\rho} =
    $K^{\omega}$. Moreover, either $\rho$ is
    infinite or $\rho$ contains some occurrence of rule in
    $\Re^{K,K^{\omega}}_{PAR}\setminus\Re_{PAR}^{K}$.
\end{itemize}

($\Rightarrow$) The result follows directly from Lemma
\ref{Lemma:From-M-To-MSEQ1}.\\

($\Leftarrow$) From hypothesis we have
\begin{enumerate}
    \item $X$ \multiderseqK{\lambda} $t$ with
    $t\in{}T_{SEQ}\setminus\{\varepsilon\}$, $last(t)=Y$ and
    \maximalseqK{\lambda} $\subseteq{}K$.
    \item $Y$ \multiderparKomega{\rho}  with \maximalparKomega{\rho} = $K$
    and
    \maximalparKomegaInf{\rho} $\cup$ \maximalparKomegaPlus{\rho} =
    $K^{\omega}$. Moreover, either $\rho$ is
    infinite or $\rho$ contains some occurrence of rule in
    $\Re^{K,K^{\omega}}_{PAR}\setminus\Re_{PAR}^{K}$.
\end{enumerate}
Since $X\in{}SEQ(X)$, from condition 1 and lemma
\ref{Lemma:From-MKSEQ-To-M}, it follows that there exists a
$s\in{}T$ such that $t\in{}SEQ(s)$ and $X$ \multider{\eta} $s$
with \maximal{\eta} $\subseteq{}K$. From condition 2 and  lemma
\ref{Lemma:From-MPAR-To-M-Inf} it follows that there exists  a
 $(K,K^{\omega})$-accepting infinite derivation  in $M$
of the form $Y$ \multider{\sigma}. Since $Y\in{}SubTerms(s)$, from
proposition \ref{Prop:Subterms1} we have that $s$
\multider{\sigma}. After all, we obtain $X$ \multider{\eta} $s$
\multider{\sigma}, that is  a
 $(K,K^{\omega})$-accepting infinite derivation  in $M$ from $X$.
 This concludes the proof.

\subsection{Proof of Theorem \ref{Theorem:Problem2.2}}

Let us assume that $K=K^{\omega}$. Given $X\in{}Var$, we have to
prove that there exists a
 $(K,K^{\omega})$-accepting infinite derivation  in $M$ from
 $X$ if, and only if,  one of the following properties is satisfied:
\begin{enumerate}
    \item There exists a variable  $Y\in{}Var$ reachable from $X$ in
    $\Re_{SEQ}^{K}$ through a $(K',\emptyset)$-accepting derivation in $M_{SEQ}^{K}$
    with $K'\subseteq{}K$, and there exists a derivation
    $Y$ \multiderparKomega{\rho}
    such that \maximalparKomega{\rho} = $K$
    and
    \maximalparKomegaInf{\rho} $\cup$ \maximalparKomegaPlus{\rho} =
    $K^{\omega}$. Moreover, either $\rho$ is
    infinite or $\rho$ contains some occurrence of rule in
    $\Re^{K,K^{\omega}}_{PAR}\setminus\Re_{PAR}^{K}$.
    \item There exists a
     $(K,K^{\omega})$-accepting infinite derivation  in $M_{SEQ}^{K}$ from
     $X$.
\end{enumerate}

($\Rightarrow$) It suffices to prove that, assuming that condition
1 does not hold, condition 2 must hold. Under this hypothesis, we
show that there exists a succession of terms
  $(t_{h})_{h\in{}N}$ in
$T_{SEQ}\setminus\{\varepsilon\}$ satisfying  the following
properties:
\begin{description}
    \item[i.] $t_{0}=X$
    \item[ii.] for all $h\in{}N\quad{}last(t_{h})$
    \multiderseqK{\rho_{h}} $t_{h+1}$ with \maximalseqK{\rho_{h}} $=K$.
    \item[iii.] for all $h\in{}N$ there exists a
    $(K,K^{\omega})$-accepting infinite derivation  in $M$ from $last(t_{h})$.
    \item[iv.] for all $h\in{}N$ $last(t_{h})$ is
    reachable from $X$ in $\Re_{SEQ}^{K}$ through a $(K',\emptyset)$-accepting
    derivation in $M_{SEQ}^{K}$ with $K'\subseteq{}K$.
\end{description}
For $h=0$ properties iii and iv are satisfied,  by setting
$t_{0}=X$. So, assume the existence of a finite  sequence of terms
$t_{0},t_{1},\ldots,t_{h}$ in $T_{SEQ}\setminus\{\varepsilon\}$
sa\-ti\-sfying  properties i-iv. It suffices to prove that there
exists a term $t_{h+1}$ in $T_{SEQ}\setminus\{\varepsilon\}$
satisfying  iii and iv, and such that  $last(t_{h})$
    \multiderseqK{\rho_{h}} $t_{h+1}$ with \maximalseqK{\rho_{h}} $=K$.
>From the inductive hypothesis, $last(t_{h})$ is reachable from
    $X$ in $\Re_{SEQ}^{K}$ through a $(K',\emptyset)$-accepting
    derivation in $M_{SEQ}^{K}$ with $K'\subseteq{}K$, and there exists a $(K,K^{\omega})$-accepting
    infinite derivation in $M$ from $last(t_{h})$.
    From lemma \ref{Lemma:From-M-To-MSEQ3} applied to  variable $last(t_{h})$,
    and the fact that
     condition 1 does not hold, it follows that there exists a term
   $t\in{}T_{SEQ}\setminus\{\varepsilon\}$ such that
    $last(t_{h})$
    \multiderseqK{\rho_{h}} $t$ with \maximalseqK{\rho_{h}} $=K$, and
    there exists a $(K,K^{\omega})$-accepting infinite derivation in $M$ from
    $last(t)$. Since $last(t_{h})$ is reachable from
    $X$ in $\Re_{SEQ}^{K}$ through a $(K',\emptyset)$-accepting
    derivation in $M_{SEQ}^{K}$ with $K'\subseteq{}K$,  it follows that  $last(t)$ is reachable
    from
    $X$ in $\Re_{SEQ}^{K}$ through a $(K,\emptyset)$-accepting
    derivation in $M_{SEQ}^{K}$. Thus, setting
    $t_{h+1}=t$, we obtain the result.\\
    Let $(t_{h})_{h\in{}N}$ be the succession of terms in  $T_{SEQ}\setminus\{\varepsilon\}$
     satisfying  properties i-iv. Since in this case $|K|>0$ (remember that $|K|+|K^{\omega}|>0$),
     we have $|\rho_{h}|>0$  for all $h\in{}N$. Then, by property 1 of  proposition
     \ref{Prop:Subterms2} we obtain that for every
     $h\in{}N$
\begin{displaymath}
\textrm{$t_{h}$
    \multiderseqK{\rho_{h}} $t_{h}\circ{}t_{h+1}$ with
    \maximalseqK{\rho_{h}} = $K$}
\end{displaymath}
>From property 2 of proposition \ref{Prop:Subterms2} we have that
for all $h\in{}N$
\begin{displaymath}
\textrm{
 $t_{0}$$\circ$$t_{1}$$\circ$$\ldots$$\circ$$t_{h}$
    \multiderseqK{\rho_{h}} $t_{0}$$\circ$$t_{1}$$\circ$$\ldots$$\circ$$t_{h}$$\circ$$t_{h+1}$
     }
\end{displaymath}
Therefore,
\begin{eqnarray}
\textrm{$X=t_{0}$ \multiderseqK{\rho_{0}} $t_{0}$$\circ$$t_{1}$
\multiderseqK{\rho_{1}} $t_{0}$$\circ$$t_{1}$$\circ$$t_{2}$
\multiderseqK{\rho_{2}} $\ldots$ \multiderseqK{\rho_{h-1}}
$t_{0}$$\circ$$t_{1}$$\circ$$\ldots$$\circ$$t_{h}$}\nonumber\\
\textrm{\multiderseqK{\rho_{h}}
$t_{0}$$\circ$$t_{1}$$\circ$$\ldots$$\circ$$t_{h}$$\circ$$t_{h+1}$
\multiderseqK{\rho_{h+1}}$\ldots$}\nonumber
\end{eqnarray}
 is an  infinite derivation in
$\Re_{SEQ}^{K}$ from $X$. Setting $\delta=\rho_{0}\rho_{1}\ldots$,
from ii and proposition \ref{Prop:Maximal} we obtain that
\begin{displaymath}
\textrm{\maximalseqK{\delta} =
$\bigcup_{h\in{}N}$\maximalseqK{\rho_{h}} = $K$.}
\end{displaymath}
\begin{displaymath}
\textrm{\maximalseqInfK{\delta} =
$\bigoplus_{h\in{}N}$\maximalseqK{\rho_{h}} = $K$ = $K^{\omega}$.}
\end{displaymath}
Hence, condition 2 holds.\\

($\Leftarrow$) At first, let us assume the condition 2 holds.
Then, since $X\in{}SEQ(X)$, the result follows directly from lemma
\ref{Lemma:From-MKSEQ-To-M}. Assume that condition 1 holds
instead. Then, we reason as in the proof of theorem
\ref{Theorem:Problem2.1}.


\end{document}